\documentclass{article}
\pdfoutput=1
\usepackage{amsmath,amssymb,gensymb,color,graphicx,array,authblk,multirow,soul}
\usepackage{natbib,hyperref,aas_macros,multirow,placeins,siunitx,textcomp}
\usepackage[margin=22mm,top=10mm]{geometry}
\usepackage[utf8]{inputenc}
\usepackage{threeparttable}
\usepackage{booktabs}
\usepackage{microtype}
\newcommand{\planck}{{\it Planck}}
\newcommand{\wmap}{{\sl WMAP}}
\newcommand{\cobe}{{\sl COBE}}
\newcommand{\code}[1]{\texttt{#1}}

\DeclareFixedFont{\ttb}{T1}{txtt}{bx}{n}{9} 
\DeclareFixedFont{\ttm}{T1}{txtt}{m}{n}{9}  

\newcommand{\dfn}[1]{\textbf{#1}}
\newcommand{\img}[2][]{\resizebox{\hsize}{!}{\includegraphics[#1]{#2}}}
\newenvironment{closetabcols}[1][0.5mm]{\setlength{\tabcolsep}{#1}}{}
\newenvironment{closetabrows}[1][0.02]{\renewcommand{\arraystretch}{#1}}{}
\newcommand{\degrange}[3]{$#1^\circ < \textrm{#2} < #3^\circ$}

\title{The Atacama Cosmology Telescope: DR5 maps of 18\,000 square degrees of the microwave sky from ACT 2008-2018 data}

\author[1]{Sigurd~Naess\footnote{snaess@flatironinstitute.org}}
\author[1]{Simone~Aiola}
\author[22]{Jason~E.~Austermann}
\author[20]{Nick~Battaglia}
\author[22]{James~A.~Beall}
\author[22]{Daniel~T.~Becker}
\author[35]{Richard~J.~Bond}
\author[13]{Erminia~Calabrese}
\author[19,20]{Steve~K.~Choi}
\author[19]{Nicholas~F.~Cothard}
\author[36]{Kevin~T.~Crowley}
\author[37]{Omar Darwish}
\author[21]{Rahul~Datta}
\author[22]{Edward~V.~Denison}
\author[29]{Mark~Devlin}
\author[19]{Cody~J.~Duell}
\author[22]{Shannon~M.~Duff}
\author[3]{Adriaan~J.~Duivenvoorden}
\author[2]{Jo~Dunkley}
\author[18]{Rolando~D\"unner}
\author[22]{Anna~E.~Fox}
\author[19]{Patricio~A.~Gallardo}
\author[31]{Mark~Halpern}
\author[7,1]{Dongwon~Han}
\author[1]{Matthew~Hasselfield}
\author[24,1]{J.~Colin~Hill}
\author[22]{Gene~C.~Hilton}
\author[8]{Matt~Hilton}
\author[23]{Adam~D.~Hincks\footnote{ORCID~0000-0003-1690-6678}}
\author[23]{Ren\'ee~Hlo\v{z}ek}
\author[3]{Shuay-Pwu~Patty~Ho}
\author[22]{Johannes~Hubmayr}
\author[32]{Kevin~Huffenberger}
\author[16]{John~P.~Hughes}
\author[30]{Arthur~B.~Kosowsky}
\author[5]{Thibaut~Louis}
\author[33]{Mathew~S.~Madhavacheril}
\author[25,26,27,28]{Jeff~McMahon}
\author[15]{Kavilan~Moodley}
\author[14]{Federico~Nati}
\author[22]{John~P.~Nibarger}
\author[19,20]{Michael~D.~Niemack}
\author[3]{Lyman~Page}
\author[4]{Bruce~Partridge}
\author[17]{Maria~Salatino}
\author[11,12]{Emmanuel~Schaan}
\author[9]{Alessandro~Schillaci}
\author[29]{Benjamin~Schmitt}
\author[37]{Blake~D.~Sherwin}
\author[7]{Neelima~Sehgal}
\author[10]{Crist\'obal~Sif\'on}
\author[1,2]{David~Spergel}
\author[3]{Suzanne~Staggs}
\author[19]{Jason~Stevens}
\author[3]{Emilie~Storer}
\author[22]{Joel~N.~Ullom}
\author[22]{Leila~R.~Vale}
\author[34]{Alexander~Van~Engelen}
\author[22]{Jeff~Van~Lanen}
\author[19]{Eve~M.~Vavagiakis}
\author[6]{Edward~J.~Wollack}
\author[29]{Zhilei~Xu}

\affil[1]{Center for Computational Astrophysics, Flatiron Institute, New York, NY, USA 10010}
\affil[2]{Department of Astrophysical Sciences, Peyton Hall, Princeton University, Princeton, NJ, USA 08544}
\affil[3]{Joseph Henry Laboratories of Physics, Jadwin Hall, Princeton University, Princeton, NJ, USA 08544}
\affil[4]{Department of Physics and Astronomy, Haverford College, Haverford, PA, USA 19041}
\affil[5]{Laboratoire de l'Acc\'el\'erateur Lin\'eaire, Univ. Paris-Sud, CNRS/IN2P3, Universit\'e Paris-Saclay, Orsay, France}
\affil[6]{NASA/Goddard Space Flight Center, Greenbelt, MD, USA 20771}
\affil[7]{Physics and Astronomy Department, Stony Brook University, Stony Brook, NY 11794}
\affil[8]{Astrophysics Research Centre, University of KwaZulu-Natal, Westville Campus, Durban 4041, South Africa}
\affil[9]{Department of Physics, California Institute of Technology, Pasadena, California 91125, USA}
\affil[10]{Instituto de Física, Pontificia Universidad Católica de Valparaíso, Chile}
\affil[11]{Lawrence Berkeley National Laboratory, Berkeley, California 94720, USA}
\affil[12]{Berkeley Center for Cosmological Physics, University of California, Berkeley, California 94720, USA}
\affil[13]{School of Physics and Astronomy, Cardiff University, The Parade, Cardiff, Wales, UK CF24 3AA}
\affil[14]{Department of Physics, University of Milano-Bicocca, Piazza della Scienza, 3 - 20126 Milano (MI), Italy}
\affil[15]{Astrophysics Research Centre and School of Mathematics, Statistics and Computer Science, University of KwaZulu-Natal, Durban 4041, South Africa}
\affil[16]{Department of Physics and Astronomy, Rutgers, The State University of New Jersey, Piscataway, NJ USA 08854-8019}
\affil[17]{Physics Department, Stanford University Kavli Institute for Particle Astrophysics and Cosmology (KIPAC) Stanford, California CA}
\affil[18]{Instituto de Astrof\'isica and Centro de Astro-Ingenier\'ia, Facultad de F\'isica, Pontificia Universidad Cat\'olica de Chile, Av. Vicu\~na Mackenna 4860, 7820436, Macul, Santiago, Chile}
\affil[19]{Department of Physics, Cornell University, Ithaca, NY 14853, USA}
\affil[20]{Department of Astronomy, Cornell University, Ithaca, NY 14853, USA}
\affil[21]{Dept. of Physics and Astronomy, The Johns Hopkins University, 3400 N. Charles St., Baltimore, MD, USA 21218}
\affil[22]{NIST Quantum Devices Group, 325 Broadway Mailcode 817.03, Boulder, CO, USA 80305}
\affil[23]{Department of Astronomy and Astrophysics, University of Toronto, 50 St. George Street, Toronto, ON M5S 3H4, Canada}
\affil[24]{Department of Physics, Columbia University, New York, NY, USA 10027}
\affil[25]{Kavli Institute for Cosmological Physics, University of Chicago, Chicago, IL 60637, USA}
\affil[26]{Department of Astronomy and Astrophysics, University of Chicago, Chicago, IL 60637, USA}
\affil[27]{Department of Physics, University of Chicago, Chicago, IL 60637, USA}
\affil[28]{Enrico Fermi Institute, University of Chicago, Chicago, IL 60637, USA}
\affil[29]{Department of Physics and Astronomy, University of Pennsylvania, 209 South 33rd Street, Philadelphia, PA, USA 19104}
\affil[30]{Department of Physics and Astronomy, University of Pittsburgh, Pittsburgh, PA, USA 15260}
\affil[31]{Department of Physics and Astronomy, University of British Columbia, Vancouver, BC, Canada V6T 1Z4}
\affil[32]{Department of Physics, Florida State University, Tallahassee FL, USA 32306}
\affil[33]{Centre for the Universe, Perimeter Institute for Theoretical Physics, Waterloo, ON, Canada N2L 2Y5}
\affil[34]{School of Earth and Space Exploration, Arizona State University, Tempe, AZ, USA 85287}
\affil[35]{Canadian Institute for Theoretical Astrophysics, 60 St. George Street, University of Toronto, Toronto, ON, M5S 3H8, Canada}
\affil[36]{Department of Physics, University of California Berkeley, Berkeley, CA 94720}
\affil[37]{Department of Applied Mathematics and Theoretical Physics, University of Cambridge, Wilberforce Road, Cambridge CB3 0WA, UK}

\begin{document}

\maketitle
\begin{abstract}
This paper presents a maximum-likelihood algorithm for combining sky maps
with disparate sky coverage, angular resolution and spatially varying
anisotropic noise into a single map of the sky. We use this to merge hundreds of
individual maps covering the 2008--2018 ACT observing seasons, resulting in
by far the deepest ACT maps released so far. We also combine the maps with
the full \planck\ maps, resulting in maps that have the best features of both \planck\
and ACT: \planck's nearly white noise on intermediate and large angular
scales and ACT's high-resolution and sensitivity on small angular scales.
The maps cover over 18\,000 square degrees, nearly half the full sky, at
100, 150 and 220 GHz. They reveal 4\,000 optically-confirmed clusters
through the Sunyaev Zel'dovich effect (SZ) and 18\,500 point source candidates
at $> 5\sigma$, the largest single collection of SZ clusters and millimeter wave
sources to date. The multi-frequency maps provide millimeter images of nearby
galaxies and individual Milky Way nebulae, and even clear detections of several
nearby stars.  Other anticipated uses of these maps include, for example,
thermal SZ and kinematic SZ cluster stacking, CMB cluster lensing and galactic
dust science. The method itself has negligible bias. However, due to the preliminary
nature of some of the component data sets, we caution that these maps should not
be used for precision cosmological analysis. The maps are part of ACT DR5, and
are available on LAMBDA at \url{https://lambda.gsfc.nasa.gov/product/act/actpol_prod_table.cfm}.
There is also a web atlas at \url{https://phy-act1.princeton.edu/public/snaess/actpol/dr5/atlas}.
\end{abstract}
\filbreak

\section{Introduction}
\FloatBarrier
\begin{figure}[htp]
	\centering
	\begin{closetabrows}[0.6]
	\begin{closetabcols}
		\begin{tabular}{>{\centering\arraybackslash}m{2mm}>{\centering\arraybackslash}m{5cm}>{\centering\arraybackslash}m{5cm}>{\centering\arraybackslash}m{5cm}}
			& \bf \planck & \bf ACT+\planck & \bf ACT \\
			\rotatebox[origin=c]{90}{\bf f090} & \img{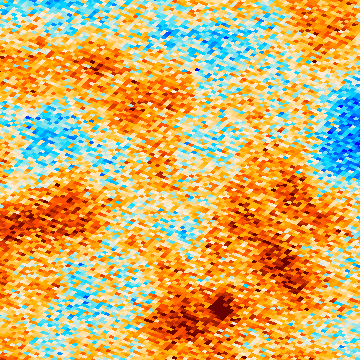} & \img{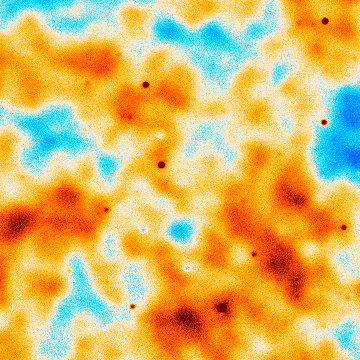} & \img{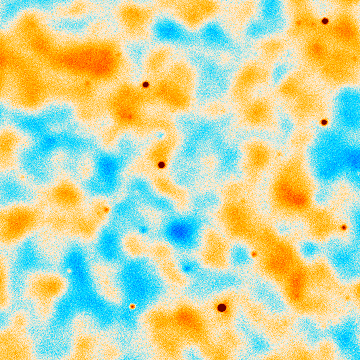} \\
			\rotatebox[origin=c]{90}{\bf f150} & \img{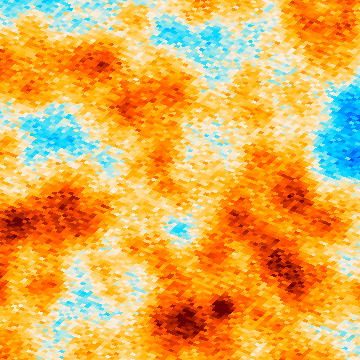} & \img{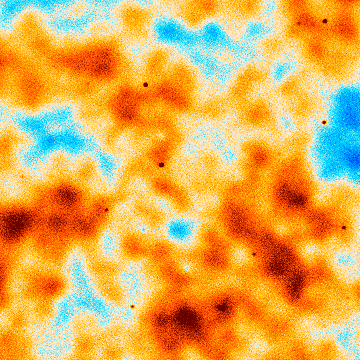} & \img{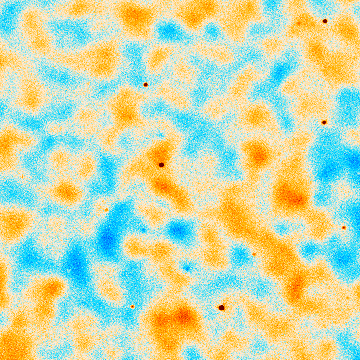} \\
			\rotatebox[origin=c]{90}{\bf f220} & \img{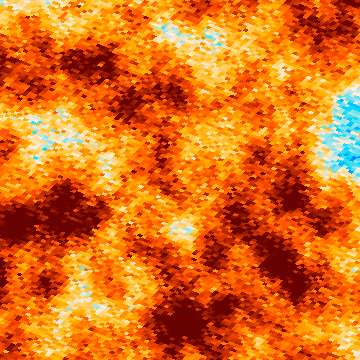} & \img{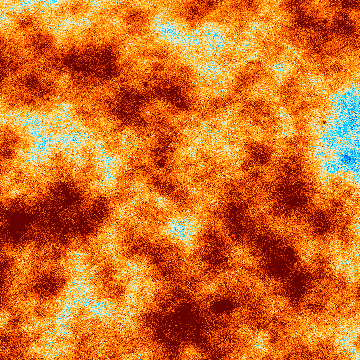} & \img{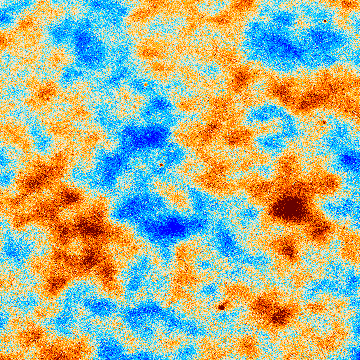}
		\end{tabular}
	\end{closetabcols}
	\end{closetabrows}
	\caption{Comparison of \planck, ACT+\planck\ and ACT-only in a $3^\circ\times3^\circ$ patch centered on
	RA = 231.5$^\circ$, dec = 16.5$^\circ$. The map of this region includes ACT daytime data.
	The ACT map depths in this region are 8/8/30 \textmu{}K-arcmin at f090/f150/f220 (see figure~\ref{fig:bands}
	for band definitions). ACT+\planck\ is a substantial improvement over \planck\ alone, both in
	resolution and depth, and captures the larger scales that ACT alone has trouble measuring.
	See figure~\ref{fig:matched-filter} for an image filtered to emphasize the point sources, clusters
	and other small-scale features.}
	\label{fig:papa-small}
\end{figure}

Over the past three decades, cosmologists have been mapping the microwave sky
with increasing precision. \cobe\ \citep{bennett/etal:1994}, \wmap\
\citep{bennett/etal:2003,bennett/etal:2013} and \planck\
\citep{planck_mission/2013,Planck-overview:2018}
have produced
multi-frequency maps with nearly white (spatially uncorrelated) noise
and with increasing angular resolution. These observations from space are complemented
by measurements with ground based telescopes that have higher sensitivity
and potentially higher resolution (e.g. SPT \citep{spt3g-2014} and ACT \citep{thornton/2016}),
but suffer from large atmospheric noise
contamination with complicated covariances at large scales.

This paper presents an approach to building tiled coadds of these heterogeneous maps.
Section \S\ref{sec:dataset} describes the \planck\ and ACT data used in constructing the heterogeneous temperature and polarization maps used in the tiled coadding process. In this analysis, we use a larger set of the ACT data than used in \citet{aiola/etal:2020} and \citet{choi/etal:2020}, including data from the 2017 and 2018 seasons as well as data from daytime observations.  Section \S\ref{sec:coadd} describes the algorithms for coadding maps.  While this paper describes applying these algorithms to the ACT and \planck\ data, our approach of producing
tiled  coadds of heterogeneous maps  can be generalized to other data sets with complex noise properties.

Section \S\ref{sec:tiles} introduces the tiled constant correlation noise model
used to approximate the noise properties of the maps. Section
\S\ref{sec:constructing} describes how individual tiled coadded maps are
calculated and then assembled into a full map. Section
\S\ref{sec:map_properties} describes the properties of the coadded maps.
Section \S\ref{sec:map_features} reveals the maps and focuses on images of
the
astronomical objects revealed by the maps, and concludes by
noting the maps' limitations.

Previous related work includes, for example, \citet{crawford/etal:2016} and \citet{chown/etal:2018} (SPT+\planck);
and \citet{pact-2019} and \citet{mat-ymap-2019} (ACT+\planck). The
methods in this paper differ in that a) we model the spatial dependence and stripiness of the noise;
b) we preserve the full resolution of the input maps; and c) we apply them to maps covering seven times the
sky area.

The main product of this paper is a set of coadded maps.
Figure~\ref{fig:papa-small} shows a small piece of the maps and provides a taste of the full data product.  The figure compares \planck\ to the ACT+\planck\ and ACT-only coadds in total intensity
in a $3^\circ\times3^\circ$ sub-patch,
showing both the much greater depth and
resolution ACT provides and the large-scale information \planck\ adds. Only one of the $\sim 20$
point sources (red dots) we see in ACT panels can be seen in \planck, and none of the clusters (blue dots,
three visible without filtering).

\begin{figure}[ht]
	\centering
	\includegraphics[width=0.8\textwidth,trim=5mm 5mm 5mm 3mm]{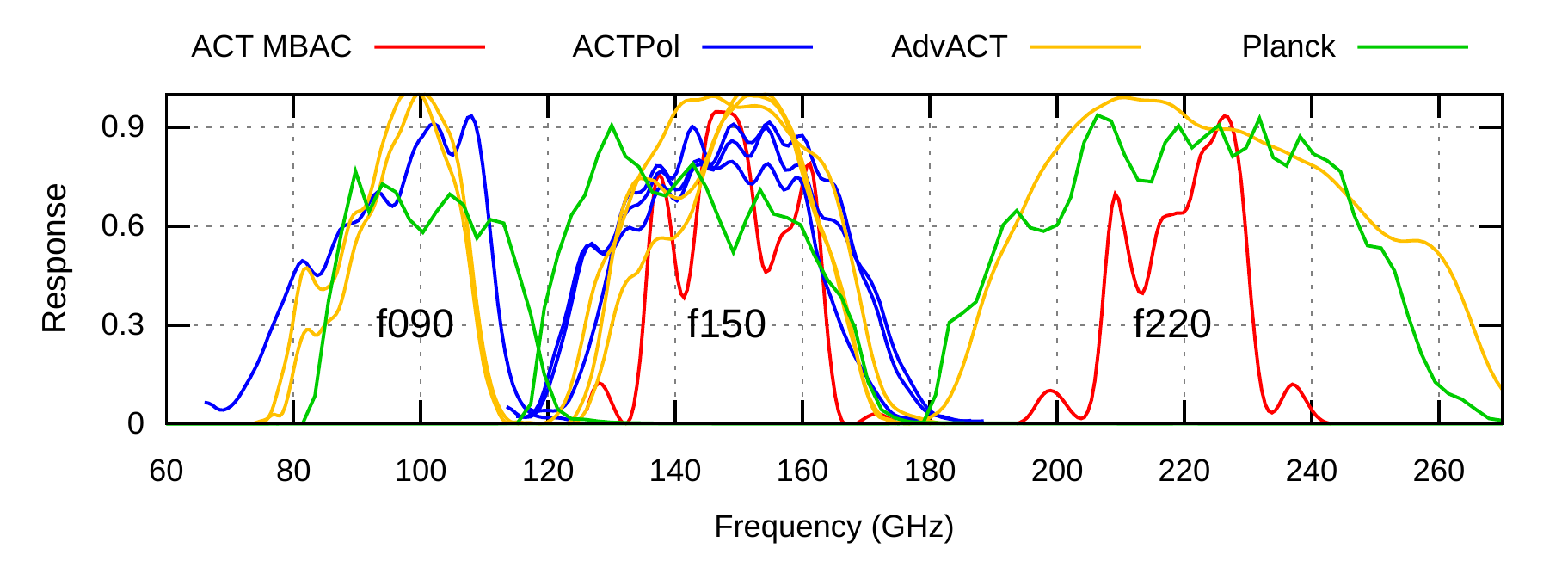}
	\caption{Comparison of the ACT and \planck\ bandpasses in the
	range 60~GHz to 270~GHz. (See Section~\ref{sec:act_data} for description of the three ACT cameras:  MBAC, ACTPol and Advanced ACTPol.) They fall into three groups, centered at roughly 90~GHz, 150~GHz and
	220~GHz. We label these bandpass-groups {\it f090}, {\it f150} and {\it f220}. In this paper we approximate all bandpasses in
	each group as being equivalent. This results in a small scale-dependence of the effective
	band center for non-CMB parts of the combined maps -- see section~\ref{sec:bandpasses}.}
	\label{fig:bands}
\end{figure}

For each of the frequency bands f090, f150 and f220, centered at roughly 98 GHz, 150 GHz and 224 GHz
(but see figure~\ref{fig:bands}), we provide all combinations of
ACT-only and ACT+\planck; day+night and night-only; and normal and source-free maps, each
of which is a (43200,10320,3) FITS image containing Stokes parameters I, Q and U in single
precision, for a total of 5.0 GB each (see appendix~\ref{sec:data-release}).
We also provide \planck\ maps reprojected to the same
pixelization, as well as inverse white noise variance maps.

The full FITS images are available at \url{https://lambda.gsfc.nasa.gov/product/act/actpol_prod_table.cfm}
(``DR5 2008-2018 Coadd maps''). A browser-based pannable, zoomable visualization of the
maps can be found at \url{https://phy-act1.princeton.edu/public/snaess/actpol/dr5/atlas}.

\section{The data}
\label{sec:dataset}
The combined sky maps presented in this article build on the data described in this section.
In total, we have 78 {\it data sets} -- sets of maps covering the same area with the same
beam and noise properties, but made from independent {\it splits} (subsets) of the data; e.g. half-mission
maps in the case of \planck \footnote{In ACT, splits are built from data taken on different days. Since
ACT does not recover signal on time-scales longer than a few minutes, this ensures that the maximum
correlation between the noise on two consecutive days is $O(10^{-3})$, and zero for non-consecutive days.
This makes the splits statistically independent for all practical purposes.}. The splits in each data set
are used to build the noise model in section~\ref{sec:noise}. In total
we have 276 such split maps, taking up a combined 260 GB for the signal
maps and 136 GB for the corresponding {\it inverse variance maps}, which are estimates of
the level of {\it uncorrelated} noise in each pixel.
With the exception of the unpolarized ACT-MBAC data (see section~\ref{sec:act_data}), each map
consists of three fields, one for each of the Stokes parameters, I (here called T), Q and U,
for a total of 748 individual fields.\footnote{Only the T component is stored in the inverse variance maps for data sets where polarization is the standard factor of two lower inverse variance than T.}
The data sets are tabulated in table~\ref{ta:data} and further described below.

These data sets fall into two categories: 1. Mature maps that are already properly calibrated.
This includes the \planck maps, the old ACT MBAC maps, and the ACT DR4 maps. The calibration of
these products is described in their corresponding papers. 2. Preliminary maps, i.e. ``Prelimnary
Advanced ACTPol'' and ``Preliminary ACT daytime data''. These did not have mature calibration and
so some work had to be done to calibrate them. This procedure is described in sections \ref{sec:advact-dataset}
and \ref{sec:daybeam} (see also figure \ref{fig:daybeam}).

\begin{table}[htp]
\begin{center}
\begin{threeparttable}
\caption{Data descriptions}
\begin{tabular}{llr!{--}rr!{--}rlrr}
	\toprule
	\bf Survey & \bf Patch & \multicolumn{2}{c}{\bf RA ($^\circ$)} & \multicolumn{2}{c}{\bf dec ($^\circ$)} & {\bf Datasets}\,\tnote{1} & {\bf nsplit}\,\tnote{2} & {\bf total}\,\tnote{3} \\
	\midrule
	\planck & \planck & 0 & 360 & -90 & 90 & \planck\ f090+f150+f220 & 2 & 6 \\
	ACT-MBAC & South  & -114 & 147 & -57  & -48 & AR1+AR2 2008--2010 & 4 & 24 \\
	ACT-MBAC & Equ    & -250 &  65 & -2.4 & 2.6 & AR1+AR2 2009--2010 & 4 & 16\\
	ACT DR4 & D1  & 140 & 161 &  -5 &   6 & PA1 2013 & 4 & 4 \\
	ACT DR4 & D5  & -19 &  13 &  -7 &   6 & PA1 2013 & 4 & 4 \\
	ACT DR4 & D6  &  19 &  48 & -11 &   1 & PA1 2013 & 4 & 4 \\
	ACT DR4 & D56 & -23 &  54 & -10 &   7 & PA1+PA2 2014--2015, PA3 2015 & 4 & 24 \\
	ACT DR4 & D8  & -12 &  18 & -52 & -32 & PA1+PA2+PA3 2015 & 4 & 16 \\
	ACT DR4 & BN  & 102 & 257 &  -7 &  22 & PA1+PA2+PA3 2015 & 4 & 16 \\
	ACT DR4 & AA  &   0 & 360 & -62 &  22 & PA2+PA3 2016 & 2 & 6 \\
	AdvACT  & AA  &   0 & 360 & -62 &  22 & PA4+PA5+PA6 2017--2018 & 2 & 24\\
	ACT day & BN     & 102 & 257 &  -7 &  22 & PA1+PA2 2014--2015, PA3 2015 & 4 & 24 \\
	ACT day & Day-N   & 162 & 258 &   3 &  20 & PA2+PA3 2016, PA4+PA5+PA6 2017--2018 & 4 & 60 \\
	ACT day & Day-S   & -25 &  60 & -52 & -29 & PA4+PA5+PA6 2017--2018 & 4 & 48\\
\bottomrule
\end{tabular}
\label{ta:data}
\begin{tablenotes}
\item [1] Naming conventions for bands and arrays are in Section~\ref{sec:dataset}
\item [2] The number of splits for each of the listed data sets
\item [3] The total number of maps, combining frequencies, years and splits
\end{tablenotes}
\end{threeparttable}
\end{center}
\end{table}

\subsection{\planck}

In this paper, we use \planck\ HFI maps measured at close to the same central frequencies as the ACT maps:  the \planck\ 100 GHz map for f090, 143 GHz for f150 and 217 GHz for f220. See figure~\ref{fig:bands} for
a comparison of the \planck\ and ACT bandpasses.
We use a hybrid of the 2015 (PR2,  \cite{PLANCK-DR2}) and 2018 (PR3, \cite{Planck-overview:2018} \planck\ data releases, extracting the  temperature (T) maps
from 2015 and the polarization ($P$) maps from 2018. As in \citet{mat-ymap-2019},
we refrain from using the 2018 release maps as the effective intensity bandpasses become
component-dependent due to the \planck\ polarization systematics-cleaning procedures.
However, since we do not account for the detailed bandpasses inside each band in this analysis
it would have made little difference to use the 2018 \planck\ maps for both T and P (polarization).
Because of this hybridization,
the combined maps should not be used for $TE$ cross-spectrum analysis at scales with significant
\planck\ contribution (see figure~\ref{fig:actfrac}).

The maps were transformed from HEALPix $N_\textrm{side} = 2048$ maps\footnote{https://healpix.jpl.nasa.gov} \citep{healpix}
to the same 0.5 arcmin resolution Plate Carreé (CAR) projection we use for the other maps.
For the sky maps themselves this was done by performing a Spherical Harmonics Transform (SHT)
to get the multipole coefficients $a_{lm}$, rotating these from galactic to equatorial coordinates
using the \texttt{rotate\_alm} function from \texttt{healpy}, followed by an inverse SHT onto
the target CAR pixels using the \texttt{libsharp} \citep{libsharp} wrapper in \texttt{pixell}\footnote{https://github.com/simonsobs/pixell}. This harmonic
reprojection preserves power on all scales in the input map up to its band limit of about $\ell=6100$,
beyond which the map has no power. Section \ref{sec:reprojection} describes our treatment of the \planck\ maps
beyond that scale.

For the inverse
variance we simply used direct nearest-neighbor lookups
at the HEALPix pixels corresponding to the coordinates of the CAR pixels, including a correction
for the change in pixel size. This simpler interpolation scheme was used because inverse variance
maps should be strictly non-negative, and more accurate interpolation schemes like harmonic or
spline interpolation can introduce faint negative values at the boundaries of the exposed area
in the form of ``ringing''. In any case, the  inverse variance maps do not
change quickly enough to make the sub-pixel accuracy these methods provide necessary.
With two half-mission maps for splits, we use a total of six maps, as seen in table~\ref{ta:data}.

\subsection{Atacama Cosmology Telescope}
\label{sec:act_data}

The Atacama Cosmology Telescope (ACT), described in \citet{fowler/etal:2007},
observes the millimeter-wave sky from northern Chile with arc-minute resolution.  Its
primary goal is to make maps of the CMB temperature anisotropy and polarization
at angular scales and sensitivities that complement those of the {\sl WMAP} and
{\sl Planck} satellites.   ACT is a 6~m off-axis aplanatic Gregorian telescope
that scans in azimuth as the sky drifts through the field of view. There have
been three generations of receivers: MBAC \citep{swetz/etal/2011} which
observed at 150, 220, and 277 GHz;  ACT's first polarization-sensitive receiver,
ACTPol \citep{thornton/2016}, which observed at 90 GHz and 150 GHz; and the
Advanced ACTPol (AdvACT) receiver which is currently configured with detector
arrays at 30, 40, 90, 150, and 220 GHz. ACT has had a series of data releases (DR), described below.

\subsubsection{ACT-MBAC}

ACT-MBAC consists of data taken from 2008 to 2010 with the
polarization-insensitive MBAC
comprising three detector arrays \citep{swetz/etal/2011}: AR1 (f150),
AR2 (f220) and AR3 (f280). DR1  covered a southern region (``South," centered on RA$=60^\circ$ dec$=-52.7^\circ$) in 2008 at 148 GHz \citep[e.g.,][]{dunner/etal:2013, dunkley/etal:2011}.  DR2  covered the South and the SDSS stripe 82 equatorial region (``Equ")  in 2008--2010, and added 217 GHz and 277 GHz  \citep[e.g.,][]{sievers/etal:2013,das/etal:2011,gralla/etal:2019}.   Only the first two (f150 and f220) were used in this analysis because
no f280 data are available in later ACT seasons.
The ACT-MBAC data for the two regions yield a total of 40 maps\footnote{
These are counted by summing up the number of splits for each array for each year. For example, for
MBAC South, there are three years (2008--2010), with two arrays active, and four splits each, for a total of
$3\cdot 2\cdot 4 = 24$ maps.
}. These were downloaded from the MBAC directory on LAMBDA\footnote{http://lambda.gsfc.nasa.gov}. For each map we also
need its white noise inverse variance per pixel. MBAC did not release these, but provided hitcount maps
that are proportional to the inverse variance per pixel to good accuracy. We determined the factor of
proportionality from the observed small-scale variance in the maps.\footnote{
"This constant of proportionality varies slightly by year, but was typically about 2500/pK$^2$
at f150 and 1200/pK$^2$ at f220. Given the 400 Hz MBAC sample rate, this can be reintepreted as
mean per-detector sensitivities of about 1000 \textmu{}K$\sqrt{\textrm{s}}$ at f150 and
1450 \textmu{}K$\sqrt{\textrm{s}}$ at f220.}

The MBAC maps come in a slightly different pixelization than the ACTPol and AdvACT maps.
The former are in cylindrical equal-area (CEA) with 0.495 arcminute pixels conformal on dec = $0^\circ$
for the Equ patch and $-69^\circ$ for the South patch, while the latter are in 0.5 arcminute
Plate Carreé (CAR) conformal on the equator. For this analysis we standardize on the latter, so all MBAC
maps were repixelized to CAR using bicubic spline interpolation, including an area rescaling
correction for the inverse variance maps. This interpolation, which only applies to the MBAC data set,
introduces a small transfer function which we handle in section~\ref{sec:reprojection}.

\subsubsection{ACT DR4}

ACT data release 4 (DR4) consists of data taken from 2013 to 2016 with the polarization-sensitive
ACTPol camera consisting of three polarized-detector arrays (PAs): PA1 (f150), PA2 (f150) and the dichroic PA3 (f090, f150).
These separate arrays of NIST-fabricated MoCu TES detectors \citep{grace/etal:2014,Datta_2016} are each contained in a separate ``optics tube" with its own set of filters and lenses. PA3, added in the 2015 season (s15), is dichroic, which means it simultaneously measures polarizations in the f090 and f150 frequency bands
at the output of one feed horn.  \citet{aiola/etal:2020} and \citet{choi/etal:2020} use the DR4 data in their analyses.
\citet{aiola/etal:2020} also describes the DR4 map-making methodology. ACT DR4 covers seven patches, described in table~\ref{ta:data}, for a total of 74 maps.

\subsubsection{Preliminary Advanced ACTPol}
\label{sec:advact-dataset}
The preliminary AdvACT data used here were taken from 2017 to 2018 with the polarization-sensitive
Advanced ACTPol camera consisting of three detector arrays, all of which are dichroic: PA4 (f150, f220),
PA5 (f090, f150) and PA6 (f090, f150). The Advanced ACT data will eventually include coverage at
five frequency bands from 28 to 230 GHz \citep{henderson/etal:2016}.  This data set covers just a single patch (AA, which was also observed by ACTPol), for a total of 24 maps.

These maps were made using the same methodology as ACT DR4,
but have not yet been well enough tested to use for cosmological analyses. They should overall be of good quality, though with the following
caveats.
\begin{enumerate}
	\item The polarized near sidelobes were not subtracted. These are
individually weak (-35 dB or less) approximately beam-shaped sidelobes with near 100\% T$\rightarrow$P
leakage and an integrated power of up to 0.3\% of the main beam. They are offset by roughly 0.5$^\circ$
from the main beam. The main effect of leaving these in is a slight excess of $TE$ for $\ell \lesssim 1000$.  See the discussion of the beams in
\citet{choi/etal:2020}.
	\item The Conjugate Gradient iteration to find the Maximum-Likelihood maps was not run all the
way to the end, but stopped after 300 steps to avoid impacting the computer time available for
the main DR4 analysis. Additionally, we did not compensate for the bias introduced
by applying the noise model to the same data it was measured from.\footnote{This is normally done
in a multi-pass procedure where the noise model of each pass is estimated from data where the
best estimate of the sky from the previous pass has been subtracted.} Together, these omissions
effectively introduce a gentle low-pass filter for $\ell \lesssim 750/1200/1750$ at f090/f150/f220
for the T maps, with a much smaller impact on polarization.
	\item The 2018 data were mapped with detector response time constants fixed at 1~ms instead
of the proper per-detector values, as these had not been measured yet. The effect of this is
a very slight beam broadening which was measured using the same techniques as for the
daytime beams described in the next section.
	\item Due to the lack of mature gain calibration, the gain of each map was estimated by
fitting the model $[C_\ell^{PP}, C_\ell^{AP}, C_\ell^{AA}] =
[1,g,g^2] C_\ell (1 + (\ell/\ell_\textrm{knee})^{-\alpha}) + [\beta_{PP},0,\beta_{AA}]$
to the \planck--\planck\ ($C_\ell^{PP}$), ACT-\planck\ ($C_\ell^{AP}$) and ACT-ACT ($C_\ell^{AA}$)
split TT cross-spectra. Here $g$ is the ACT gain deficiency relative to \planck, which we
measure while marginalizing over the nuisance parameters $C_\ell$ (the sky angular power spectrum),
$\ell_\textrm{knee}$ and $\alpha$ (the shape of the low-$\ell$ lack of power (see item 1),
and $\beta_{PP}$ and $\beta_{AA}$ (Poisson tails\footnote{
	We do not include a Poisson tail for the ACT-\planck\ cross-spectrum because having three free
	Poisson amplitudes for three observed spectra would cause a degeneracy with the $C_\ell$
	parameters. We allow separate Poisson power for ACT-ACT and \planck-\planck\  to account for
	point source variability and differences in ACT and \planck's very different beams' interaction
	with the galactic and point source mask.} -- the high-$\ell$ contribution to the angular
power spectra caused by unsubtracted point sources).
\end{enumerate}
Because of these limitations, the maps produced in this paper should not be used for precision cosmological analyses.

\subsubsection{Preliminary ACT daytime data} \label{sec:daybeam}
The preliminary ACT daytime
data were collected during the day from
2014--2018 with ACTPol and AdvACT PAs. These data are  challenging to work with due to the time-dependent deformation
of the telescope mirror caused by the Sun's heat, which results in large pointing offsets
and beam deformations that change on time scales of hours. These data have not yet been used in cosmological analyses.

We correct the pointing offsets using the same method as we use for night-time
data in DR4: by measuring the observed positions of bright quasars that fall
within each 10-minute chunk of CMB observations.

The beam deformation is much more difficult to measure and correct for. Its time-variability
results in a position-dependent effective beam in the maps, though repeated exposures
reduce this effect somewhat. In this analysis we measure only the {\it average} beam
for each patch by cross-correlating night-time and daytime observations while masking
out the brightest point sources to reduce bias from point source variability.

\begin{figure}[h]
	\centering
	\hspace*{-2mm}\begin{tabular}{cc}
		\includegraphics[height=4.4cm,trim=10mm 3mm 3mm 6mm,clip]{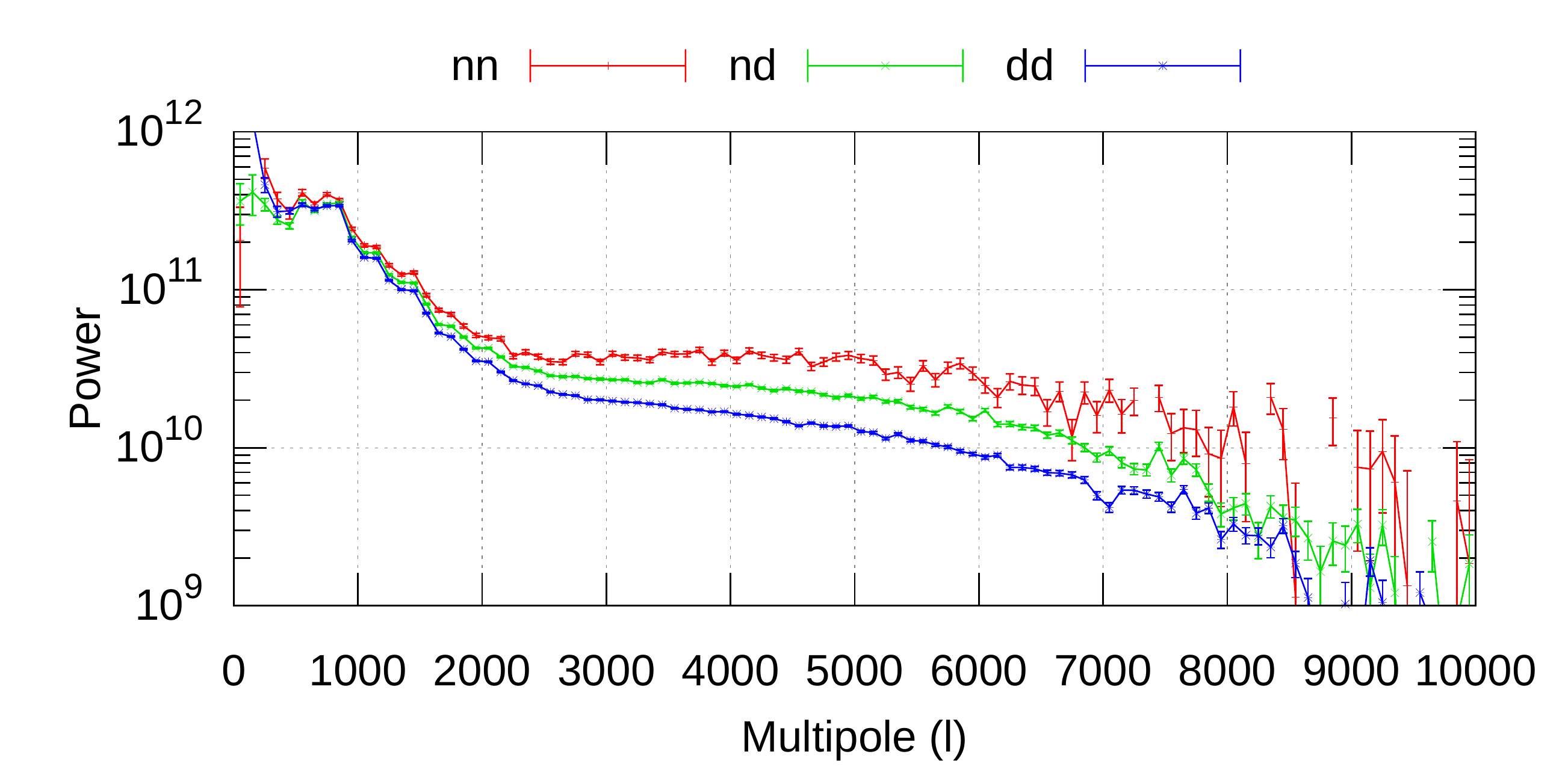} &
		\includegraphics[height=4.2cm,trim=20mm 3mm 5mm 0]{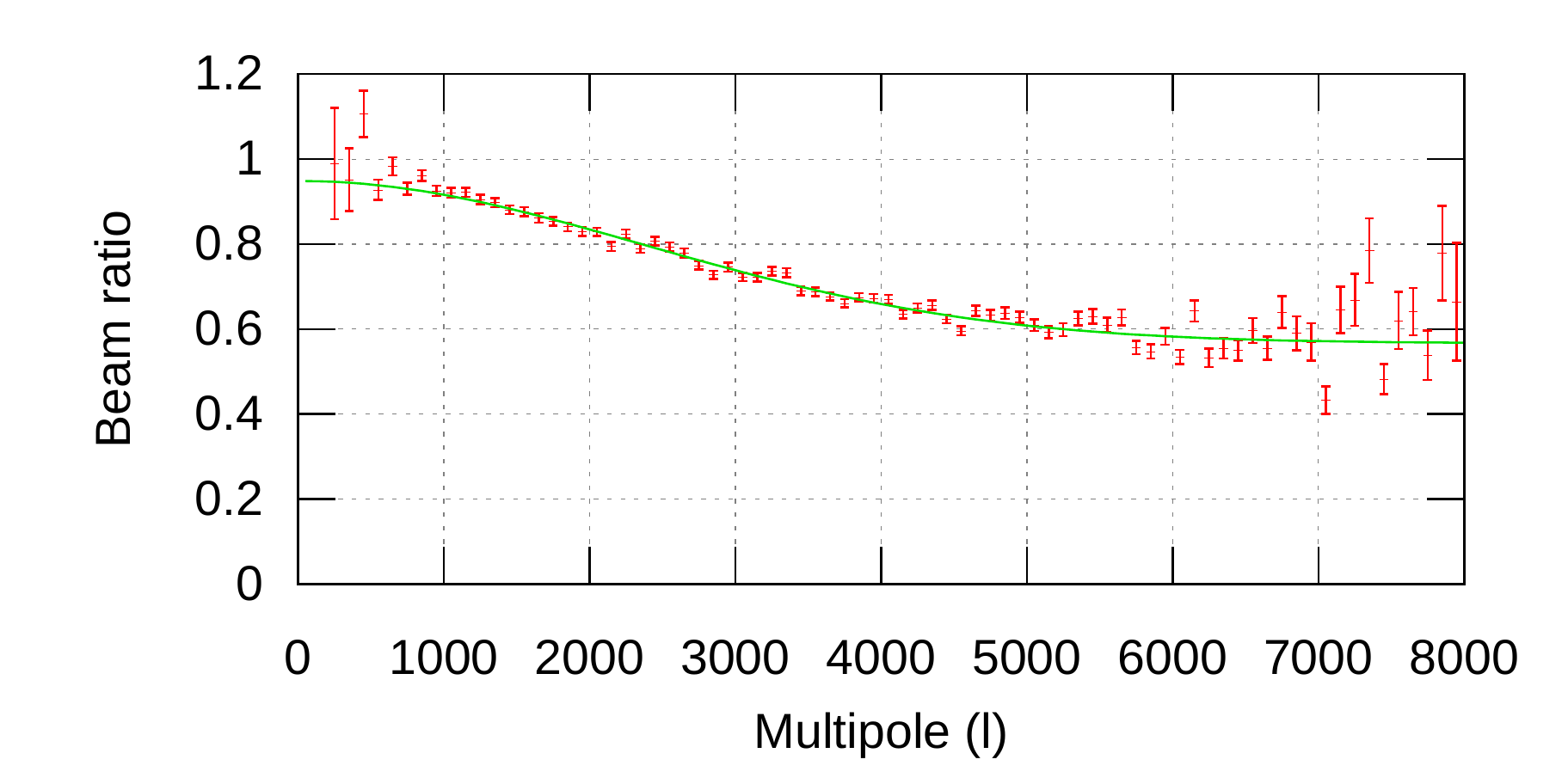}
	\end{tabular}
	\caption{\dfn{Left}: Night-night (red), night-day (green) and day-day (blue) cross pseudo-spectra
	as a function of multipole
	for s18 PA5 f090 for the Day-N patch. The daytime loss in power on small scales is clearly visible.
	\dfn{Right}: The day/night relative beam inferred from the three spectra in each bin (red), and the
	smooth three-parameter model fit to it (green). See section~\ref{sec:daybeam} for details.}
	\label{fig:daybeam}
\end{figure}

Given the
daytime ($d$) and night-time ($n$) maps for a given patch, array and frequency, we compute
the day-to-night ratio $\alpha$ in bins of multipoles as the maximum-likelihood
solution of the equation
\begin{align}
	[\textrm{Cross}(n,n), \textrm{Cross}(n,d), \textrm{Cross}(d,d)] &= [1, \alpha, \alpha^2]P + \textrm{noise},
\end{align}
while marginalizing over the unknown noise-free signal power $P$ in each bin,
with Cross being a noise-bias free covariance estimator based on map splits. This results in a noisy
set of $\alpha$ values, one for each bin, to which we fit a smooth, three-parameter model:
$\beta(\ell) = A + (B-A)\exp(-\frac12 \ell^2\sigma^2)$.
The parameters represent a daytime mean gain error
($A$), a high-$\ell$ loss of power (B) and a Gaussian transition between the two ($\sigma$).
This resulted in a better fit than the more physically motivated two-parameter model with
$A=0$, and avoids predicting overly large ratios between the day and night beam at
high multipoles ($\ell > 10\,000$) where the error bars are too large for a good measurement.
Figure~\ref{fig:daybeam} shows an example of these fits.

This approximate
treatment of the daytime beam means that this data collection
is of lower quality than the others. However, due to its high depth over
a moderately large area it is still valuable for use cases that can accept
O(10\%) beam errors. The combined maps presented in this paper therefore come in two
variants - night-only and day+night.
The preliminary ACT daytime data cover three patches\footnote{Day-S actually
extends all the way to RA = $95^\circ$, but the area above $60^\circ$ was cut due to the poor quality of its daytime beams.} as described in table~\ref{ta:data}, for a total of 132 maps.

\section{Co-adding maps}
\label{sec:coadd}
In principle, optimally coadding a set of maps is straightforward. We model
the maps as noisy, transformed versions of a single underlying sky, $m$:
\begin{align}
d &= Pm + n,
\label{eq:model}
\end{align}
where $d$ is a column vector containing the pixels from all the observed
maps, $P$ is a response matrix that can encode beams, pixel windows or frequency
differences, $m$ is a column vector containing the sky signal sampled at each pixel,
and $n$ is the map noise, which we assume to be Gaussian with
covariance $N$. Here, we will
assume that each map has independent noise and that the only response
difference we need to worry about is the beam. This results in the
block-diagonal equation set:
\begin{align}
	\begin{pmatrix} m_0 \\ m_1 \\ \vdots\end{pmatrix} &=
	\begin{pmatrix} B_0 \\ B_1 \\ \vdots\end{pmatrix} B_\textrm{out}^{-1} m + n, &
	N &= \begin{pmatrix}N_1 & 0 & \cdots\\ 0 & N_2 & \cdots\\ \vdots & \vdots & \ddots\end{pmatrix},
\end{align}
where $B_i$ is the beam of map $m_i$, and $B_\textrm{out}$ is the beam we want
the output map $m$ to have.\footnote{In principle any output beam size could be chosen,
but an output beam significantly smaller than the smallest input beams would result in
an output map with very high noise at high $\ell$.}
The maximum-likelihood solution $\hat m$ to this equation system
is given by:
\begin{align}
	(P^T N^{-1} P) \hat m &= P^T N^{-1} d,
\end{align}
or equivalently
\begin{align}
	\sum_i \bar B_i^T N_i^{-1} \bar B_i \hat m &= \sum_i \bar B_i^T N_i^{-1} m_i,
	\label{eq:maxlik}
\end{align}
where the relative beam is defined as $\bar B_i = B_i B_\textrm{out}^{-1}$.
In the next subsections, we describe the approximation used to estimate
$B_i$ (\S \ref{sec:beam}) and $N_i^{-1}$ (\S \ref{sec:noise}).

\subsection{Beam Model}
\label{sec:beam}
\begin{figure}[h!]
	\centering
	\begin{tabular}{ccc}
		&\bf Raw & \bf Regularized \\
		\raisebox{30mm}{\rotatebox[origin=c]{90}{Beam transfer function}} &
		\includegraphics[height=6cm,trim=5mm 5mm 5mm 0]{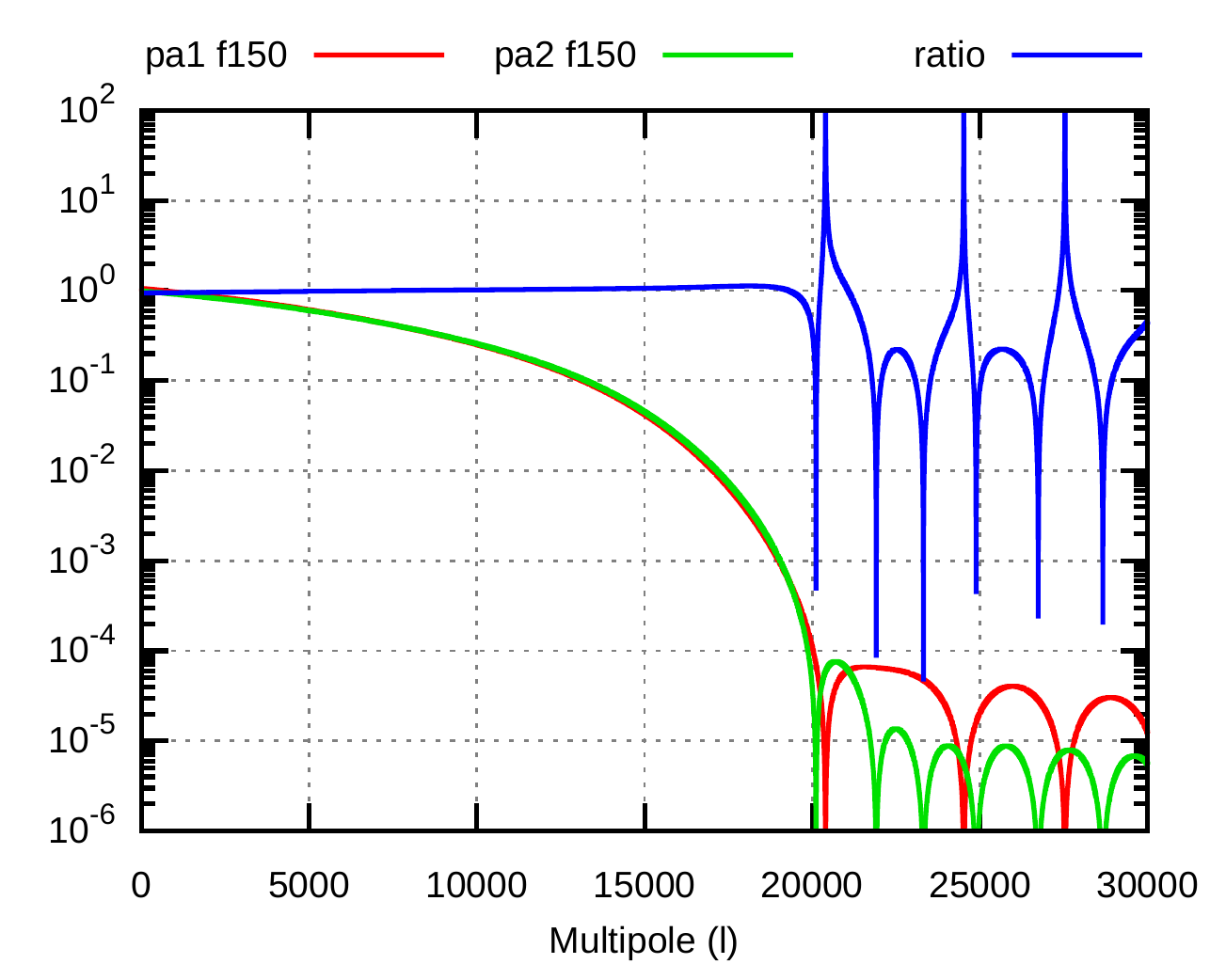} &
		\includegraphics[height=6cm,trim=5mm 5mm 5mm 0]{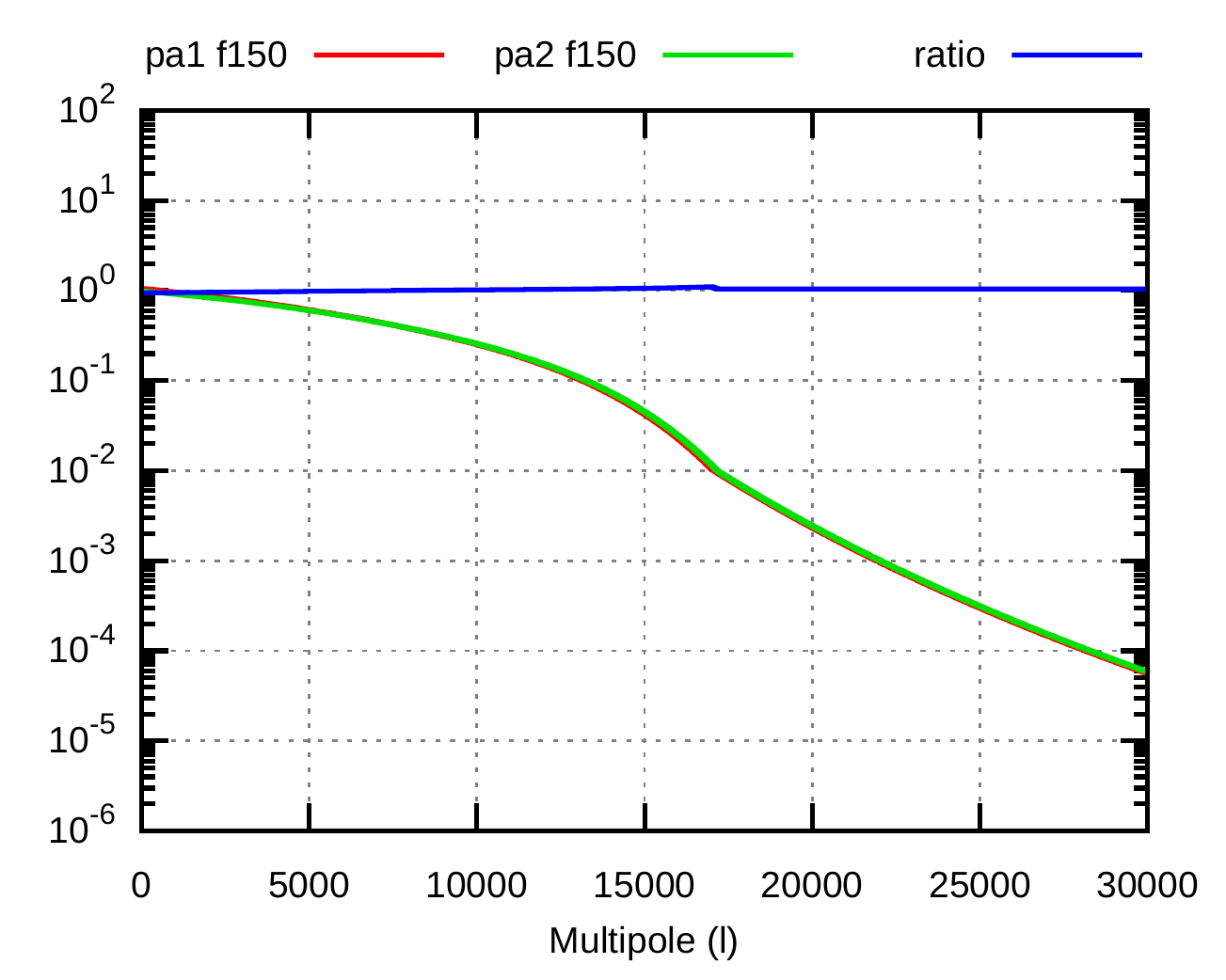}
	\end{tabular}
	\caption{\dfn{Left}: The ACT PA1 and PA2 beams at f150, and their ratio. The ratio is
	quite constant up to $\ell = 20000$, where the beam model breaks down and the ratio
	starts swinging wildly. \dfn{Right}: The same beams after regularizing them by replacing the
	values after they fall to 0.01 of the peak with a smooth function that preserves the beam ratio.}
	\label{fig:beam-reg}
\end{figure}

The \planck\ beams are slightly elliptical and slightly position-dependent \citep{Planck-beams/2013,Planck-beams/2016},
but in the frequency
range considered here they are reasonably well approximated as Gaussian, especially for $\ell < 1500$
($\ell < 3000$ for f220) where the \planck\ data are relevant for this coadd\footnote{
The Gaussian approximation is accurate to
better than 1.4\%/0.6\%/1.3\% at these frequencies in the multipole range where \planck\ contributes
significantly to the combined map. For comparison, the \planck\ solid angle varies
across the sky with a standard deviation of 0.3\%/0.4\%/1.0\% at f090/f150/f220.}. We used the following
Gaussian FWHM beams from the \planck\ 2018 explanatory supplement\footnote{\planck 2018 explanatory
supplement section on effective beams: \url{https://wiki.cosmos.esa.int/planck-legacy-archive/index.php/Effective_Beams}}:
$9.66'$ for f090, $7.22'$ for f150 and $4.90'$ for f220. These beams include the effect of the HEALPix
pixel window to within the accuracy of the Gaussian approximation.

The ACT beam model is based on planet measurements and physical models of the optical system.
\citet{hasselfield/etal/2013}
describes the basic approach used for the maps.
\citet{choi/etal:2020} describes recent improvements in removing the atmospheric contribution to the planet map
and the inclusion of  a scattering term from the primary surface deformations.  These beam models are used as inputs.

These beam models break down at  very high $\ell$ where the response becomes very low. This is illustrated in the left panel of
figure~\ref{fig:beam-reg}, where two very similar beams and their ratios are plotted.
The beams fall off smoothly as we approach $\ell=20\,000$, but then start oscillating messily
around zero. If used directly, this would lead to an unphysical and wildly fluctuating beam
ratio, which would translate into the map in question having its weight fluctuate by orders
of magnitude from one multipole to another.  As these oscillations occur on very small angular scales where the beam has suppressed
all sky signal, we use a more well-behaved function: we replace
the parts of the beam after the point $\ell^*$ where it falls to a fraction $v^* = 10^{-2}$
of its peak value $B_\textrm{max}$ with $v^* B_\textrm{max} (\ell/\ell^*)^{2\log(v^*)}$.
This extrapolation has the property that it preserves the ratio between two beams and
matches both the value and first derivative of a Gaussian at the transition point.\footnote{
Since the ACT beams are not Gaussian the beam extrapolation has a kink at the transition point,
but it is still continuous. The exact form of the extrapolation does not matter, as long as it
does not lead to excessive ratios between beams at high $\ell$.} The result
of applying this regularization to the beams is shown in the right panel of figure~\ref{fig:beam-reg}.

After regularizing each beam, we evaluate them at each pixel in 2D Fourier space using
linear interpolation, and, in the case of MBAC, multiply it by the vertical ($f_y$) and
horizontal ($f_x$) transfer functions (see section~\ref{sec:reprojection}).
The resulting 2D beams are divided by the
desired 2D output beam to form the final relative beams $\bar B_i$.

\subsection{Choosing the Noise Model}
\label{sec:noise}

The inverse noise matrix $N^{-1}$ in equation~\ref{eq:maxlik} serves as the
weight when averaging together the different data sets. Ideally it would be a
full $N_\textrm{pix}$ by $N_\textrm{pix}$ matrix (with $N_\textrm{pix} \approx 10^9$)
describing the full noise
behavior, but due to the large number of degrees of freedom of such a matrix,
this would be both hard to estimate and computationally infeasible to represent. We must therefore
in practice approximate $N^{-1}$ using some simplifying assumptions. Thankfully,
equation~\ref{eq:maxlik} does not rely on the value $N^{-1}$ for producing a
bias-free map $\hat m$, only for its optimality, so we have considerable room
for approximations.

\subsubsection{Uncorrelated noise model}
\label{sec:uncorr-noise}
For maps with nearly white noise (e.g., WMAP and \planck), a good approximate description of the
map's inverse noise covariance is the
\textbf{uncorrelated} noise model, where $N^{-1}$ is modeled as diagonal in pixel
space, $N^{-1} \approx W^{-1}$, where W is a pixel-diagonal matrix representing the
white noise variance of the map, which is often available as a mapmaker output,
or can be estimated from the hitcount map. This noise model captures point-to-point
changes in noise level, but cannot handle correlated noise such as that produced by the atmosphere.

\begin{figure}[h]
	\centering
	\begin{tabular}{>{\centering\arraybackslash}m{5cm}>{\centering\arraybackslash}m{5cm}>{\centering\arraybackslash}m{5cm}}
		\bf uncorrelated & \bf const covariance & \bf const correlation \\
		\img{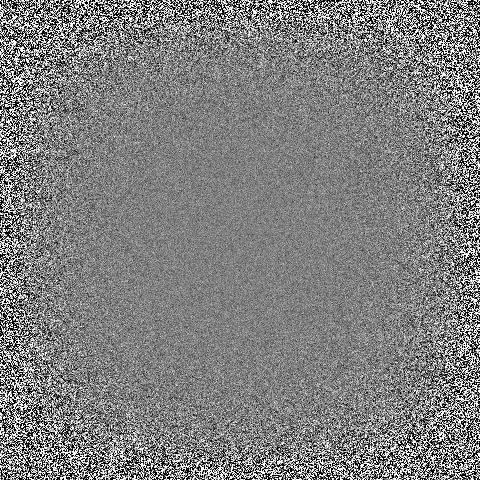} &
		\img{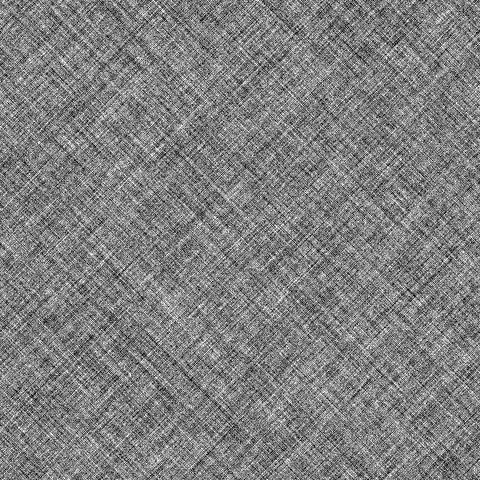} &
		\img{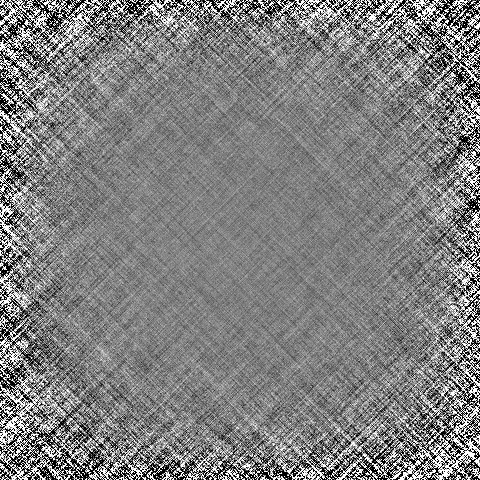}
	\end{tabular}
	\caption{The {\it uncorrelated} noise model (\dfn{left}) can represent spatially inhomogeneous
	noise, but ignores all spatial correlations. The {\it constant covariance} noise model (\dfn{middle})
	is the opposite, capturing complex spatial correlations but having no concept of position-dependence.
	The {\it constant correlation} noise model (\dfn{right}) combines these two models to allow for both
	correlation and inhomogeneity.}
	\label{fig:threemodels}
\end{figure}

\subsubsection{Constant covariance noise model}
If maps have uniform but non-white noise spectra, then a reasonable approximation
is the \textbf{constant covariance} noise model with
$N$ approximated as diagonal in Fourier space, representing a position-independent
2D noise power spectrum.\footnote{A 2D noise power spectrum represents the power in a map
in terms of both horizontal ($\ell_x$) and vertical ($\ell_y$) Fourier modes, which we
can index by the 2D wavenumber $\vec \ell \equiv (\ell_x, \ell_y)$. The advantage of
a 2D power spectrum over a simpler 1D one (which would only depend on $\ell=|\vec \ell|$),
is that it can handle stripy anistropic noise, which is usually present in ground-based
surveys.} This model handles correlated and stripy noise well, but does not treat spatial
variations in depth.

\subsubsection{Constant correlation noise model}
For a survey like ACT with spatially varying correlated noise, a better model for the noise is to describe it as a
 \textbf{constant correlation} pattern
modulated by the inverse variance level, $N^{-1} \approx W^{-\frac12} C^{-1} W^{-\frac12}$,
where $C^{-1}$ is a Fourier-diagonal matrix representing the 2D inverse
correlation spectrum. Examples of these first three models are compared in figure~\ref{fig:threemodels}.

\begin{figure}[h]
	\centering
	\begin{tabular}{>{\centering\arraybackslash}m{5cm}>{\centering\arraybackslash}m{5cm}}
		\bf curved & \bf fade \\
		\img{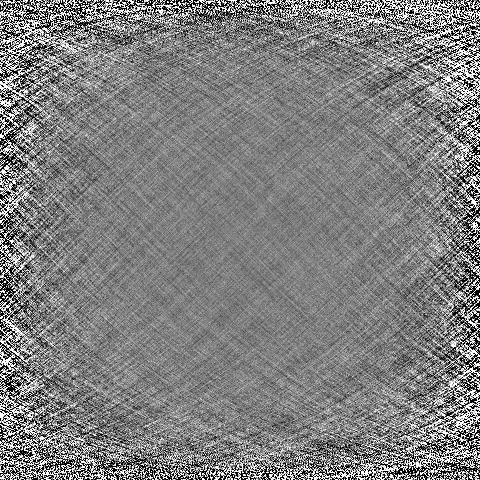} &
		\img{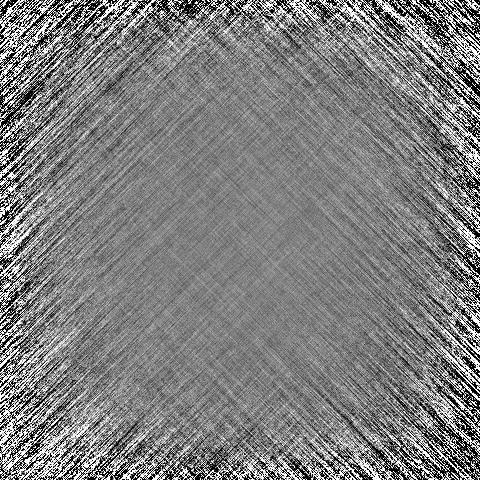}
	\end{tabular}
	\caption{Examples of more complicated combinations of position-dependence and correlation
	patterns that cannot be represented by the constant correlation noise model. \dfn{Left}:
	Here the direction of the stripes is not constant, becoming shallower as we move up or
	down from the middle of the image. \dfn{Right}: Here the image goes from being dominated by
	/-stripes to \textbackslash-stripes as we move from left to right.}
	\label{fig:curved-fade}
\end{figure}

The constant correlation approximation works well over small to medium size areas, like the 600 square degree
ACT D56 patch, but it breaks down when the amount or direction of noise stripiness changes.
This can happen due to scanning pattern variation (``fade" as illustrated in figure~\ref{fig:curved-fade}),
or simply due to the sky's curvature (``curved"). For example,
in equirectangular cylindrical projection maps in equatorial coordinates,
constant elevation scans trace out sine wave segments in the sky leading to a declination-dependent
noise stripiness direction.

\subsubsection{Tiled constant correlation noise model}
\begin{figure}[h]
	\centering
	\includegraphics[width=0.4\textwidth]{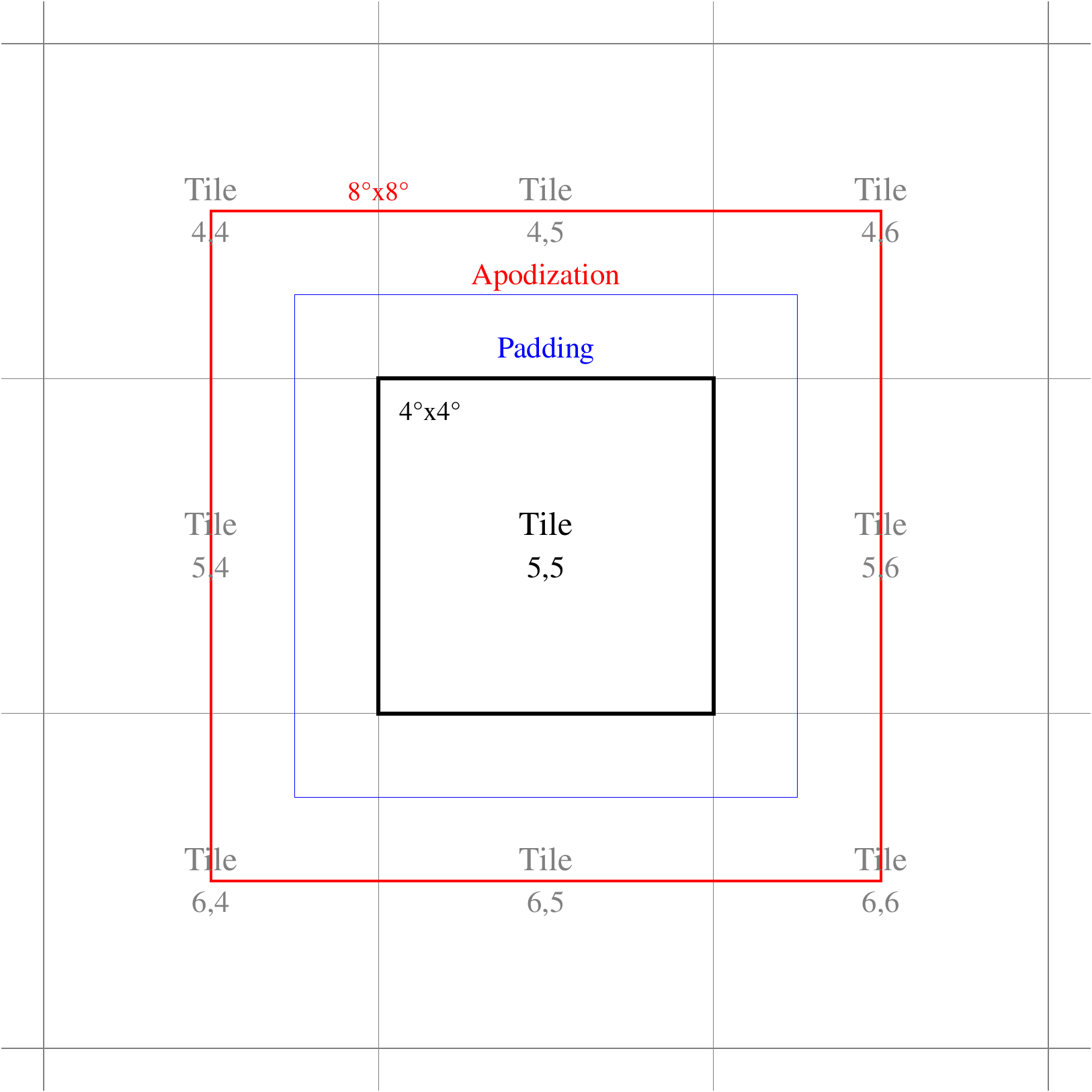}
	\caption{Illustration of our tiling scheme. The sky is tiled into $4^\circ\times4^\circ$ degree
	tiles, but to ensure continuity in the solution and to make each tile Fourier-friendly we
	apply $1^\circ$ of padding (blue) and $1^\circ$ of apodization (red) to each tile, resulting in a set of
	overlapping $8^\circ\times8^\circ$ tiles.}
	\label{fig:tiling}
\end{figure}

For this work, we extend the constant correlation noise model to large area maps so that we can model the
 position-dependent
correlation pattern. The \textbf{tiled correlation pattern} approach used in this paper involves three steps:
\begin{enumerate}
	\item Split the maps into equal-sized, overlapping tiles that are small enough that the 	constant correlation approximation is reasonably accurate.  We chose a tile size of $4^\circ\times4^\circ$
	with $1^\circ$ of overlapping padding and a further $1^\circ$ of apodization to avoid implied wrap-around
in the Fourier transforms. The resulting $8^\circ\times8^\circ$ full tile and its overlap with neighboring
tiles is illustrated in figure \ref{fig:tiling}.  Smaller tiles are better able to respond to fast changes in noise properties,
while larger tiles let us model longer distance noise correlations and give us more
statistical weight for building the per-tile noise model; $4^\circ\times4^\circ$ is a compromise.
	\item Solve equation~\ref{eq:maxlik} independently for each tile.
	\item Use the overlap to seamlessly merge the coadded tiles into a single map (see section~\ref{sec:constructing} and figure~\ref{fig:merge}).
\end{enumerate}

\FloatBarrier

\begin{figure}[h]
	\centering
	\includegraphics[width=0.8\textwidth]{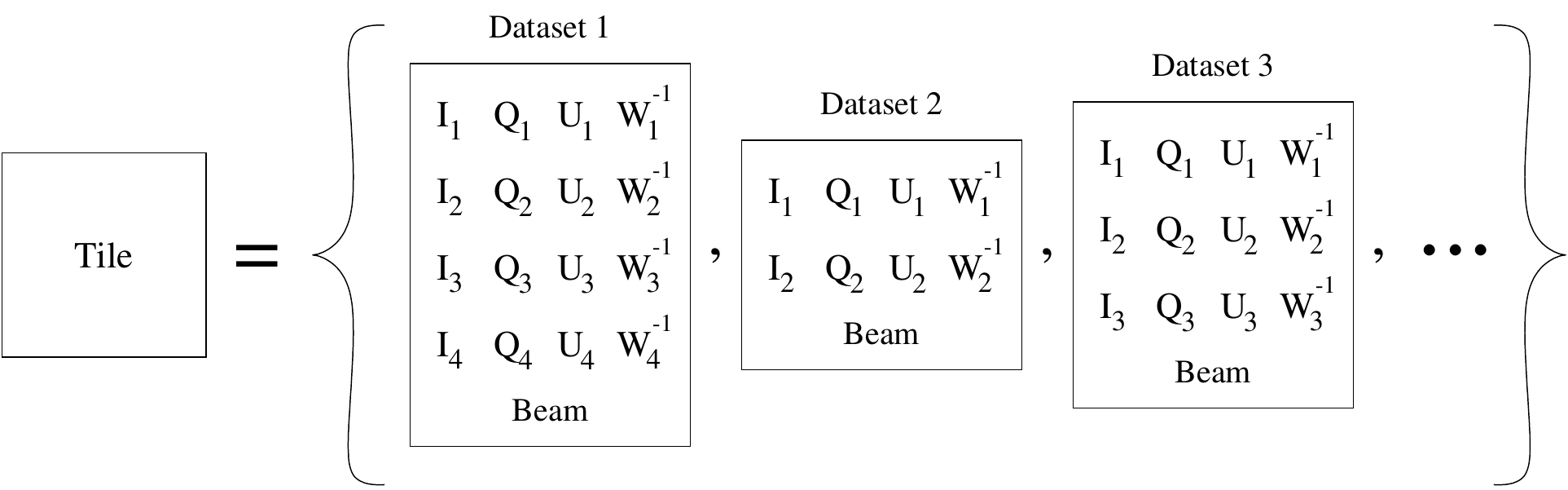}
	\caption{The input data sets that go into building the combined sky model for a given tile
	consist of a variable number of Stokes $IQU$ maps with corresponding inverse variance maps $W^{-1}$,
	all of which share a per-data-set beam. At the noise model estimation stage these are joined by
	a per-data-set inverse correlation matrix.}
	\label{fig:datasets}
\end{figure}

\begin{figure}[h]
	\centering
	\begin{closetabcols}[0.5mm]
	\begin{tabular}{ccc}
		raw map & model & cleaned \\
		\includegraphics[width=5.5cm]{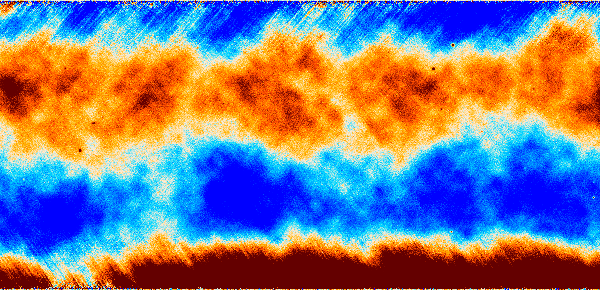} &
		\includegraphics[width=5.5cm]{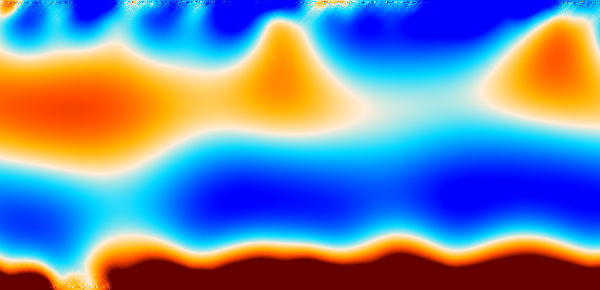} &
		\includegraphics[width=5.5cm]{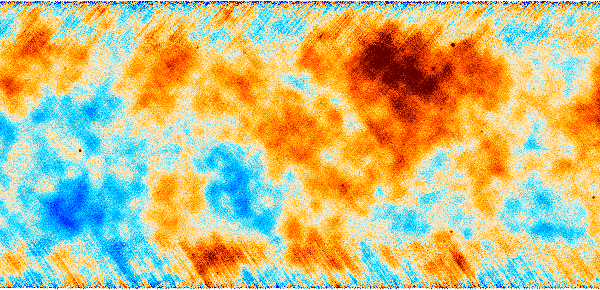}
	\end{tabular}
	\end{closetabcols}
	\caption{\dfn{Left}: A 10$^\circ$x5$^\circ$ section from the ACT MBAC f150 map showing an example
	of strong ground pickup, which shows up as mostly horizontal stripes that are usually stronger near the
	edge of the map. \dfn{Middle}: The maximum likelihood value for the pickup in the interior given the
	values in the outermost 60 pixels (half a degree) in the image. \dfn{Right}: The residual after subtracting
	this model. The pickup is almost completley removed, at the cost of some of the larger CMB scales.}
	\label{fig:ground}
\end{figure}

\section{Estimating the Constant Correlation Noise Model for Each Tile}
\label{sec:tiles}
\begin{figure}[h]
	\centering
	\includegraphics[width=\textwidth]{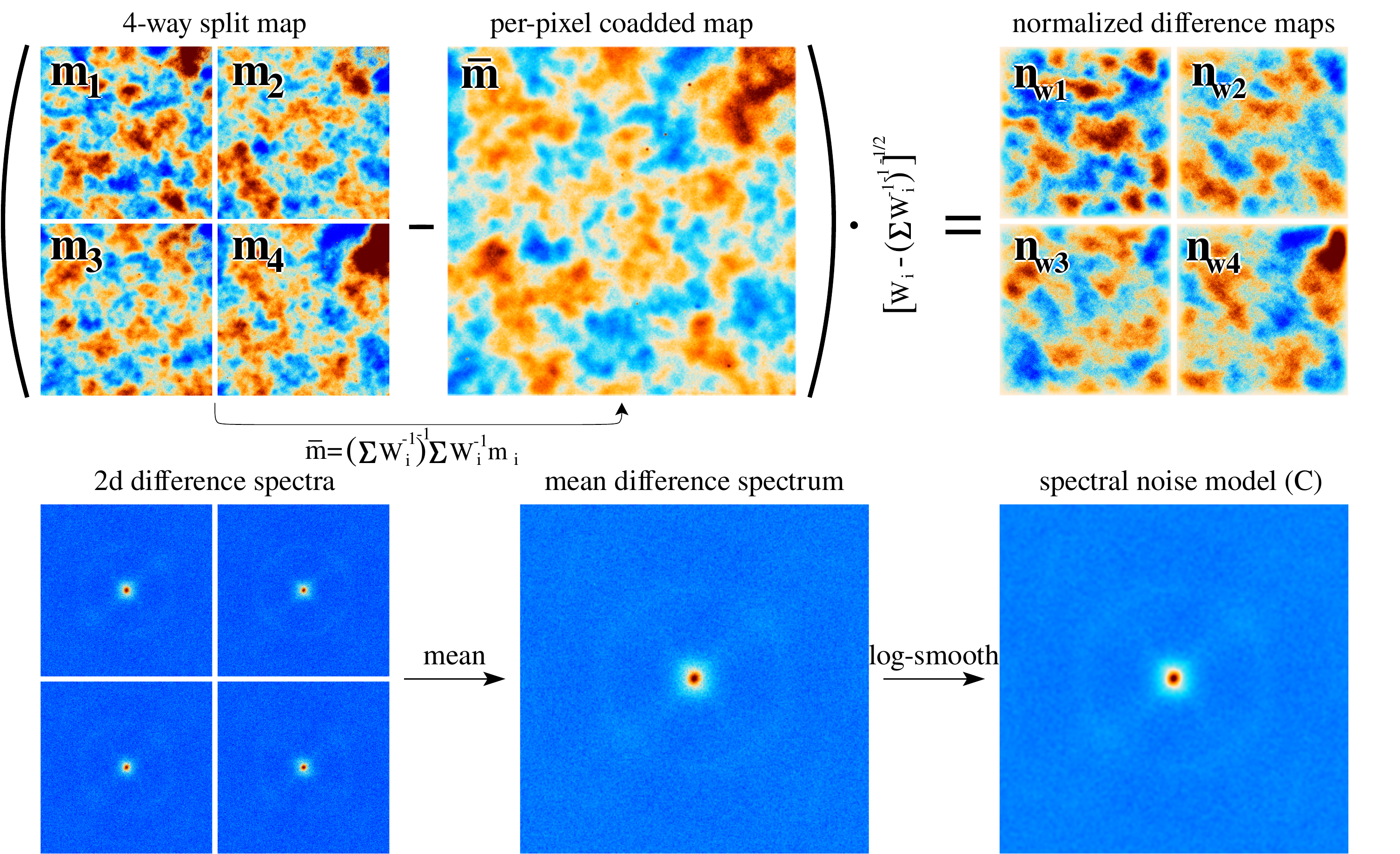}
	\caption{How the 2D noise correlation model $C(\vec \ell)$ is built for an individual $8^\circ \times 8^\circ$ tile.
	\dfn{Top}: The inverse
	variance weighted average $\bar m$ of a set of map splits $\{m_i\}$ is subtracted from each split and
	then normalized by their expected white noise levels, resulting
	in a set of noise-only maps $\{m_{wi}\}$. The 4-way split map and per-pixel coadd shown here cover the same tile, but
	look quite different due to atmospheric noise. \dfn{Bottom}: The 2D power spectra of these noise-only maps
	are averaged together and smoothed to produce the final 2D noise correlation model. The
	horizontal and vertical axes of the plots in the bottom row are the horizontal wavenumber $\ell_x$
	and the vertical wavenumber $\ell_y$, with the origin at the center. The central bright spot
	represents the atmosphere-dominated region at $\ell \lesssim 2000$, while the edge is at $\ell = 21\,000$.
	Both the vertical and horizontal axes are linear.
	The color scale is linear at low values (blue), but becomes logarithmic at higher values (turquoise
	to red) to represent the large span in power.}
	\label{fig:noisemodel-build}
\end{figure}

After reading in the data for all data sets in a tile (see
figure~\ref{fig:datasets}), some preparation is necessary before we can build
their noise models and solve for the combined maps:

\begin{enumerate}
	\item We expand all maps and inverse variances to full $TQU$ maps. For the T-only MBAC maps,
		the polarization inverse variance is set to zero, ensuring that they get zero weight, while
		the polarization signal maps are filled with white noise for convenience -- having a non-zero
		signal here allows us to
		handle these maps on the same footing as the polarization maps from the other data sets without
		any special cases.
	\item For all maps but \planck\ we apply a gentle detrending filter to remove excessive
		ground pickup from the edge of the maps (see figure~\ref{fig:ground}). We assumed that
		the edge of the tile is dominated by some smoothly varying contaminant, and
		used this to in-paint the interior of the tile by solving for the maximum-likelihood
		value for the interior pixels given the value of the edge pixels. This was done by
		solving the system
		\begin{align}
			(V^{-1} + M^{-1})v = M^{-1}m,
		\end{align}
		where $m$ is the data map and $v$ is the ground estimate we wish to construct. $M^{-1}$
		is a pixel-diagonal matrix which is effectively a mask that selects the edge pixels
		as the reference values for the interpolation. It has value 0 in the interior and
		$1/(1$\textmu{}\mbox{K}$^2)$ at the edge (though any value $\gg \textrm{max}(V^{-1})$ will work).
		$V^{-1} = \langle W^{-1} \rangle_\textrm{pix} /(1 + ((\ell+0.5)/1000)^{-3.5}$ represents
		a smoothly varying signal, with the smoothness governed by the constants 1000 and $-3.5$
		in the expression. $W$ was defined in section~\ref{sec:uncorr-noise}. The result is quite insensitive to their exact values. The ones chosen here
    were based on the behavior of the ACT noise, but any spectrum with large correlations
		on the tile scale would work. Subtracting this ground estimate ($m\rightarrow m-v$)
		greatly reduces the ground pickup,
		at the cost of introducing a bias by removing some signal power. This is mostly
		on the scale of the tile size, i.e. 4$^\circ$--8$^\circ$, corresponding
		to $\ell \lesssim 100$, but smaller levels of bias extend up to higher $\ell$,
		falling below 0.5\% at $\ell > 1000$ (see appendix~\ref{sec:sims}).
		This ground subtraction was especially necessary for the MBAC maps. For the ACT+\planck\
		maps, this loss of power at low $\ell$ is reduced due to the dominance of \planck\ there.\footnote{
The ground subtraction is done per tile. Each tile has an apodization region at the edge, which overlaps
with the area covered by other tiles and is discarded before stacking the tiles into the final image.
The ground subtraction procedure strongly biases the tile data in the region we use as input for the
maximum likelihood estimation of the ground, but since that region is in the tile apodization area
which will be discarded anyway, most of this bias is avoided. Despite only using data in the
apodization region to constrain the ground signal, this method can still clean the interior
of the tile due to the ground's correlation structure. For example, a horizontal stripe going through the tile
would extend into both the left and right margins of the tile. The maximum-likelihood value for
the ground field given the data in those margins is a stripe that connects them through the interior
of the tile, since we assume that the ground has strong spatial correlations. Subtracting this model
removes the stripe in the interior of the tile without having directly looked at the data there.
This will also remove CMB signal on tile-sized scales, so it does introduce a bias, but on those
		scales \planck{} (which is not subject to this filtering) dominates.}
		In the future we will avoid this bias by replacing the ground filter with maximum-likelihood
		downweighting of the contaminated modes.
	\item We clip the signal map values to the $\pm 100$ K range to avoid issues with extreme pixels.
		No real signal in the maps should be bright enough to be affected by this.
	\item For each inverse variance map we compute $\gamma_1$, the median of its non-zero
		values, and $\gamma_2$, the median of the subset of its values that are larger
		than $10^{-5} \gamma_1$,\footnote{This
		two-step process is done to avoid being influenced by large areas of zeros
		or very low values.} and cap the
		inverse variance to $20 \gamma_2$. The purpose of this is to avoid giving undue weight
		to (very rare) glitched pixels with unrealistically high inverse variance.\footnote{The factor 20 is high
		enough to avoid affecting any realistic values in the maps, but the particular value is
		somewhat arbitrary and, e.g., 100 would also work without noticeable effect on the maps.}
	\item The edges of the survey areas see rapid changes in the noise properties which are difficult
		to model. Since these areas are quite noisy and do not contribute much information, we choose
		to suppress the lowest inverse variance
		areas via the transformation $W^{-1} \rightarrow W^{-1} \textrm{min}(1, W^{-1}/(0.2 \gamma_1))^5$.
		Here $W^{-1}$ is the inverse white noise variance map, and the power of 5 was chosen to
		rapidly but smoothly suppress areas with too low exposure. This leaves areas with inverse variance
		greater than $0.2 \gamma_1$ (effectively one fifth of the median of the exposed area) unchanged
		while rapidly damping lower values.
	\item Fourier-space operations assume that periodic data inside each tile, but
		the real data contain power on scales larger than the tiles. If used directly in the
		per-tile Fourier transforms, this power would alias into every other multipole, which
		would show up as ringing patterns after applying any Fourier-space weighting operations.
		The standard solution to this problem is to smoothly taper off the data towards the
		edge of the map, in a process called apodization. We reserved a $1^\circ$ border
		at the edge of each tile for this purpose (see figure~\ref{fig:tiling}), and use it
		to apply a 60-pixel ($0.5^\circ$
		at the equator) cosine taper to the edge of both the data map and $W^{-1}$
		in each tile.
		We also apply a 60-pixel cosine taper at the edge of the exposed area in the case of
		data sets that stop part-way through a tile.
	\item Any data set that ends up being empty after these steps is discarded for this tile.
\end{enumerate}
As described in section \ref{sec:noise}, we model the map $m$ as having the inverse noise matrix
$N^{-1} = W^{-\frac12}C^{-1}W^{-\frac12}$ in each tile. $W^{-1}$ is diagonal in pixel space
and represents the inverse variance of the white part of the noise. This is approximately proportional
to the number of times each pixel was observed, and was provided together with the sky maps for
the data sets we analyze here. $C$ is taken to be diagonal in 2D Fourier space, i.e. $C_{\vec \ell\vec \ell'}
= \delta_{\vec \ell\vec \ell'} C(\vec \ell)$, with $C(\vec \ell)$ being the 2D noise power spectrum of $m$
after factorizing out $W^{\frac12}$. That is, $C(\vec \ell)$ is the 2D noise power spectrum of
$m_w = W^{-\frac12}m$, which we will refer to as the {\it normalized} map.

In order to estimate $C(\vec \ell)$ we construct noise-only maps by subtracting the inverse-variance weighted mean
map $\bar m = (\sum_i W^{-1}_i)^{-1} \sum_i W^{-1} m_i$ from each split $m_i$,
resulting in noise maps $n_i = m_i - \bar m$. This is why we require several splits
with independent noise in each data set. After taking into account the covariance
of each map with the weighted mean map, we see that these difference maps have white noise variance
$W_i - (\sum W_i^{-1})^{-1}$, allowing us to construct normalized noise maps
$n_{wi} = \left[W_i - (\sum_j W_j^{-1})^{-1}\right]^{-\frac12} n_i$.
This procedure is illustrated in the top row of figure~\ref{fig:noisemodel-build}.

We then estimate the 2D noise power spectrum of each split as
\begin{align}
	C_i(\vec \ell) &= \Big| \int d\vec x e^{-2\pi \vec \ell \cdot \vec x} n_{wi}(\vec x) \Big|^2 / g_i.
\end{align}
The extra correction factor $g_i$ is there to handle cases where some maps only partially cover
the tile, leaving the rest of the tile empty. Normally variations in a data set's depth across
the tile would not be a worry at this stage, as we are working with the normalized noise maps $n_w$
where variations in depth have already been factored out. However, this fails for areas that
have exactly zero depth ($W = 0$). To avoid division by zero, any such unexposed areas are
left as zero in $n_w$. However, this leads to a deficit of power in $C(\vec \ell)$, which,
if left alone, would result in data sets that only barely extend into a tile being
given a disproportionally high weight. $g_i$ is a measure of the fraction of split $i$ that is
not empty, and dividing by this undoes the effect of not being able to normalize unexposed areas.
To be precise, we estimate $g$ as
$g = \textrm{min}\Big(\langle W^{-1} > \gamma_2/100\rangle_\textrm{pix}, \langle\alpha_i\rangle_\textrm{pix}\Big)$,
where $\alpha_i$ is the total damping of $W^{-1}$ that was applied to split $i$ in steps 5 and 6,
and where $\langle\rangle_\textrm{pix}$ denotes the mean over the tile pixels.

To summarize: $C_i(\vec \ell)$ is the 2D noise power spectrum of the non-empty parts of the
noise-only map $n_i$ after factorizing out variations in exposure time into $W_i$.
Because we normalized the maps, these noise power
spectra are dimensionless with values approaching unity in the white noise region.

We will assume that all splits of a data set have the same correlation structure, and only
differ somewhat in their white noise properties\footnote{This is generally a good approximation,
but may become inaccurate in the shallowest areas of the map where it is hard to spread the data
evenly between the splits. A less accurate noise model in these areas would result in a less optimal
(higher noise) combined map in that region, but it would not introduce any bias.}. This lets us
reduce sample variance in the
noise power spectrum by averaging them, resulting in $C(\vec \ell) = \frac{1}{N_\textrm{split}}\sum_i C_i(\vec \ell)$.

\subsection{Smoothing the spectrum}
In order to suppress sample variance in our noise power spectrum estimate,
we apply a Butterworth low-pass filter to the 2D power spectrum
at a characteristic length scale of $\Delta \ell=400$. This corresponds to the last step
in figure~\ref{fig:noisemodel-build}.\footnote{
It might sound weird to apply a low-pass-filter to a Fourier-space quantity, but
there is nothing special about Fourier space -- the 2D noise power spectrum is just
a 2D image that can be Fourier-transformed and filtered like any other.
} Because the noise spectrum is a very steep function of $\ell$, we
apply this smoothing in log-space
and correct for the difference between normal averaging and log-averaging on the noise.
We estimate this using simulations, and find that the log-smoothed power spectrum must be multiplied by
1.31 for a 2-way split and 1.14 for a 4-way split.\footnote{We could have avoided using this suboptimal smoothing
procedure  by collecting statistics over a larger area of the map than just a single
tile.  However, this would require  selecting tiles that are likely to have the same 2D
noise power spectrum.}

\subsection{Down-weighting ACT at low $\ell$}
The ACT maps are known to be missing power at low $\ell$ due to ground pickup filtering;
bias from measuring the noise model from the same data it will be applied to
\footnote{We mitigate this by making the maps in multiple passes, subtracting the  best sky estimate
from the previous pass when estimating the noise model for the next pass, but at very low $\ell$
this process converges too slowly.}; and bias from stopping the iterative solution of the maps
before the largest scales have converged \citep{aiola/etal:2020}. These effects all mainly affect $\ell \lesssim 500$ in
T and $\ell \lesssim 200$ in $P$, with the exception of the preliminary AdvACT maps, which
currently have a few-percent loss of total intensity power for $\ell \approx 750/1200/1750$ at f090/f150/f220,
with a gradually increasing loss below that.
This is also the $\ell$-range where our noise spectrum becomes unreliable
due to the smoothing performed in the previous section\footnote{The noise spectrum changes much
more quickly at low $\ell$, making a loss of Fourier-resolution relatively more serious there.}.
In total intensity this hardly matters
as \planck\ completely dominates by that point, but in polarization ACT is sensitive enough that
some of these unreliable scales could slip through. To avoid this, we down-weight all ACT data
by multiplying the inverse noise spectrum by $(1 + (\ell/200)^{-10})^{-1}$. The effect of this
is clearly visible in figure~\ref{fig:actfrac} as a cutoff of ACT at $\ell=200$.

\subsection{Correcting for reprojection effects}
\label{sec:reprojection}
Finally, we correct for the high-$\ell$ loss of power from reprojection. For \planck, we fill $\ell>6000$
with the mean noise power in the region $4500 < \ell < 6000$, emulating what the \planck\ noise power
would have looked like if it hadn't been truncated by being mapped at low resolution first.

We estimated the transfer function from the bicubic spline interpolation used when reprojecting
MBAC by simulating a set of white noise maps for the two MBAC patches and applying the same
interpolation to these. Because bicubic spline interpolation can be separated into independent
vertical and horizontal interpolation steps, we can factorize this transfer function into a
vertical and horizontal component $f_y(\ell_y)$ and $f_x(\ell_x)$, which we measure as the
horizontal and vertical average of the square root\footnote{We take the square root here because we want the transfer function that applies to the maps, not the power spectrum.} of the mean power spectra of these
interpolated simulations. These transfer functions are generally quite small, and only become
noticable at high $\ell$. They deviate from 1 by 0.01\% at $\ell = 4\,000$, 1\% at $\ell = 10\,000$ and
10\% at $\ell = 16\,000$.

We divide out these transfer functions from $C$ by applying the transformation $C(\ell_y,\ell_x)
\rightarrow C(\ell_y,\ell_x) / (f_y(\ell_y)f_x(\ell_x))$ and also incorporate them into the 2D beam for MBAC.
However, to avoid excessive deconvolution in MBAC-only areas ($<1\%$ of the survey area),
we cap the transfer functions
at 0.7 (which occurs at $\ell > 20\,000$) and decrease the statistical weight of MBAC by a factor of 100 at multipoles where the
transfer function was originally smaller. The effect of this is that any bias from ignoring part
of the transfer function has no effect in the areas with any other ACT data present, while allowing
a small bias in the form of extra smoothing in the small MBAC-only areas.

\subsection{Example noise models}
\begin{figure}[h]
	\centering
	\begin{closetabcols}
	\begin{tabular}{r>{\centering\arraybackslash}m{3cm}>{\centering\arraybackslash}m{3cm}>{\centering\arraybackslash}m{3cm}>{\centering\arraybackslash}m{3cm}>{\centering\arraybackslash}m{3cm}}
		\rotatebox[origin=c]{90}{\bf 2D noise} &
		\img{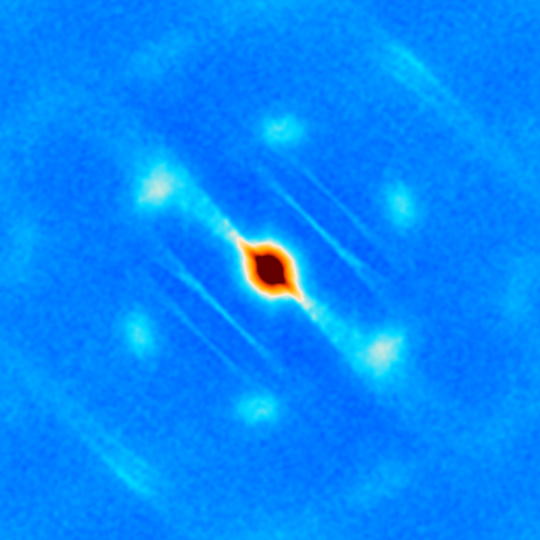} &
		\img{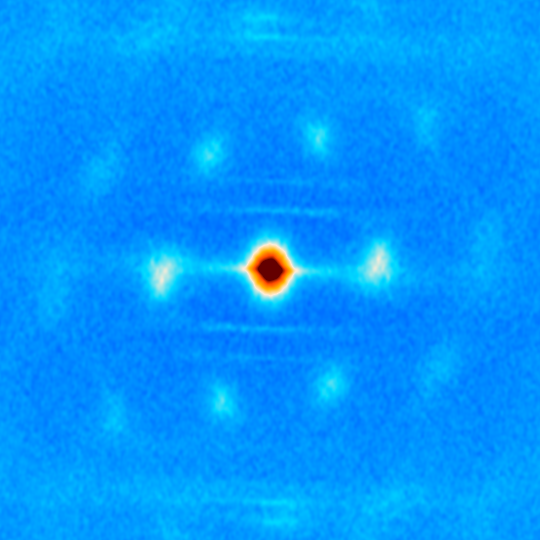} &
		\img{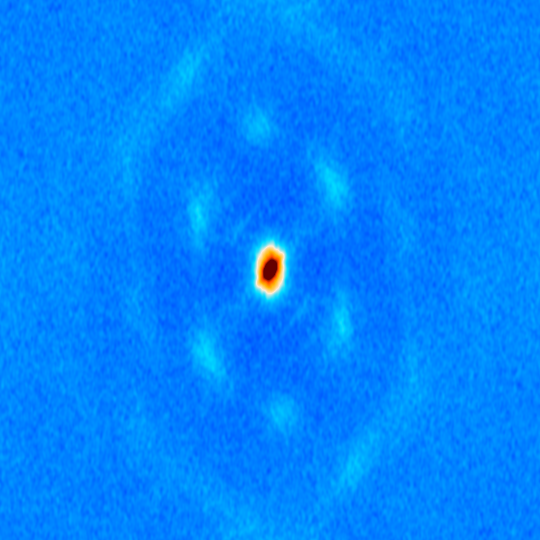} &
		\img{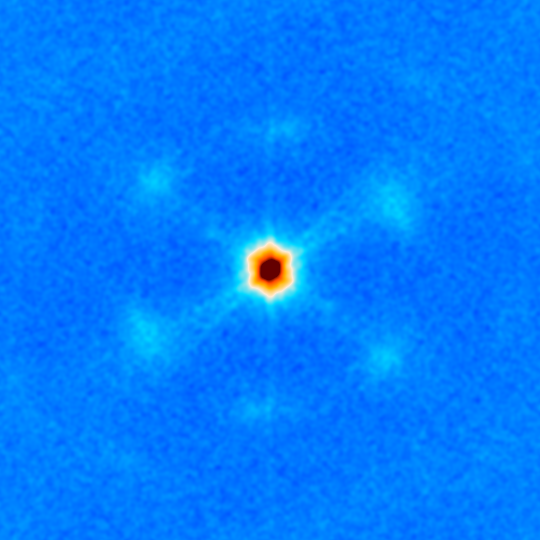} &
		\img{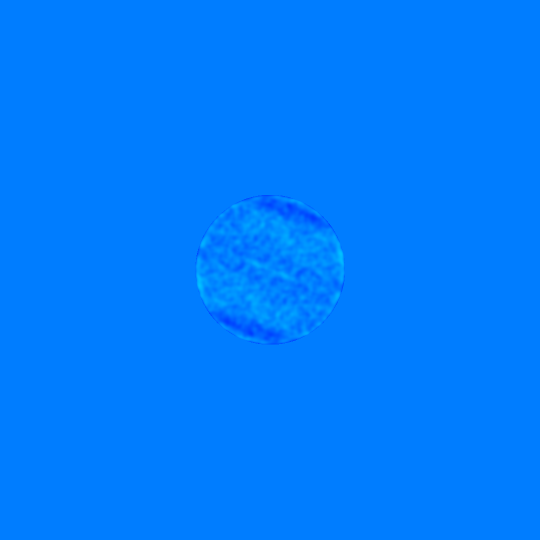} \\
		\rotatebox[origin=c]{90}{\bf anisotropy} &
		\img{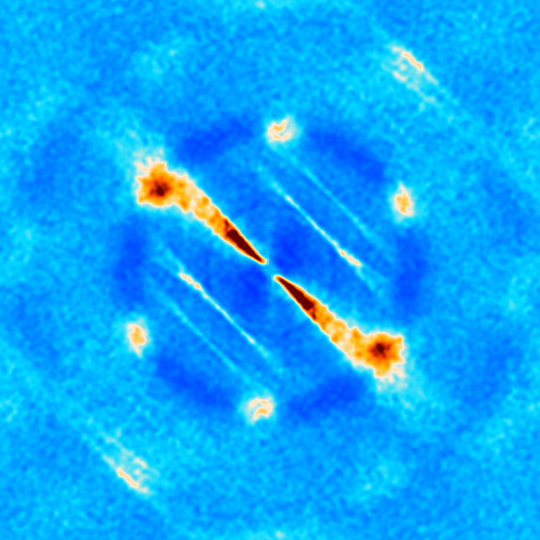} &
		\img{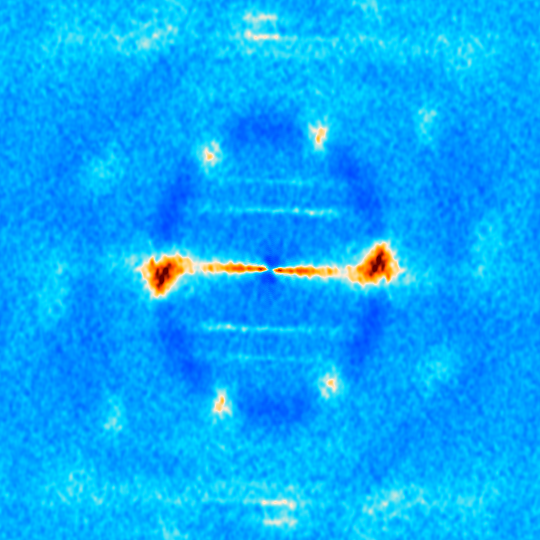} &
		\img{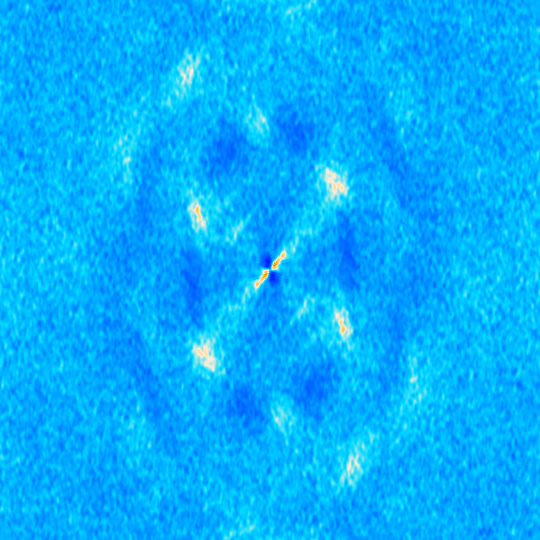} &
		\img{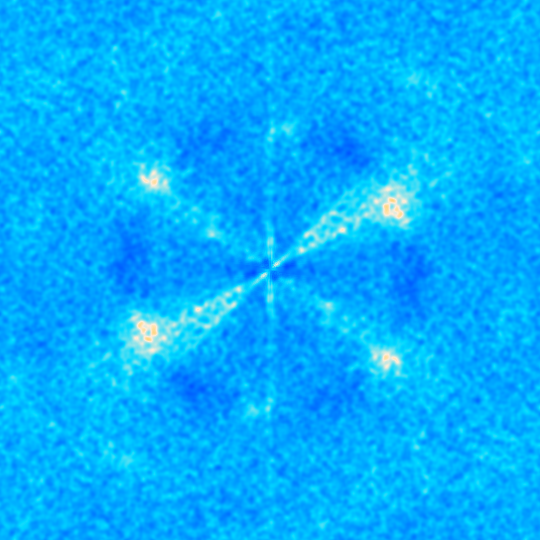} &
		\img{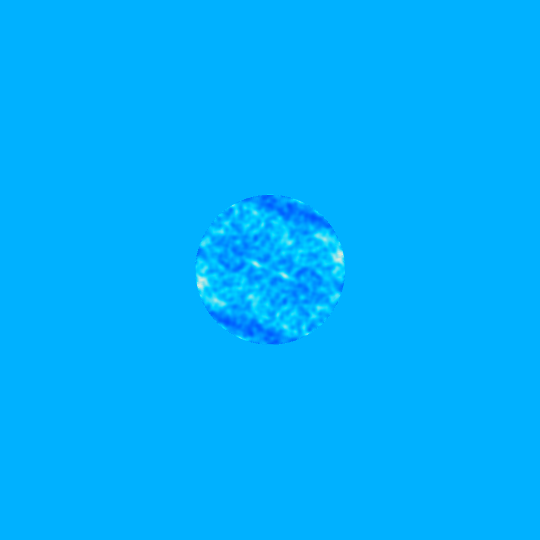}
	\end{tabular}
	\end{closetabcols}
	\caption{\dfn{Top}: Examples of the 2D noise correlation model $C$ that is built
	in section~\ref{sec:tiles}. The horizontal and vertical
	axes of each image corresponds to horizontal ($\ell_x$) and vertical ($\ell_y$) Fourier-modes
	respectively, with the origin at the center. Both axes are linear. From left to right these
	correspond to examples 1--5 in figure~\ref{fig:1d-corr}. The left-most two show examples of
	cases with very stripy noise due to low crosslinking; the next two have average levels of
	crosslinking, with roughly equal exposure in two roughly orthogonal scanning directions;
	and the last shows an example of \planck's 2D noise power. \dfn{Bottom}: As the top, but
	with the radial average subtracted in order to highlight the anisotropy.}
	\label{fig:2d-corr}
\end{figure}

\begin{figure}[h]
	\centering
	\includegraphics[width=0.6\textwidth]{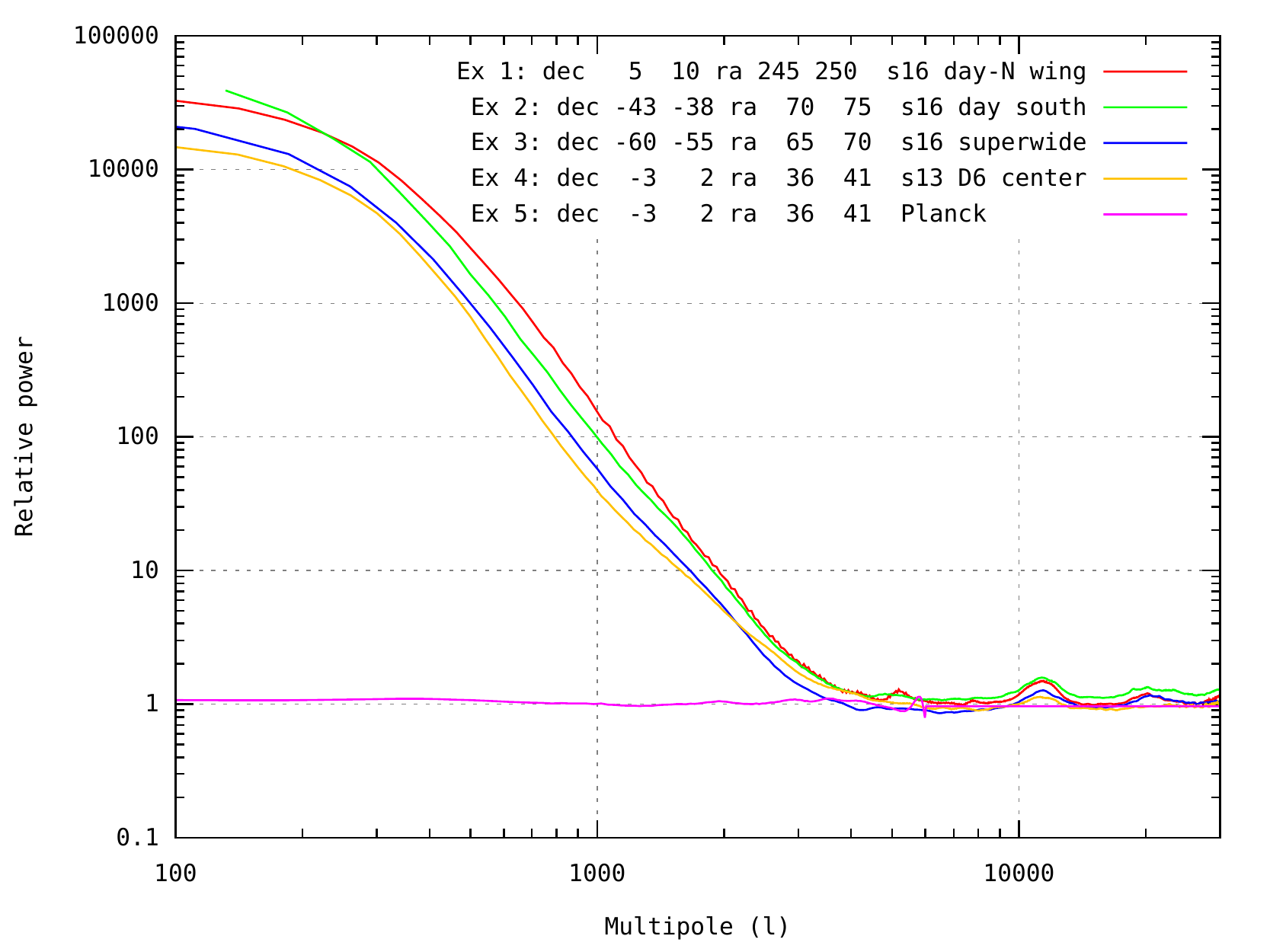}
	\caption{Radial averages of the total intensity 2D noise correlation model $C$ for the same cases as
	in figure~\ref{fig:2d-corr}. Example 1 (red) and 2 (green) have low crosslinking,
	leading both to anisotropic noise (as shown in figure~\ref{fig:2d-corr}) and higher
	large-scale power, as seen by these two curves being 2--3 times higher than the others
	for $\ell < 3000$. Example 3 (blue) and 4 (orange) are both well crosslinked, but differ
	greatly in depth. Since the depth factorizes out into the $W$ term in the noise model,
	the depth is not distinguishable in $C$. Example 5 is \planck, which is almost completely
	flat due to the absence of atmospheric emission.}
	\label{fig:1d-corr}
\end{figure}

Figure~\ref{fig:2d-corr} shows examples of the 2D correlation model $C$ for various data sets, showing the variety of
noise correlation patterns. The central red spot is the atmosphere dominated region $\ell < 3000$.
Most of the survey area is well ``crosslinked,'' meaning that the telescope scans across
each pixel in multiple different directions. In regions where this does not happen, the central
atmospheric region stretches out to higher $\ell$ in a direction perpendicular to the scanning
direction, as can be seen in column~1 of the figure.
The hexagonal detector layout in the ACT focal plane manifests as a hexagonal pattern of spots in the
2D power spectrum because nearby detectors see nearby parts of the atmosphere, and hence become
strongly correlated. The polarization power spectra, which are not shown here, are much flatter.

Radial averages for the same cases are shown in figure~\ref{fig:1d-corr},
illustrating the many order of magnitude increase in noise in the atmosphere-dominated range
in ACT, compared to the almost white noise of \planck.

\FloatBarrier

\section{Constructing the Maps}
\label{sec:constructing}

After preparing the maps, noise model and beams in a tile, we are ready to solve equation~\ref{eq:maxlik}
for the maximum-likelihood sky model. After inserting the form of the constant correlation noise model
$N^{-1} = W^{-\frac12}C^{-1}W^{-\frac12}$ into the equation, it takes the form
\begin{align}
	\sum_i \bar B_i^T W_i^{-\frac12}C_i^{-1}W_i^{-\frac12} \bar B_i \hat m &= \sum_i \bar B_i^T W_i^{-\frac12}C_i^{-1}W_i^{-\frac12} m_i.
\end{align}
This is a simple linear system that can be relatively efficiently solved using
Preconditioned Conjugate Gradient iteration (PCG) \citep[e.g.][]{pcg}. With the Fourier-diagonal
preconditioner $M = \Big[\sum_i \bar B_i^2 \langle W_i^{-1} \rangle_\textrm{pix} C_i^{-1}\Big]^{-1}$
we found the solution to converge in 15--50 steps, depending on which maps go into the tile.

After solving for the maximum-likelihood sky map in each tile, we are left with merging these into
a single consistent sky map. Ideally this would be as simple as cutting out the central, non-overlapping
$4^\circ\times4^\circ$ region of each tile and tiling them next to each other. This does work, but because
each tile has its own noise model and hence weights the input maps slightly differently from its neighbors,
this simple cropping+tile approach carries the risk of small discontinuities at the tile borders.

\begin{figure}[h]
	\centering
	\includegraphics[width=\textwidth]{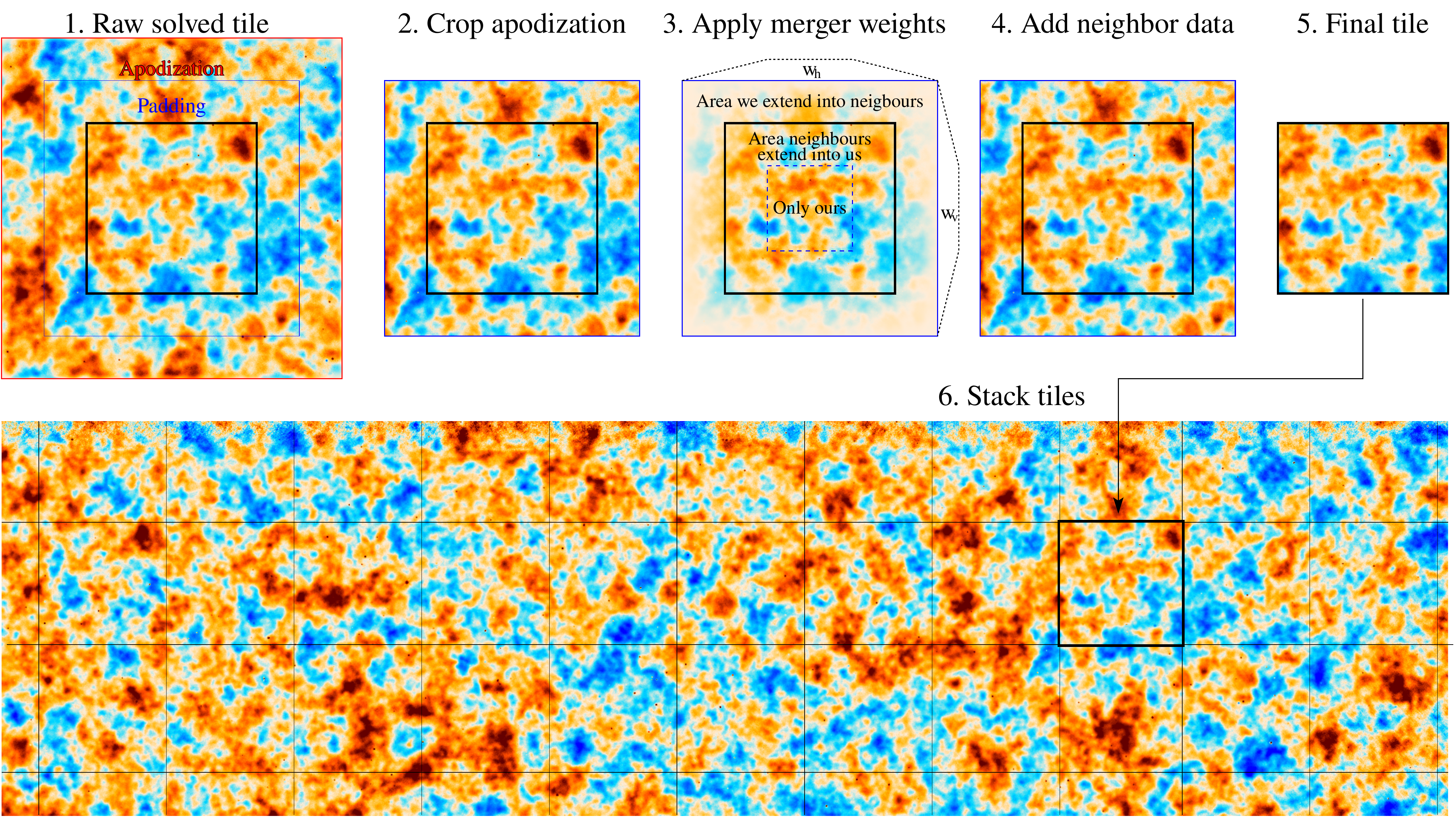}
	\caption{The tile merging procedure. Each raw solved tile contains a target $4^\circ\times4^\circ$
	degree region that would tile the sky, surrounded by $2^\circ$ of padding and $2^\circ$ of
	extra apodization, both of which overlap with neighboring tiles. We merge
	the tiles into a unified sky model by first cropping the unreliable apodization region, and
	then forming a weighted average of the overlapping regions. Finally, the central $4^\circ\times4^\circ$
	regions are stacked next to each other to form the full map. This procedure avoids tiny discontinuities
	that would arise from directly stacking the central $4^\circ\times4^\circ$ of tiles that were solved with slightly
	different noise models.}
	\label{fig:merge}
\end{figure}

To avoid such discontinuities, we instead only remove the $2^\circ$ apodization region of each tile,
leaving us with the central $6^\circ\times6^\circ$ region which still has substantial overlap with
neighboring tiles. The overlap is resolved via bilinear crossfading: Each pixel in a tile is
assigned a weight $w(\Delta x,\Delta y) = \omega(\Delta x) \omega(\Delta y)$ where $\Delta x$, $\Delta y$
is the offset from tile center, and where $\omega(x) = \textrm{min}(1, (3^\circ - |x|)/1^\circ)$.
This has the effect of assigning a weight of 1 to the central $2^\circ\times2^\circ$ region of each tile (which does
not overlap with any data from neighbors), and then linearly decreasing weight
until it reaches 0 at the tile edge,
which is a distance $3^\circ$ away from the tile center. The merged value of each tile
is then the weighted average of it and the overlapping parts from its neighbors. This process is
illustrated in figure~\ref{fig:merge}. See appendix~\ref{sec:sims} for a validation of this procedure
on simulations.

\FloatBarrier

\section{Map Properties}
\label{sec:map_properties}
The final combined maps cover the area \degrange{0}{RA}{360}, \degrange{-62}{dec}{22} in
total intensity and linear polarization at f090, f150 and f220. Each map has variants
with and without \planck, with and without daytime data and with and without point source
subtraction. The maps cover a total of $43200\cdot 10320 = 446\cdot 10^6$ CAR pixels at 0.5
arcmin resolution, corresponding to
26\,400 square degrees, of which roughly 70\%
falls within the ACT survey. The beam profiles of these maps are provided as part of the data
release. They were chosen to be similar to the ACT beams to keep the amount of reconvolution
low, and are therefore not Gaussian. They have full-widths at half-maxima of
2.1/1.3/1.0 arcmin at f090/f150/f220.

\begin{figure}[h]
	\centering
	\includegraphics[width=0.7\textwidth]{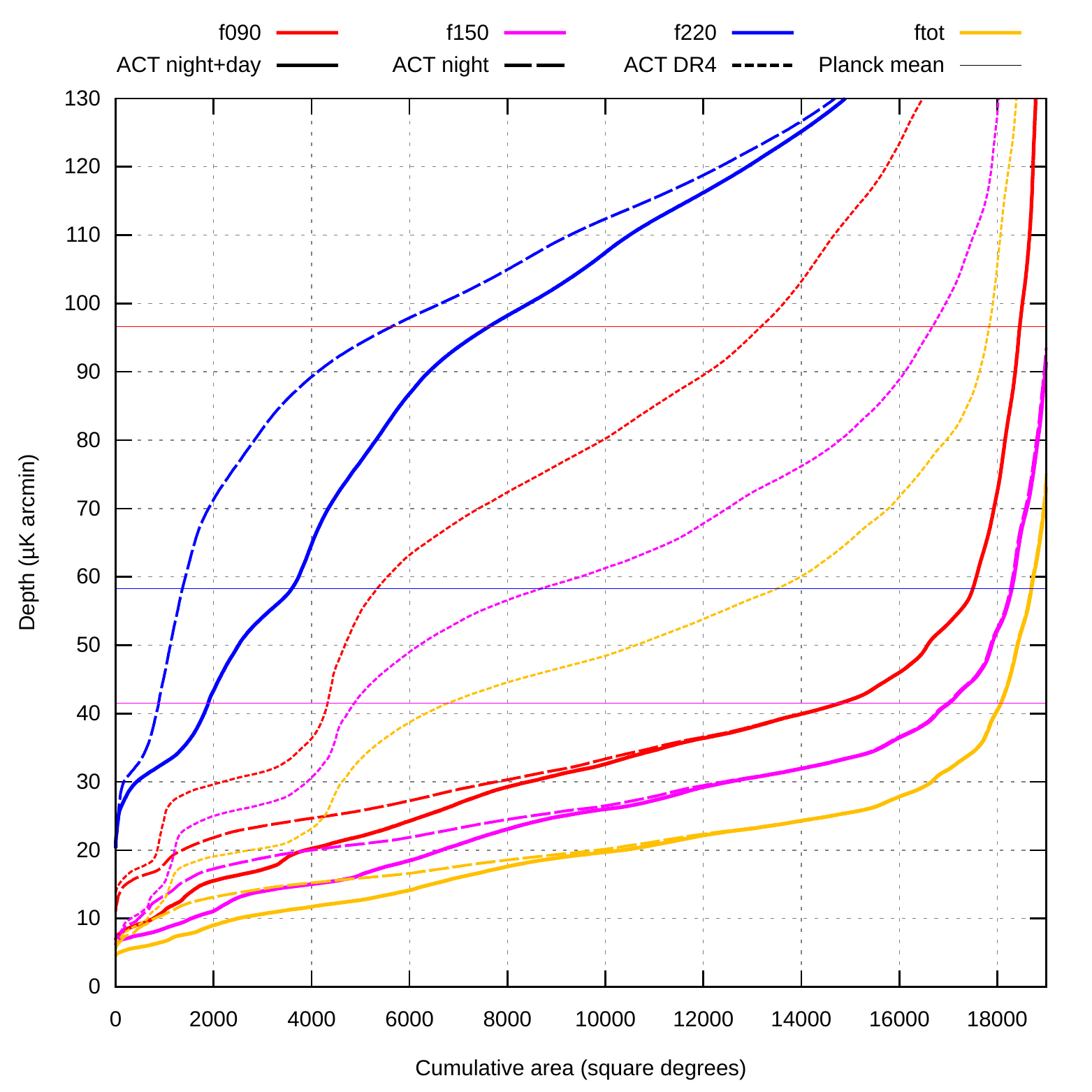}
	\caption{The white noise levels for the ACT-only coadds presented here, compared
	to that of ACT DR4 (dotted) and the average \planck\ map depths (horizontal lines). The curves show
	the total area of the survey with noise at a given level or lower.
	For example, these maps have 12\,000 square degrees where the noise level does not
	exceed 30 \textmu{}K-arcmin at f150. Alternatively, this graph can be interpreted as a
	cumulative distribution function for the map depth if the axes are transposed.
	The day+night (solid) and night-only (dashed) curves converge at the largest areas because
	the daytime data only contribute to a small subset of the survey area.}
	\label{fig:depth1d}
\end{figure}

\begin{figure}[p]
	\centering
	\hspace*{-10mm}\begin{tabular}{>{\centering\arraybackslash}m{2mm}>{\centering\arraybackslash}m{5.5cm}>{\centering\arraybackslash}m{5.5cm}>{\centering\arraybackslash}m{5.5cm}}
		& \bf f090 & \bf f150 & \bf f220 \\
		\rotatebox[origin=c]{90}{\hspace{10mm} \bf D6} &
		\img[trim=8mm 3mm 10mm 4mm]{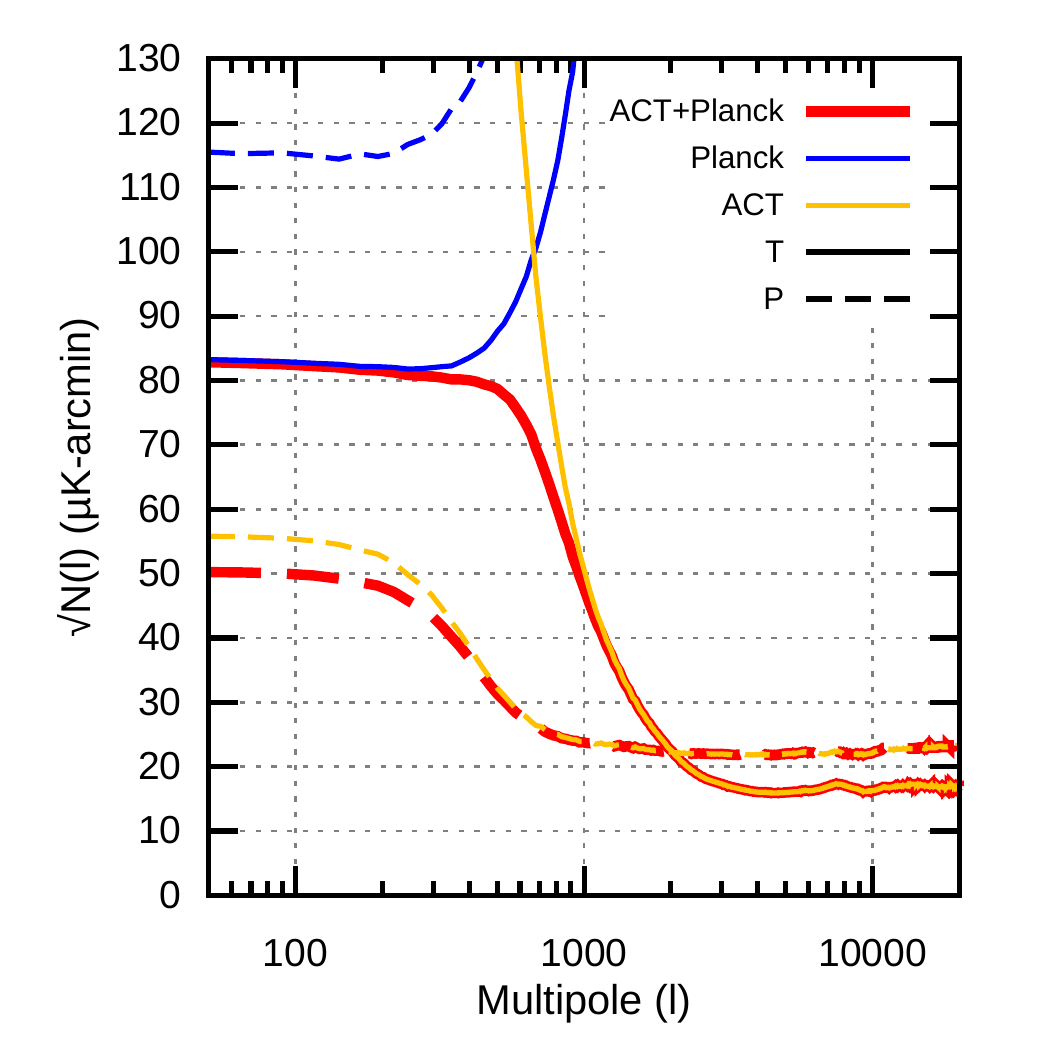} &
		\img[trim=8mm 3mm 10mm 4mm]{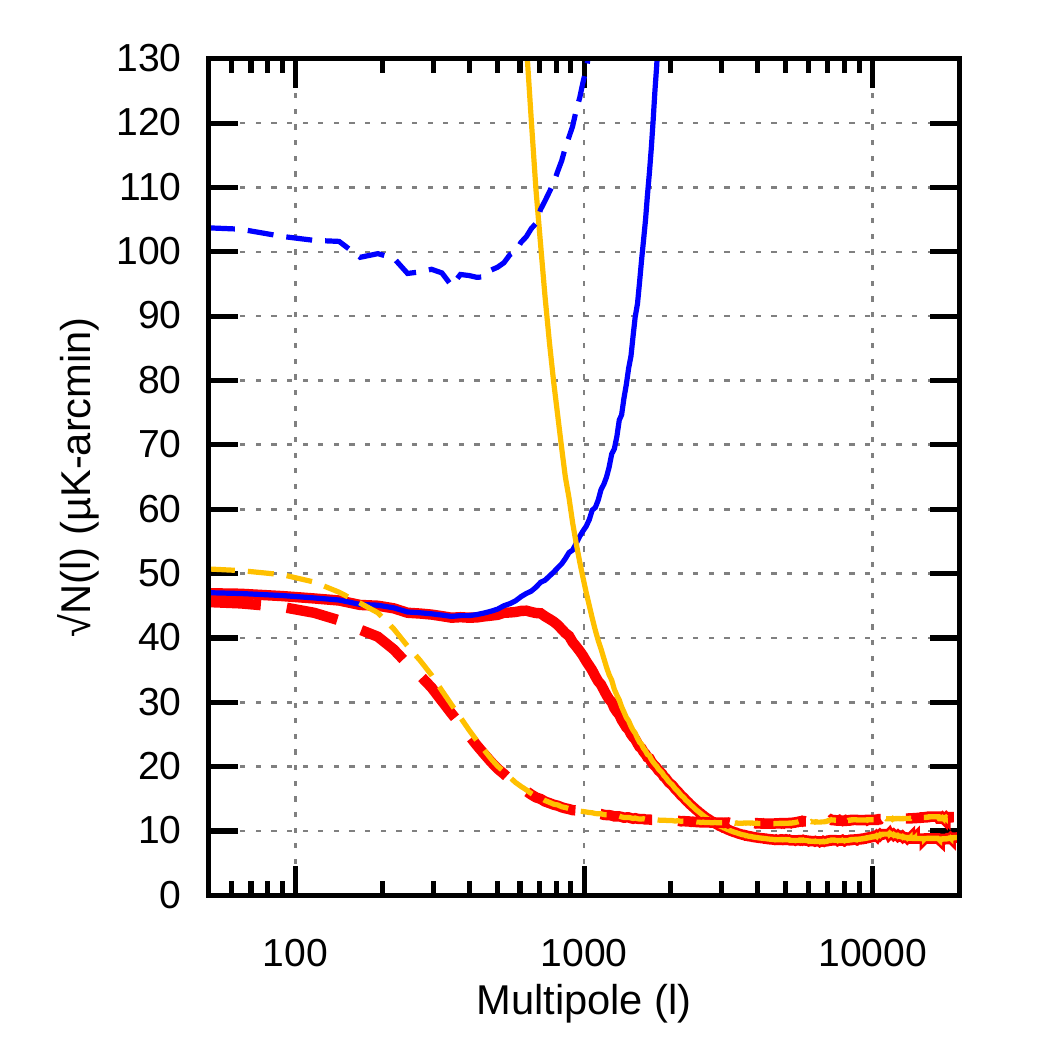} &
		\img[trim=8mm 3mm 10mm 4mm]{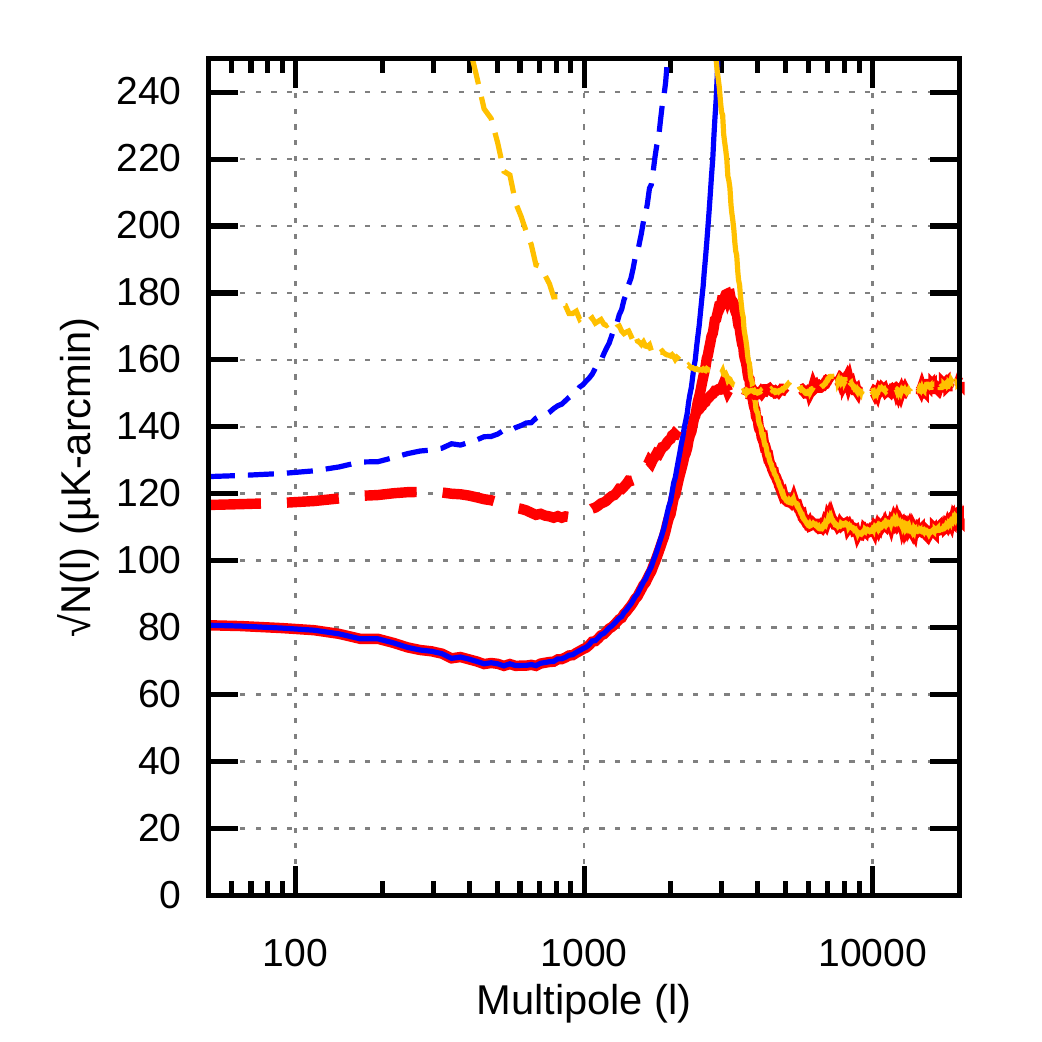} \\
		\rotatebox[origin=c]{90}{\hspace{10mm} \bf Day-N} &
		\img[trim=8mm 3mm 10mm 4mm]{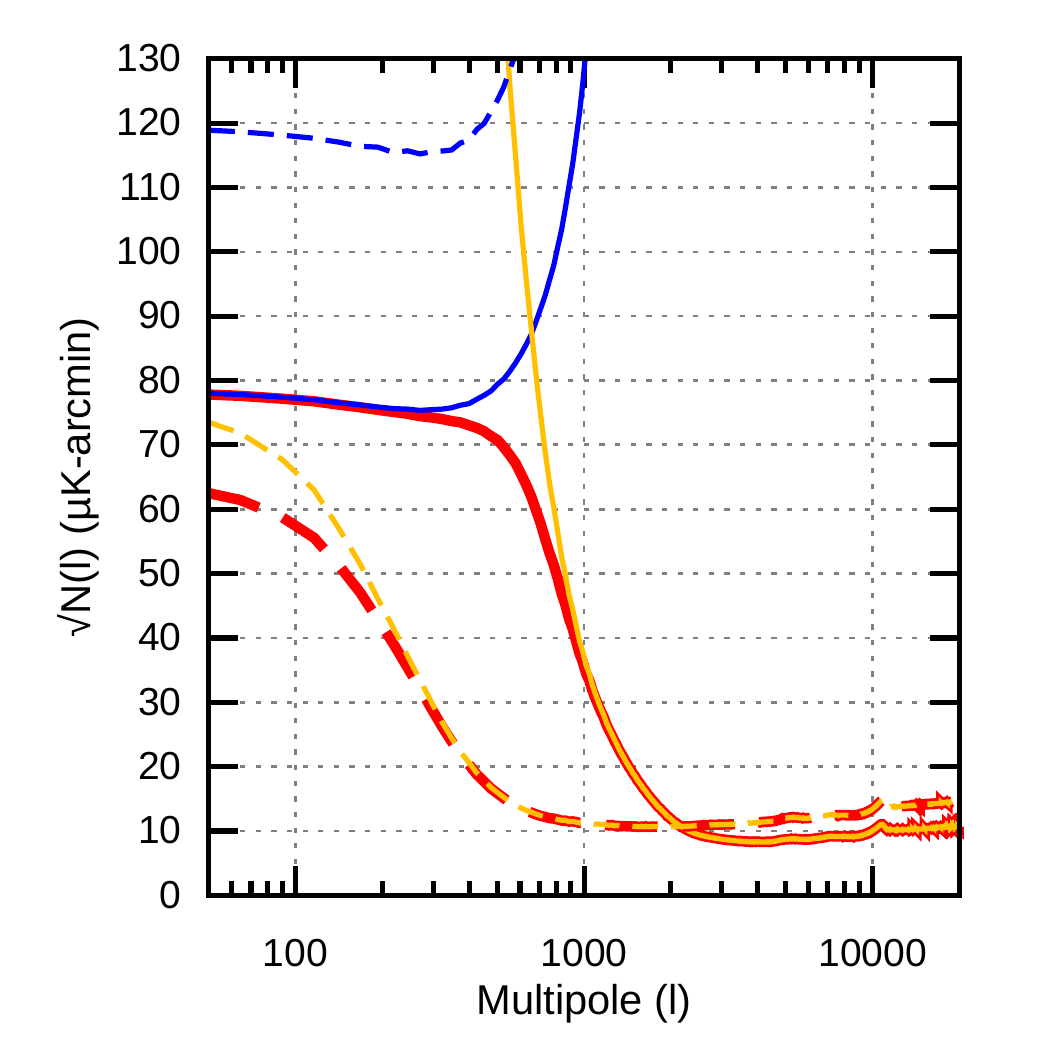} &
		\img[trim=8mm 3mm 10mm 4mm]{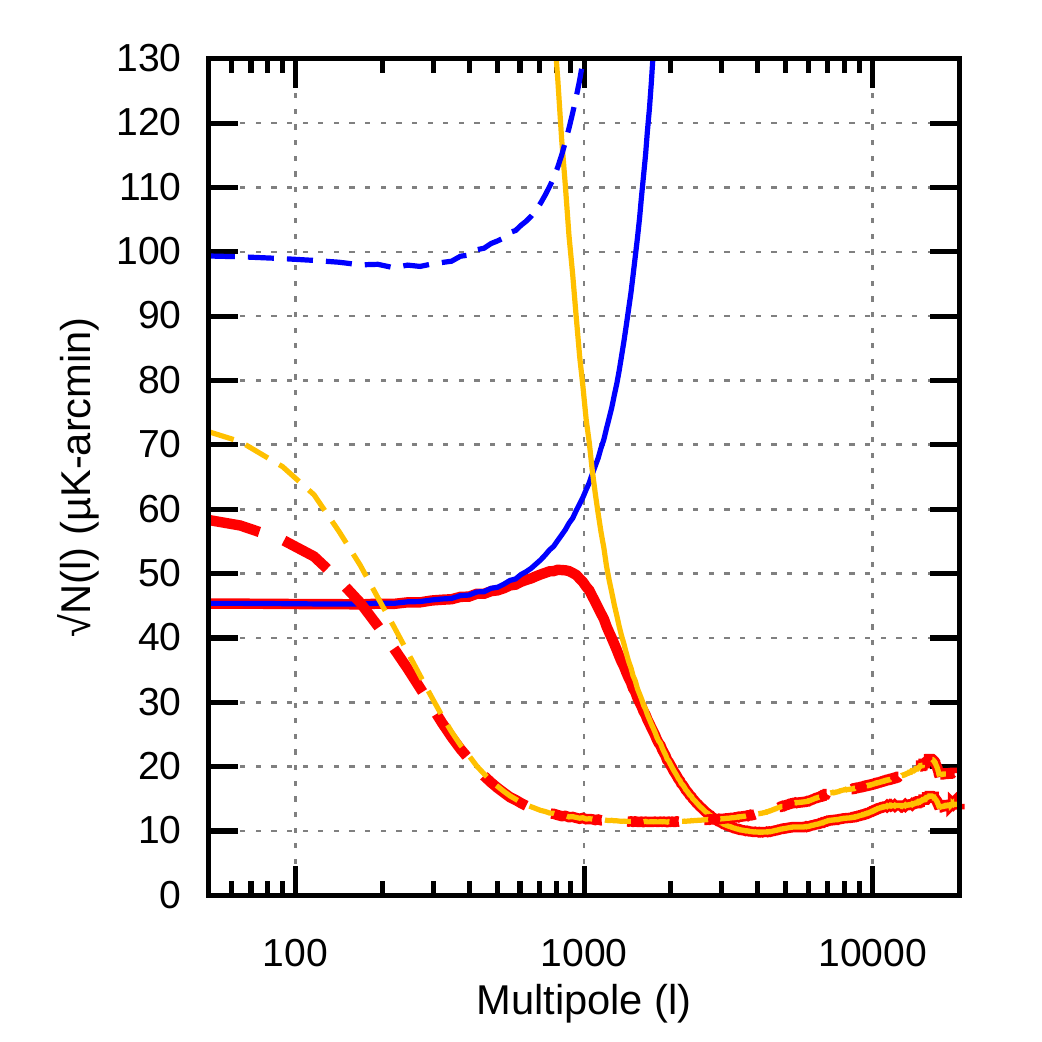} &
		\img[trim=8mm 3mm 10mm 4mm]{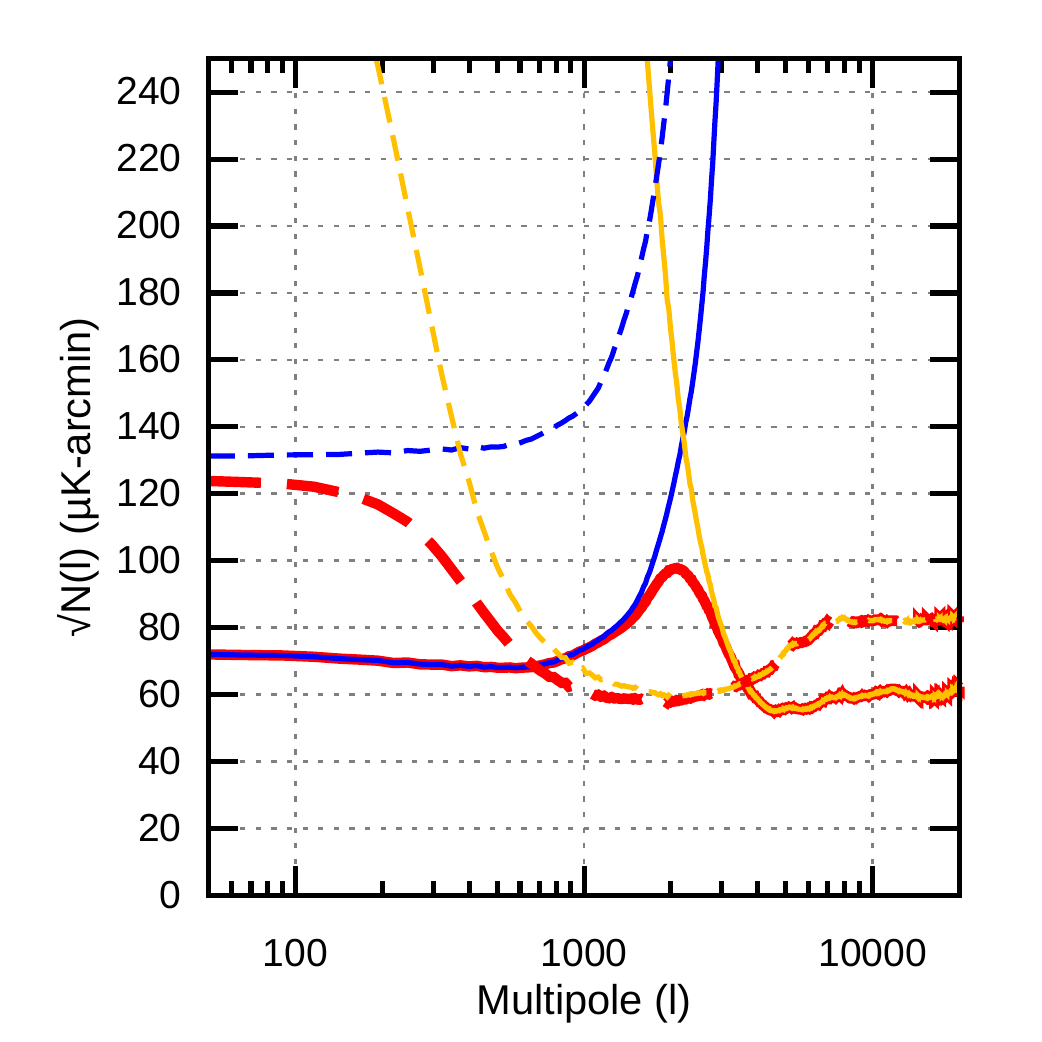} \\
		\rotatebox[origin=c]{90}{\hspace{10mm} \bf AA} &
		\img[trim=8mm 3mm 10mm 4mm]{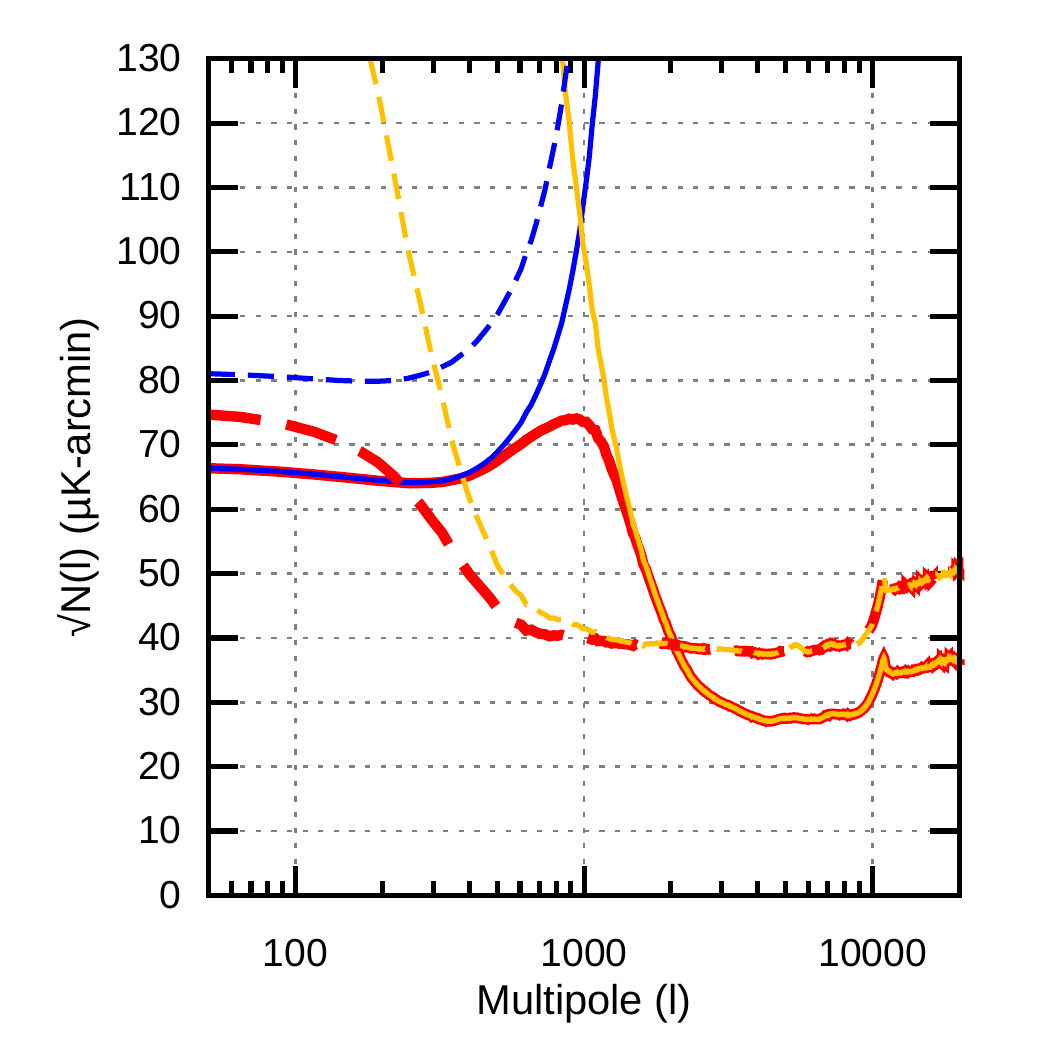} &
		\img[trim=8mm 3mm 10mm 4mm]{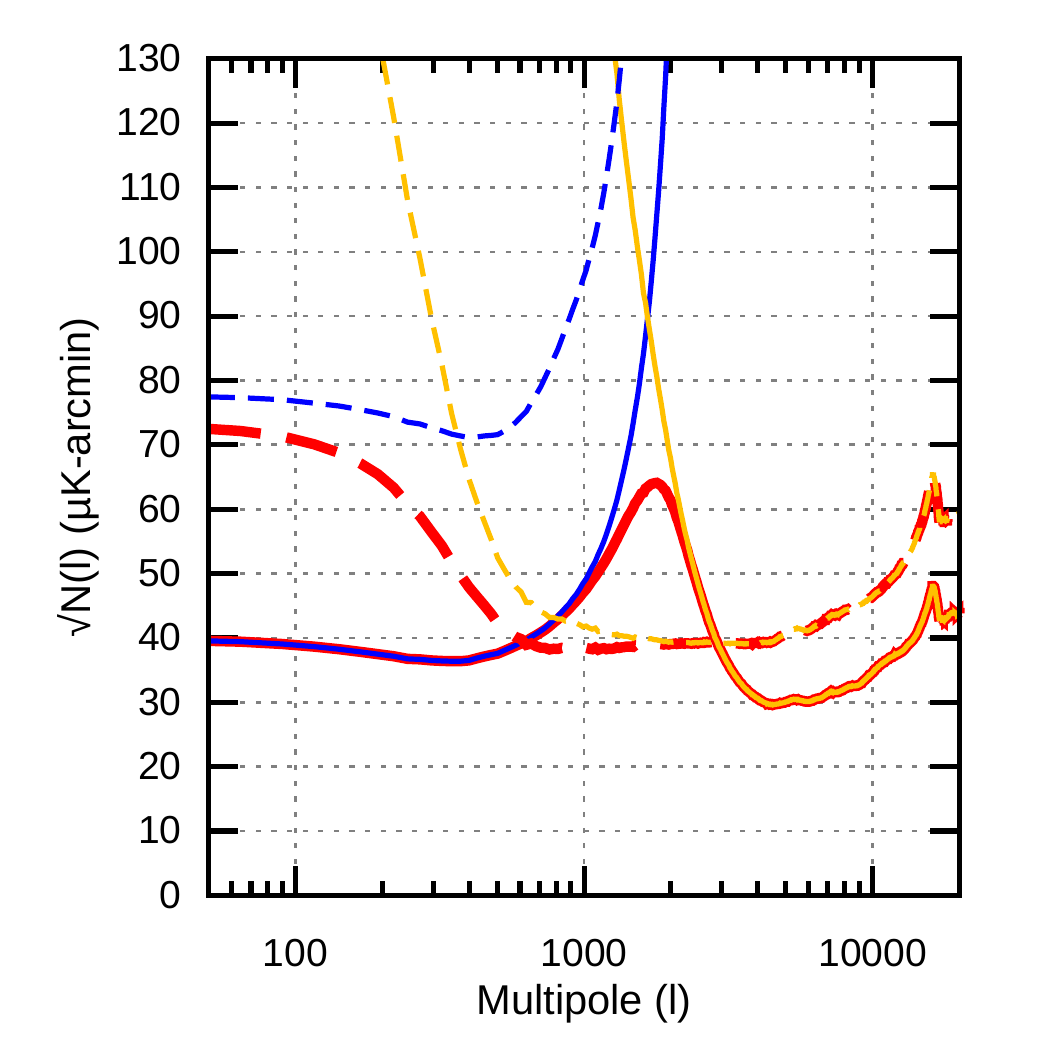} &
		\img[trim=8mm 3mm 10mm 4mm]{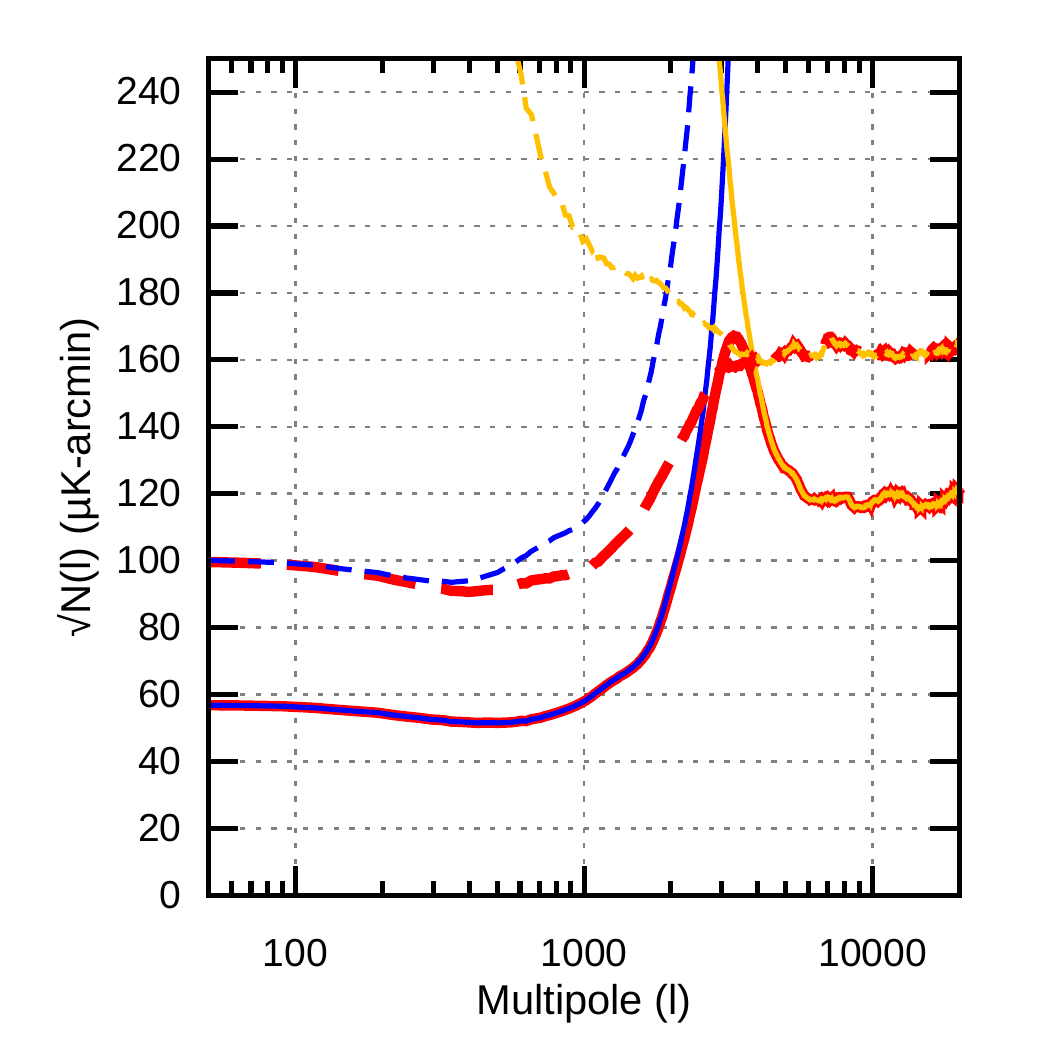}
	\end{tabular}
	\caption{The angular noise spectra of ACT+\planck\ (thick red), \planck\ only (blue) and ACT only (yellow) in
	total intensity (solid) and polarization (dashed). The columns represent the f090, f150 and f220 frequency
	bands from left to right. Each row corresponds to a 10$^\circ$x10$^\circ$ patch
	centered on different locations on the sky.
	\dfn{D6}: $32^\circ < \textrm{RA} < 38^\circ$, $-8^\circ < \textrm{dec} < 2^\circ$, the middle of the deep,
	night-time only patch ``D6" from ACT DR2-DR4.
	\dfn{Day-N}: $230^\circ < \textrm{RA} < 240^\circ$, $8^\circ < \textrm{dec} < 18^\circ$, part of the
	deep northern daytime patch.
	\dfn{AA}: $-5^\circ < \textrm{RA} < 5^\circ$, $-25^\circ < \textrm{dec} < -15^\circ$, an area representative
	of the wide, shallow Advanced ACT survey area. The noise curves shown are the square root of the
	noise spectra in units \textmu{}K arcmin to make it easier to compare with the depth maps. The \planck\
	noise blows up at high $\ell$ due to its large beam. The ACT noise blows up at low $\ell$ due to the atmosphere --
	this happens at lower $\ell$ in polarization due to strong suppression of the atmosphere there.
	The high-$\ell$ upturn seen in some of the plots is caused by significant contribution
	from data with larger beam than the output beam, resulting in net deconvolution.}
	\label{fig:noisespecs}
\end{figure}

Figure~\ref{fig:depth1d} shows the cumulative total intensity white noise depth distribution of the ACT-only maps.
This is the noise level appropriate at high $\ell$ where atmospheric fluctuations are sub-dominant
(see figure~\ref{fig:noisespecs}). The maps span a wide range of depths,
with 2500 square degrees deeper than 10 \textmu{}K-arcmin, 6500 square degrees deeper than 15 \textmu{}K-arcmin,
10\,000 square degrees deeper than 20 \textmu{}K-arcmin, 15\,000 square degrees deeper than 25 \textmu{}K-arcmin
and 19\,000 square degrees deeper than 70 \textmu{}K-arcmin when combining the three frequencies. Over most
of the survey area, these maps are $2.3/2.6/\infty$ times deeper than ACT DR4
at f090/f150/f220 (see appendix~\ref{sec:coadd-dr4});
and they are deeper than the mean \planck\ depth over 19k/17k/4k square
degrees. The polarization noise level is approximately $\sqrt{2}$ higher.\footnote{This
is in contrast to \planck\, where the polarization to intensity noise ratio varies from
$\approx \sqrt{2}$ (f220) to $>2$ (f090).}

\begin{figure}[h!]
	\centering
	\begin{closetabrows}
	\hspace*{-18mm}\begin{tabular}{m{2mm}m{19.2cm}}
		\rotatebox[origin=c]{90}{f090} & \img[trim=0 5.7mm 0 0.0mm,clip]{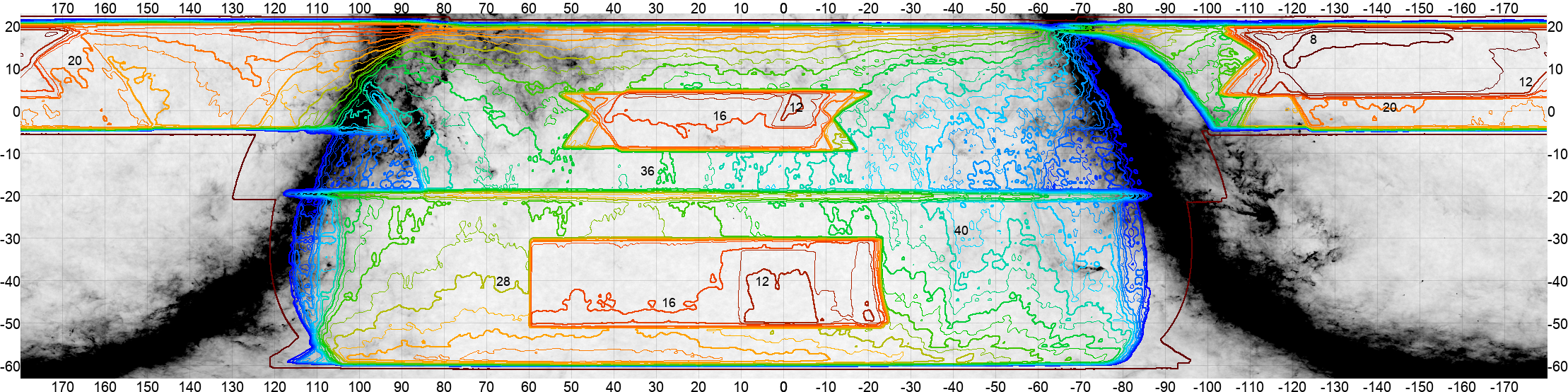} \\
		\rotatebox[origin=c]{90}{f150} & \img[trim=0 5.7mm 0 6.8mm,clip]{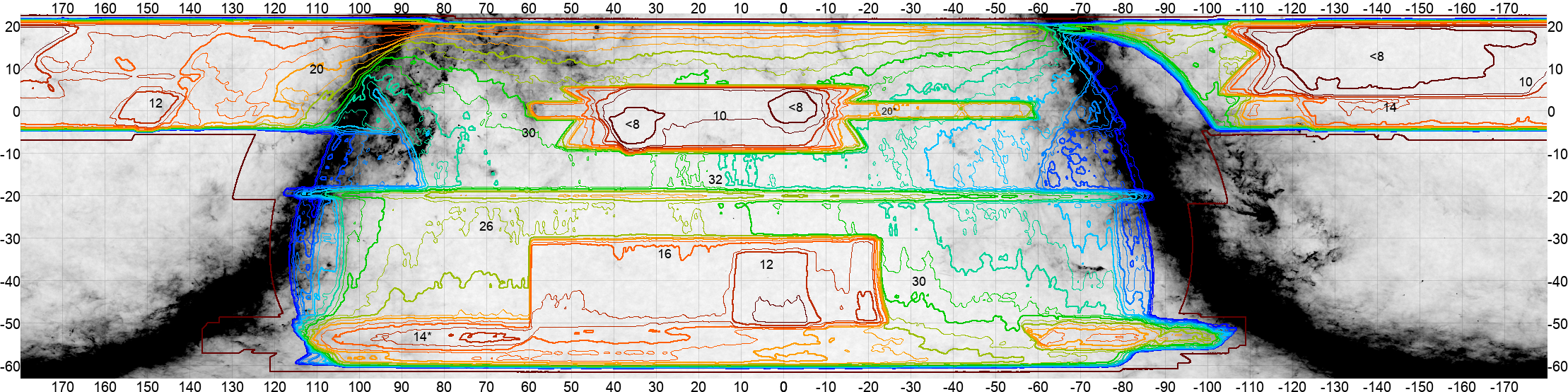} \\
		\rotatebox[origin=c]{90}{f220} & \img[trim=0 0.0mm 0 6.8mm,clip]{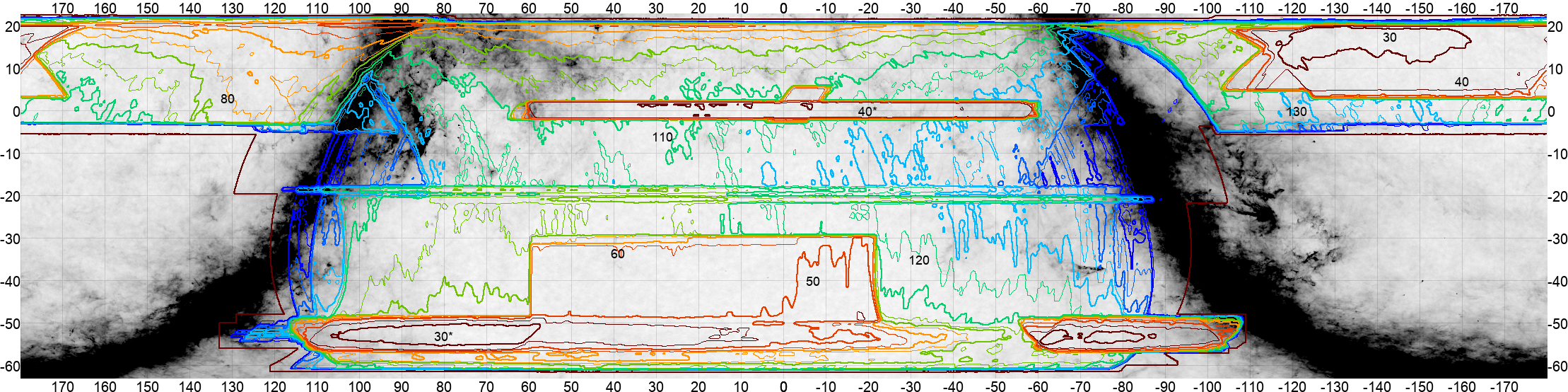}
	\end{tabular}
	\end{closetabrows}
	\caption{Spatial distribution of ACT-only map depths for the three different ACT frequencies,
	as shown by iso-depth contours that go from 8/8/30 \textmu{}K-arcmin (red) to 60/50/180 \textmu{}K-arcmin (blue)
	with contour intervals of 2/2/10 \textmu{}K-arcmin for f090/f150/f220 respectively. This covers most
	of the depth range in the maps, though there are areas that are somewhat deeper or shallower.
	The outermost red curve shows the full outline of the survey area. Selected contour lines
	are labeled with their depth to make it easier to read off values. These are T noise levels;
	Q and U are approximately $\sqrt{2}$ higher. They also only describe the noise level at small scales.
	The background grayscale map is the dust-dominated \planck\ 353 GHz map.}
	\label{fig:depth2d}
\end{figure}

Figure~\ref{fig:depth2d} shows the spatial distribution of the map depths overlaid on the
dust-dominated \planck\ 353 GHz map. The deepest region is the Day-N region centered on
RA = -145$^\circ$, which is deeper than 8 \textmu{}K-arcmin at both f090 and f150; followed by the
night-only regions D5, D6 and D56 that were the focus of ACT DR2 and ACT DR3. The ACT
observing strategy does not target the galaxy, but some of it is still hit due to
limitations in the scanning pattern implementation, in particular
the region around RA = 90$^\circ$, which includes the Orion nebula.

\begin{figure}[h]
	\centering
	\begin{closetabcols}
	\begin{tabular}{ccc}
		\bf D6 & \bf Day-N & \bf AA \\
		\includegraphics[height=60mm,trim= 2mm 2mm 6mm 2mm,clip]{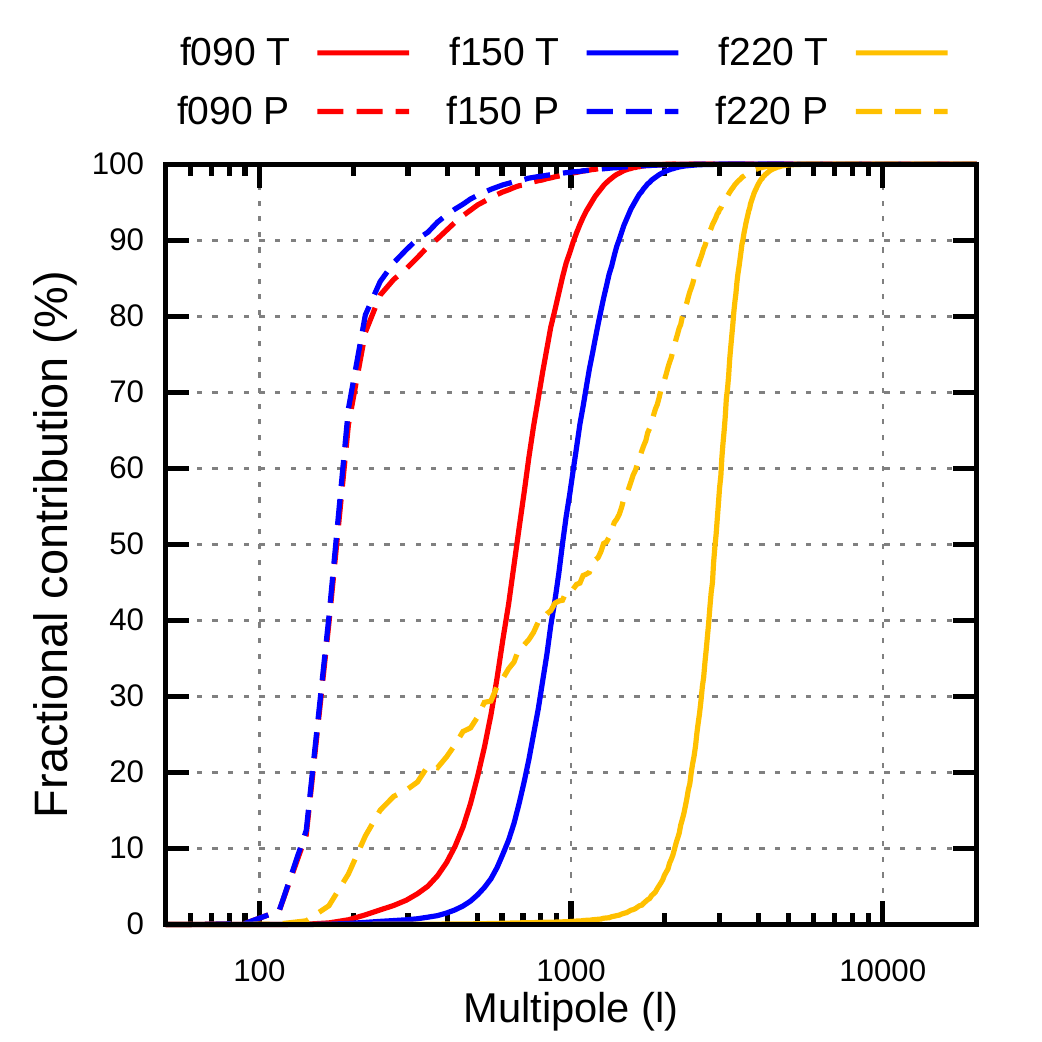} &
		\includegraphics[height=60mm,trim=16mm 2mm 6mm 2mm,clip]{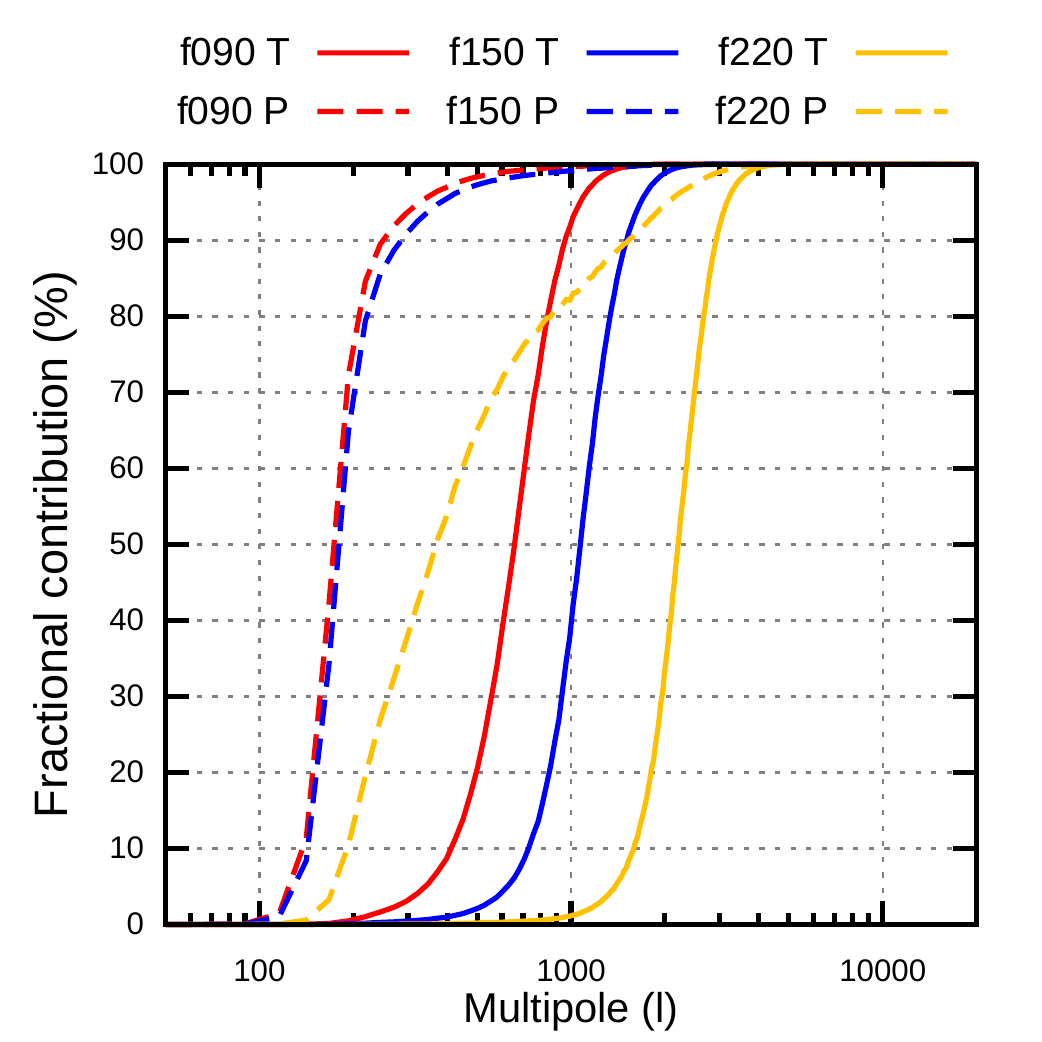} &
		\includegraphics[height=60mm,trim=16mm 2mm 6mm 2mm,clip]{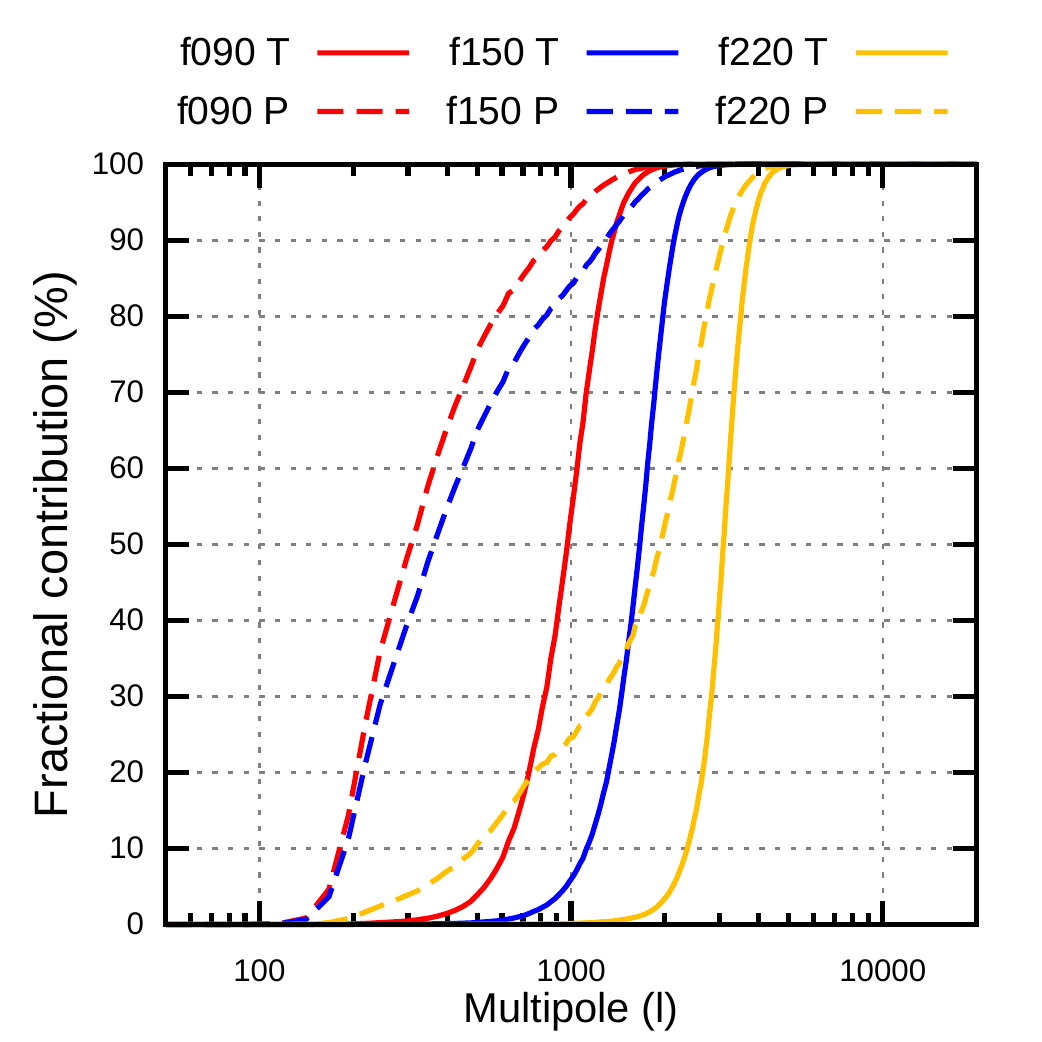}
	\end{tabular}
	\end{closetabcols}
	\caption{The fractional contribution of ACT to the ACT+\planck\ coadd as a function of angular
	scale for the same three patches as in figure~\ref{fig:noisespecs}. This is only approximate, as
	the real weights are anisotropic and position-dependent and hence can't be represented as a
	simple function of scale. The transition from \planck\ to ACT dominance is driven both by
	the \planck\ beam size and the ACT atmospheric 1/$\ell$ noise, as seen in figure~\ref{fig:noisespecs}.
	Note: D6 is much shallower in f220 than the other bands. In shallow areas, ACT dominates
	at $\ell > 1000/1700/3000$ in T and $300/360/1700$ in P, for f090/f150/f220.
	In deep areas these numbers are about 1/3 lower.}
	\label{fig:actfrac}
\end{figure}

The map white noise depth gives an incomplete picture of the map noise properties due to the
strongly scale-dependent atmospheric noise in the ACT maps, as shown in figure~\ref{fig:noisespecs}.
This noise has an $\ell_\textrm{knee}$ of about 2000/3000/4000 in total intensity and 500/500/700 in polarization
at f090/f150/f220, and increases rapidly ($N_l \propto \ell^{-3}$) below that. When coadding with \planck\
this rise stops at the \planck\ noise level, resulting in maps with much flatter noise curves.
This noise behavior is reflected in the weights ACT and \planck\ data get in the combined maps.
While the full weights are defined in a hybrid of Fourier space and pixel space, we can approximate
them as simple functions of $\ell$ in limited areas. Figure~\ref{fig:actfrac} uses this to show the approximate
fractional contribution of ACT to the ACT+\planck\ coadd as a function of angular scale in three different
areas of the sky.

\subsection{Coarse-grained noise model}
In addition to the purely spatial (figure~\ref{fig:depth2d}) and purely angular scale (figure~\ref{fig:noisespecs})
slices through the noise model, we also compute the approximate inverse noise variance $\tilde N^{-1}$
in units of $1/$\textmu{}K$^2$ per full-resolution (0.5 arcmin) pixel as a function of position in the map
(sampled at every 0.5$^\circ$ in
RA and dec), multipole (50 logarithmically spaced multipoles from 100 to 18000), detector array (e.g.
ACT MBAC AR1, ACTPol PA3 f090 or \planck\ 217) and Stokes parameter (I, Q, U). These files are labeled
``noisebox'' in the data release.

\subsection{Bandpasses}
\label{sec:bandpasses}
No attempt is made to correct for the bandpass differences between the individual
maps in each bandpass group (see figure~\ref{fig:bands}). This results in somewhat
scale-dependent effective bandpasses and band-center, with the main
feature being the transition from\planck-dominance to ACT-dominance around $\ell \sim 1000$
in the ACT+\planck\ maps (see figure~\ref{fig:actfrac}). To estimate how large an effect
this is, we first normalize the bandpasses for all detector arrays in figure~\ref{fig:bands} to units
of \textmu{}K/(MJy/sr)/GHz to make them comparable to each other:
\begin{align}
	f(\ell,\nu) &= \frac{f^\textrm{raw}(\ell,\nu)}{\int d\nu' f^\textrm{raw}(\ell,\nu') \delta B(\nu')}.
\end{align}
Here $\nu$ is the frequency of the light, $B(\nu,T)$ is the black-body spectrum of the CMB in MJy/sr,
$\delta B(\nu) = \frac{\partial B(\nu',T)}{\partial T}\Big|_{T=2.725 K}$ is the spectral response to
CMB temperature fluctuations, and the normalization is based on the
fact that all maps are defined to have unit response to these fluctuations since they are in linearized
CMB temperature units.
$f^\textrm{raw}(\ell,\nu)$ is the unnormalized bandpass for each array. This is scale-dependent (a function
of $\ell$) because the telescope beams get slightly sharper towards the high-frequency end of each bandpass.
Assuming a beam size that scales as $1/\nu$, as for a normal diffraction limited system, we get
\begin{align}
	f^\textrm{raw}(\ell,\nu) &= f^\textrm{raw}(\nu) B\Big(\ell \frac{\nu_0}{\nu}\Big)
\end{align}
where $B(\ell)$ is the instrument beam and $\nu_0 = \frac{\int d\nu f^\textrm{raw}(\nu) \delta B(\nu)}{
	\int d\nu \delta B(\nu)}$ is the effective beam band-center for CMB fluctuations.

\begin{figure}[h]
	\centering
	\begin{closetabcols}[0.2mm]
	\begin{tabular}{cc}
		\small ACT+\planck\ bandpasses & \small ACT+\planck\ band centers \\
		\includegraphics[width=9cm,trim=0 15mm 0 2mm,clip]{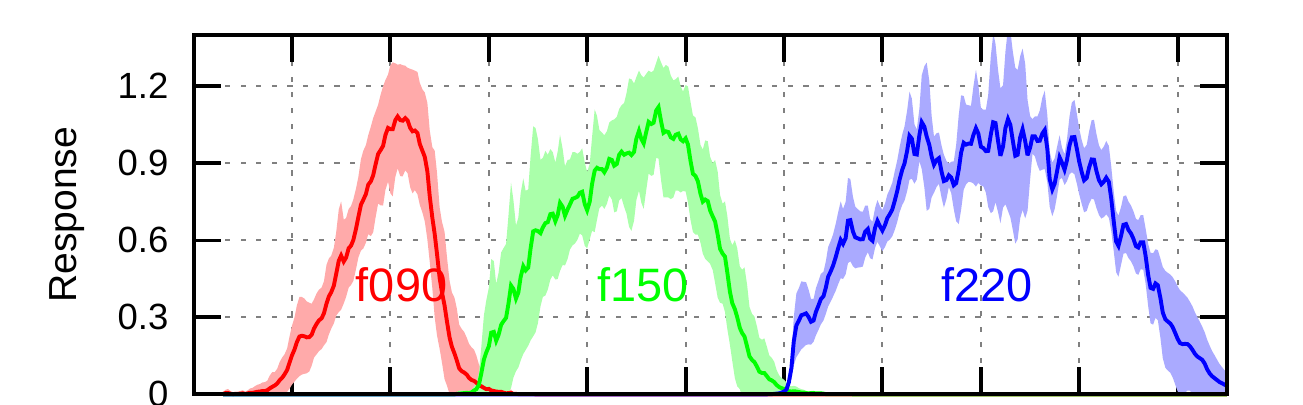} &
		\multirow{3}{*}{\includegraphics[width=9.0cm,trim=0 0 0 33mm]{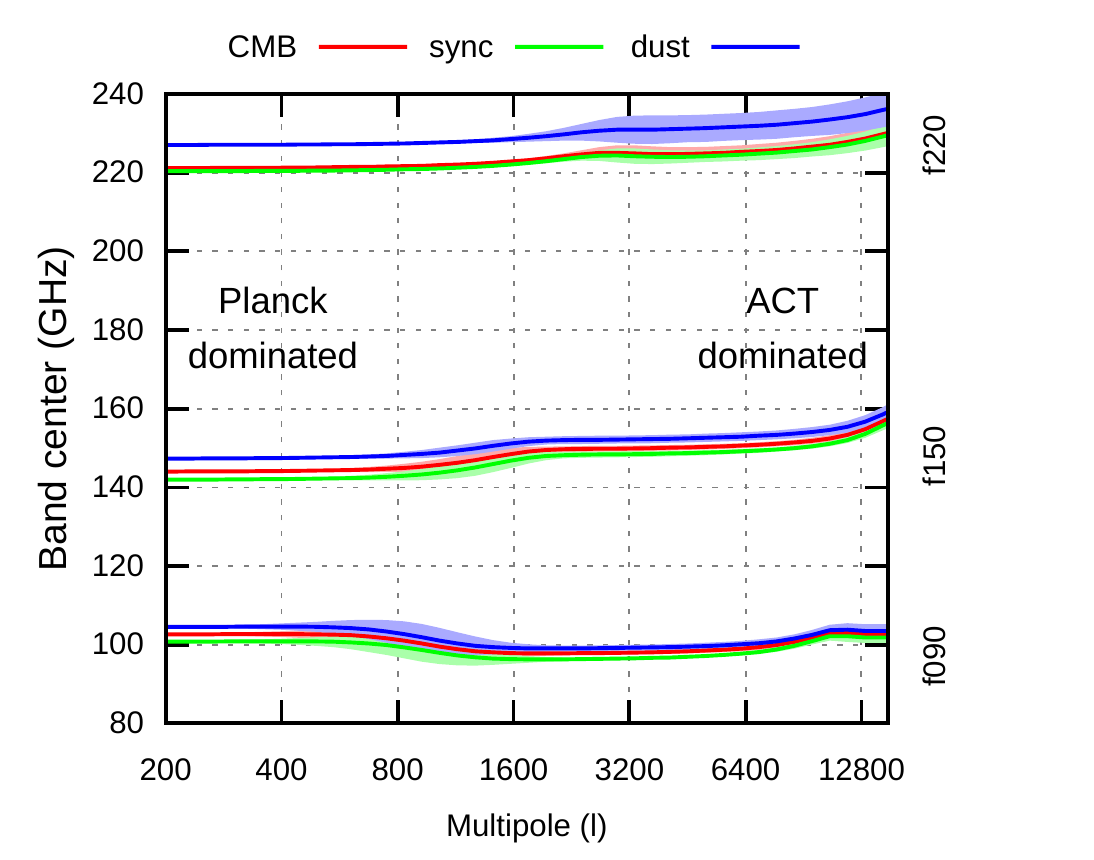}} \\
		\small ACT only bandpasses & \\
		\includegraphics[width=9cm,trim=0 0 0 4mm]{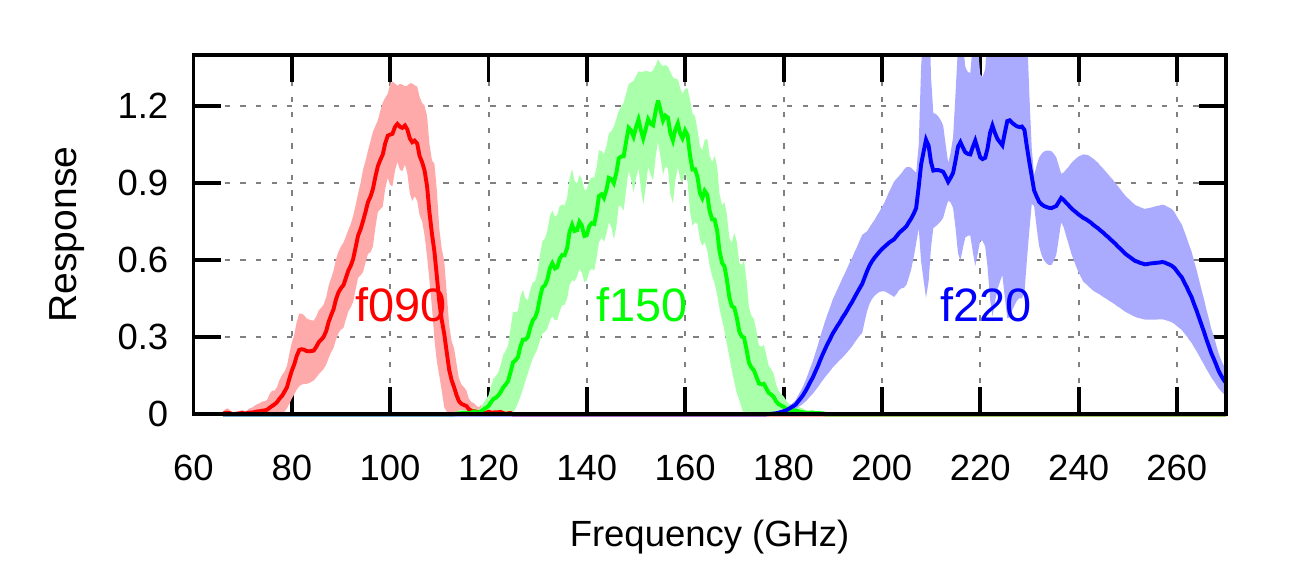} &
	\end{tabular}
	\end{closetabcols}
	\caption{\dfn{Left}: The average bandpass of the ACT+\planck\ (top) and ACT-only (bottom)
	combined maps normalized to a peak of $\approx 1$ to make them easier to plot on the same scale.
	The colored band shows the $\pm 1 \sigma$ inverse variance weighted variation of the bandpass
	across positions and scales in the maps. \dfn{Right}: The sky-averaged band-center of the ACT+\planck\
	maps as a function of multipole. The transition between \planck-dominance and ACT-dominance is clearly
	visible at $\ell \sim 1000$, and represents a 2--5\% shift. The band around each curve shows the variation
	of the bandcenter across the sky. The band-center is computed for a CMB spectrum (red), synchrotron
	spectrum (green, $\propto \nu^{-0.5}$) and dust (blue, modified blackbody with $T=19.6$K and $\beta = 1.59$).
	The frequency-dependence of the beam (slightly higher resolution at the high end of the band than
	the low) is taken into account, and leads to a slight up-tilt of the band-centers at high $\ell$.}
	\label{fig:average-bandpass}
\end{figure}

\begin{figure}[h]
	\centering
	\begin{closetabcols}[0mm]
	\begin{tabular}{ccc}
		\bf f090 & \bf f150 & \bf f220 \\
		\includegraphics[height=6.1cm,trim= 0mm 0mm 2mm 0mm,clip]{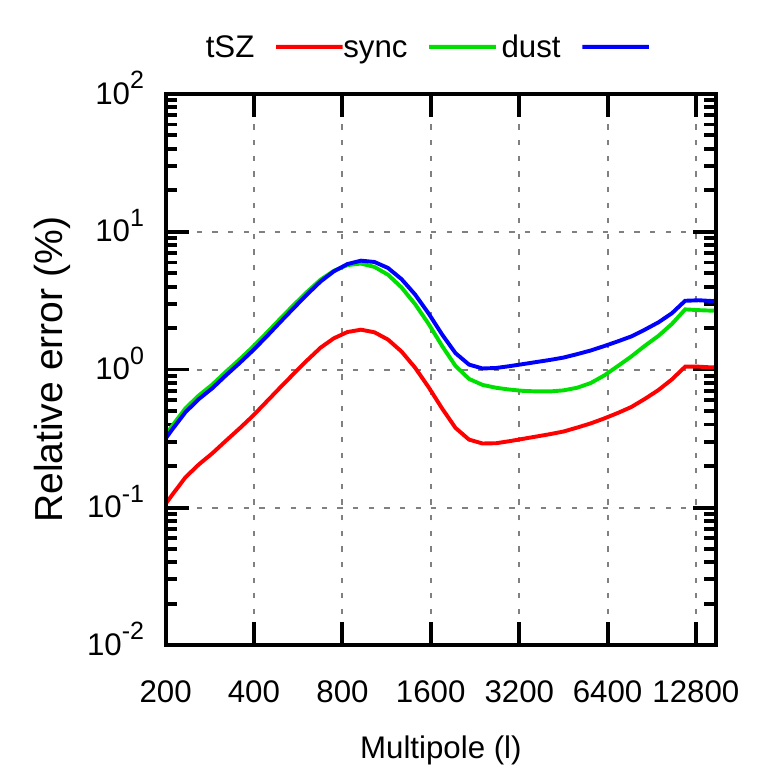} &
		\includegraphics[height=6.1cm,trim=16.7mm 0mm 2mm 0mm,clip]{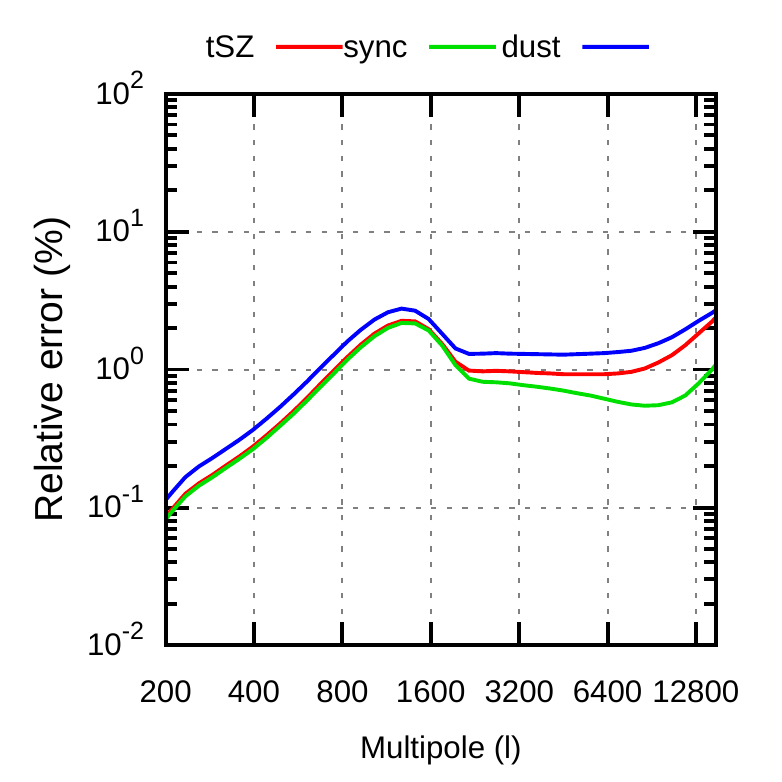} &
		\includegraphics[height=6.1cm,trim=16.7mm 0mm 2mm 0mm,clip]{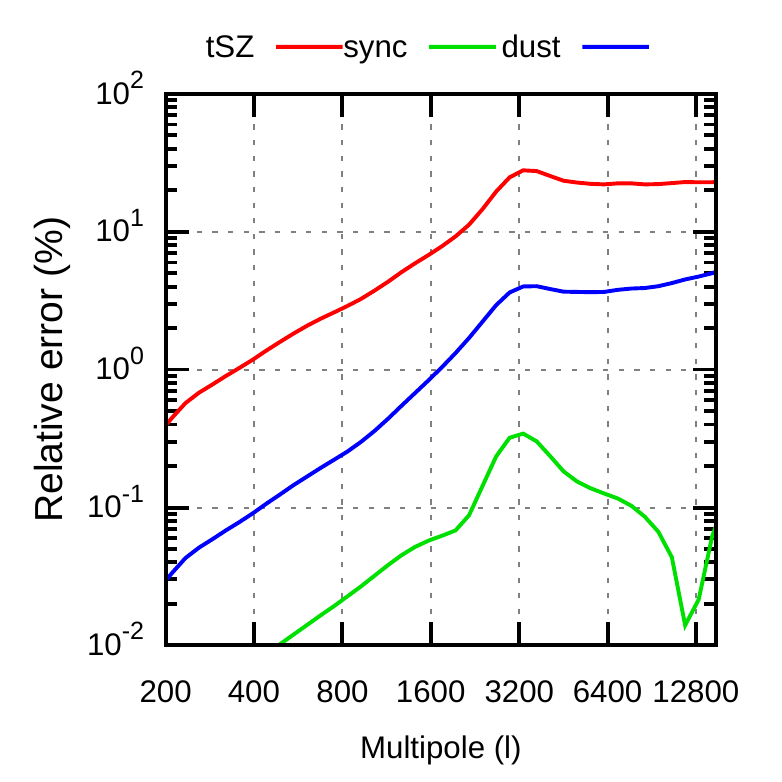}
	\end{tabular}
	\end{closetabcols}
	\caption{The relative standard deviation (standard deviation divided by the mean) of the ACT+\planck\
	map's response to the thermal Sunyaev Zel'dovich effect (red), synchrotron (green) and dust (blue)
	as a function of the multipole ($\ell$) at f090 (left), f150 (middle) and f220 (right) when averaged
	across the survey area with the inverse noise model as weights. \textbf{The response typically varies
	by about 1\% across the map.} The high value for tSZ at f220 is due to the insensitivity of f220
	to tSZ. This figure takes into account the frequency-dependence of the beam (the resolution is
	slightly higher at the high frequencies of a bandpass than the low frequencies), but does not
	include the overall calibration uncertainty of the map, which is O(1\%). The CMB itself is not
	included because the maps are calibrated to the CMB, which causes all its response uncertainty
	to be in the gain calibration rather than the bandpasses. The synchrotron and dust spectra used
	here are the same as in figure~\ref{fig:average-bandpass}.}
	\label{fig:bandpass-response}
\end{figure}

Given these normalized per-array bandpasses we can find the effective bandpass for
Stokes component $s$ at position ($\alpha$,$\delta$) and multipole $\ell$ in the map
as the inverse variance weighted average over the arrays $i$,
\begin{align}
	f^\textrm{eff}(s,\ell,\alpha,\delta,\nu) = \frac{\sum_i f_i(\ell,\nu) \tilde N^{-1}(s,i,\ell,\alpha,\delta)}{\sum_i \tilde N^{-1}(s,i,\ell,\alpha,\delta)} \label{eq:feff}
\end{align}
Figure~\ref{fig:average-bandpass} shows the mean and standard deviation of $f^\textrm{eff}$ when averaged
over the whole survey area and all scales. The bandpass varies at the $\sim 15\%$ level in the ACT-only
maps, but this only results in a $\sim 0.5\%$ variation in the band-center. For the ACT+\planck\ maps
the bandpass still varies by $\sim 15\%$ across positions and scales on the map, but there is now a
2--5\% difference between the bandpass at large ($\ell \lesssim 1000$) and small ($\ell \gtrsim 1000$) scales
(figure~\ref{fig:average-bandpass} right panel). As seen in figure~\ref{fig:bandpass-response}, the
ACT+\planck\ map's response to individual components like tSZ, synchrotron or dust has a position-dependence
of O(1\%). If more accurate control over bandpasses is needed,
they can be evaluated at any needed multipole and position using equation~\ref{eq:feff} (see appendix~\ref{sec:app-bandpass}).

\subsection{Caveats and limitations}
\begin{enumerate}
	\item These maps include preliminary data from the 2017--2018 seasons of Advanced ACT, and while a fair
		amount of work has gone into characterizing them, they have not yet been subjected to all the tests
		of a proper ACT cosmological data release, and gain/beam errors of several percent should be expected. This is doubly
		true for the daytime maps, where there is a position-dependent beam FWHM uncertainty of O(10\%).
	\item As part of the solution process, the input maps are reconvolved to a common target beam.
		For some parts of the sky this resulted in a net beam deconvolution, leading to an upwards slope of
		the high-$\ell$ noise spectrum. This is seen in the AA patch in figure~\ref{fig:noisespecs}, for example.
		This ``blue" noise does not represent an actual noise excess; it is simply a result of the choice of
		output beam. If white noise is desirable, one can simply reconvolve the maps to a slightly bigger
		beam.
	\item Many use cases benefit from having split maps made from independent subsets of the data
		in order to properly account for the noise properties. We do not provide such splits for the
		coadded maps we present here. In theory, we could make these by splitting each data set in two,
		and making one coadded map from all the first halves and another from all the second halves.
		However, because the noise model requires at least two splits, this requires us to have at
		least 4 splits for all the initial data sets, and this is not the case for large parts of our
		data set (see section~\ref{sec:act_data}).\footnote{It might seem tempting to split the data into
		two subsets {\it after} building the noise model, but this would result in the two subsets
		no longer being independent, since sample variance from each subset would leak into their
		shared noise model.} We hope to improve on this in a future combined map, either by greatly
		reducing the number of degrees of freedom in the noise model, or by coadding a set of
		noise simulations of the individual input maps.
	\item As discussed in section~\ref{sec:bandpasses}, no attempt is made to correct for the
		bandpass differences between the individual maps, resulting in a mild position dependent (0.5\%)
		and scale dependent (2--5\% if \planck\ is included, otherwise 0.5\%) effective band center.
	\item The ACT-only maps have a total intensity deficit at $\ell \lesssim 750/1200/1750$ for f090/f150/f220
		due to the contribution from the preliminary Advanced ACTPol maps, which are not yet up to the standard of
		the earlier ACT maps. This should have little impact on the ACT+\planck\ maps since \planck\ dominates for
		$\ell < 1000/1700/3000$ in the shallow areas where the preliminary Advanced ACTpol maps make up the
		dominant ACT contribution.
	\item There is a further low-$\ell$ loss of power from the ground filter we apply
		to all ACT maps (see figure~\ref{fig:ground}), which also mostly affects $\ell < 1000$
		($<0.5$\% above that, but growing to 10\% by $\ell = 350$, see appendix~\ref{sec:sims}).
		This is again mitigated, but not eliminated, by the presence of \planck\ data in the ACT+\planck\ coadds.
\end{enumerate}

\FloatBarrier

\section{Map Images and Features}
\label{sec:map_features}

The main products of this paper are multi-frequency temperature and polarization maps that have the resolution and sensitivity of ACT on small scales and the simple noise properties and sensitivities of \planck\ on large-scales.

\begin{figure}[p]
	\centering
	\begin{closetabrows}
	\begin{tabular}{m{2mm}m{15cm}}
		T & \img[width=12cm,trim=0 5.5mm 0 0.0mm,clip]{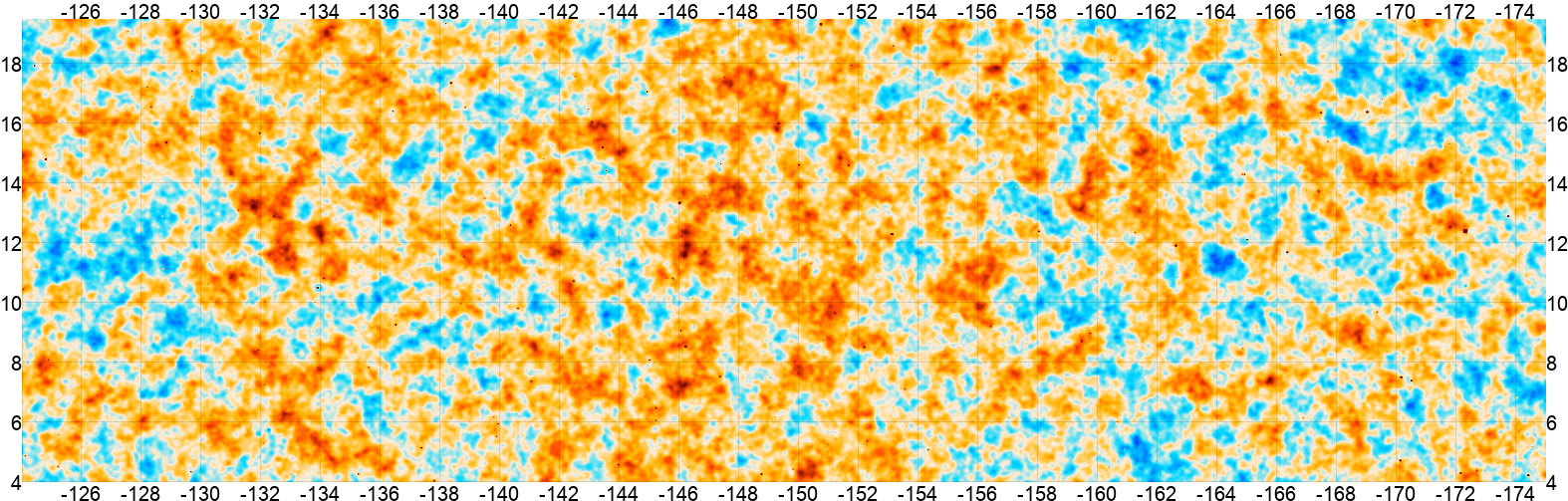} \\
		Q & \img[width=12cm,trim=0 5.5mm 0 6.5mm,clip]{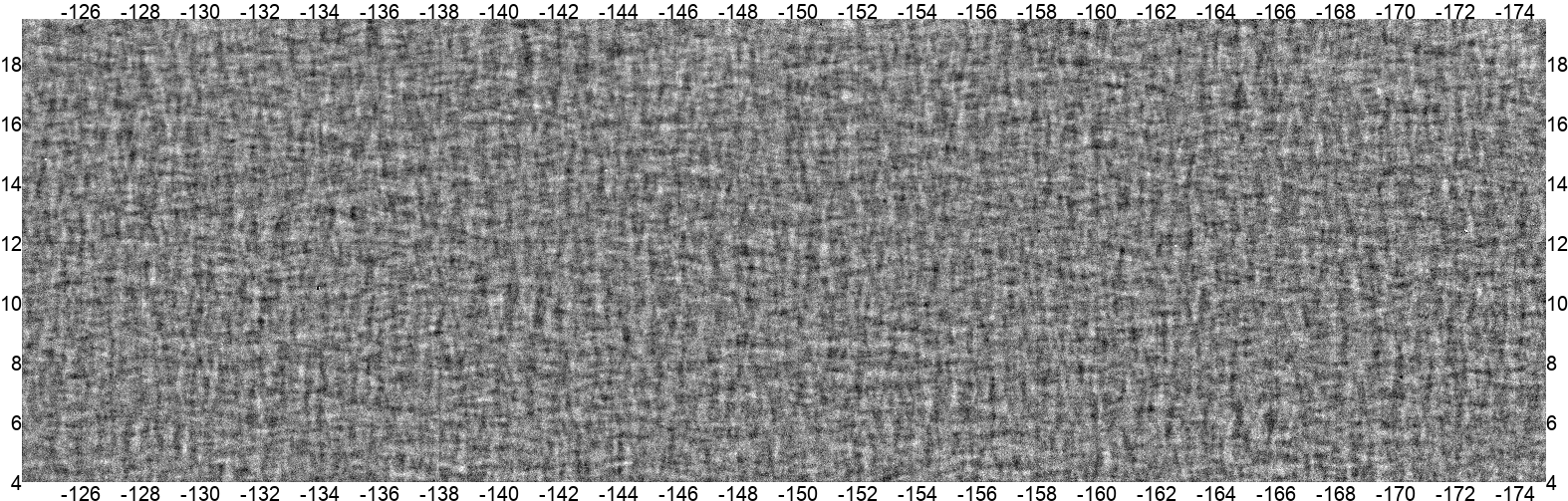} \\
		U & \img[width=12cm,trim=0 5.5mm 0 6.5mm,clip]{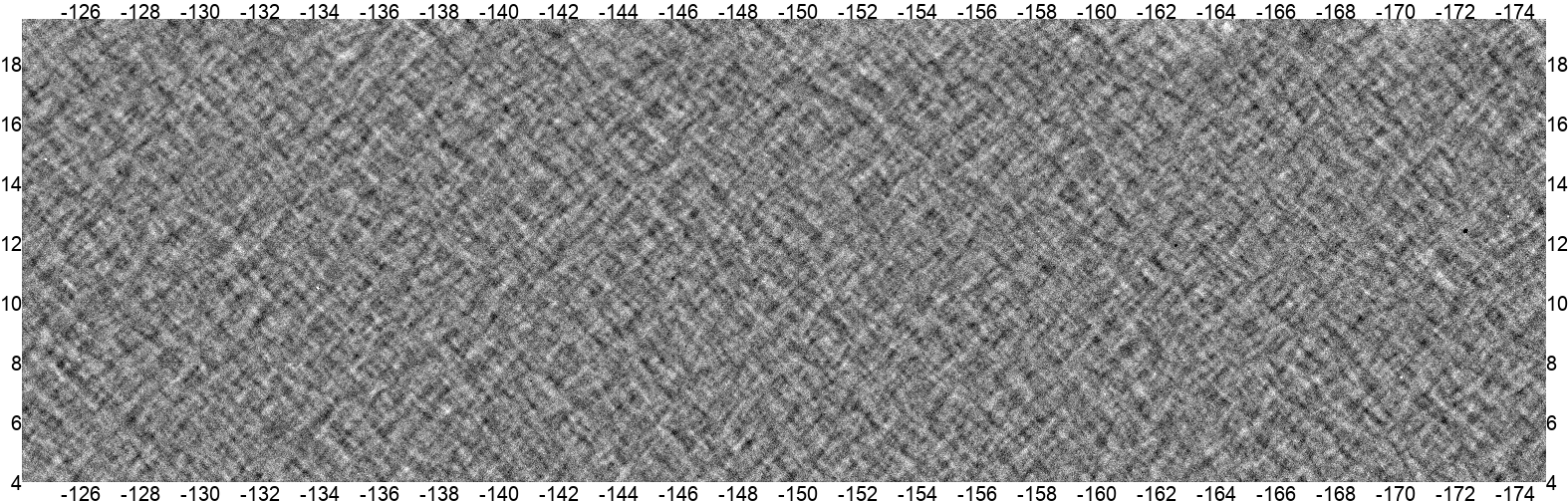} \\
		E & \img[width=12cm,trim=0 5.5mm 0 6.5mm,clip]{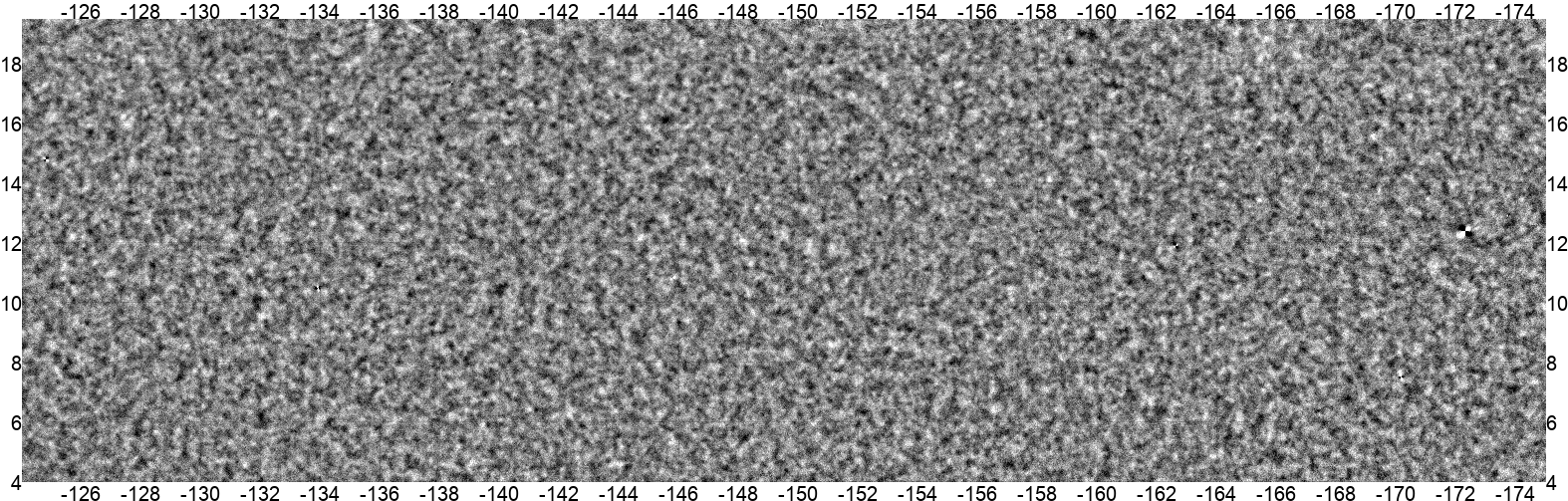} \\
		B & \img[width=12cm,trim=0 0.0mm 0 6.5mm,clip]{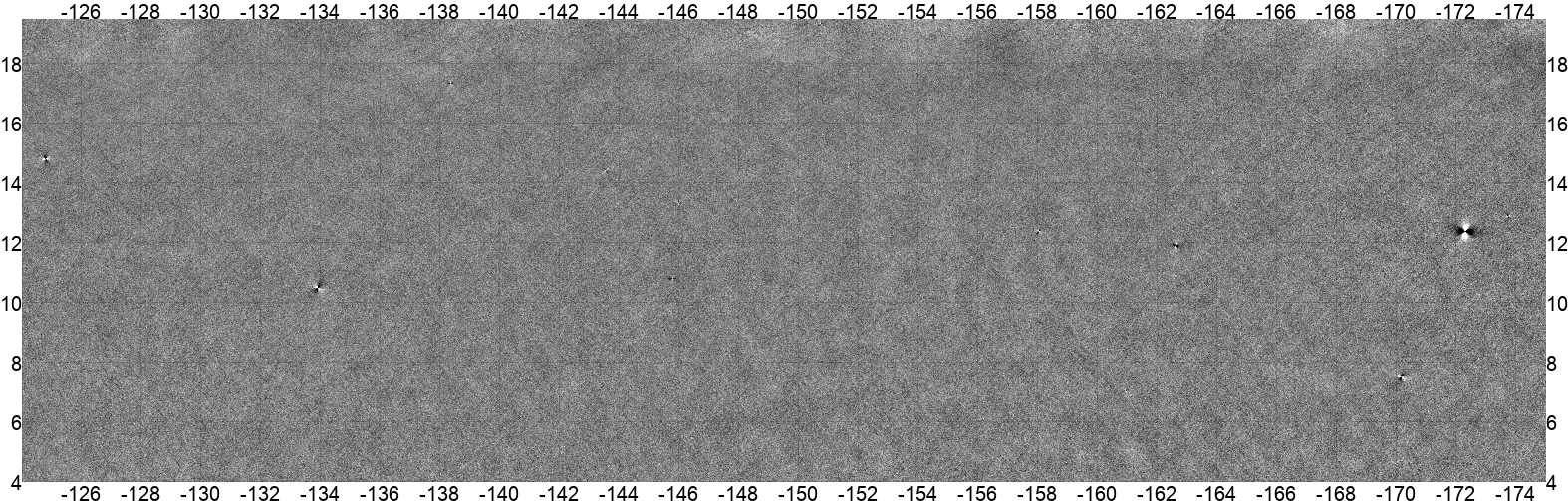}
	\end{tabular}
	\end{closetabrows}
	\caption{A 772 square degree patch covering $-175 < \textrm{RA} < -124$ and $4 < \textrm{dec} < 19$,
	one of the deepest areas in ACT. From top to bottom the panels show the temperature T, linear polarization
	Q and U, and the corresponding polarization gradient and curl fields E and B for a day+night
	coadd of f090 and f150. \planck\ is included in T, where it mainly affects scales larger than about
	half a degree. The other panels are ACT-only, but would not be noticeably different with \planck.
	The E map is clearly signal-dominated, while B is consistent with
	noise, aside from a few polarized point sources that show up as small quadrupoles.
	The color range is $\pm 500$ \textmu{}K in T (top) and $\pm 20$ \textmu{}K in the others.}
	\label{fig:tqueb-deep}
\end{figure}

Figure~\ref{fig:tqueb-deep} shows the total intensity (T), linear polarization (Q, U) and divergence (E) and
curl fields (B) \citep{zaldarriaga/seljak/1997} for a simple pixel-space average of the f090 and f150 maps in a
772 square degree patch covering most of the deep Day-N field. The effective noise level in this patch is
7.1 \textmu{}K-arcmin. The polarization Q/U maps show the clear +/x pattern indicative of high S/N E-modes, which is confirmed in the
E-mode map itself. The B map is consistent with noise (and small amounts of ground pickup near the top)
with the exception of a few polarized point sources that show up as quadrupoles due to the non-local nature
of E and B.

\begin{figure}[p]
	\centering
	\begin{closetabrows}
	\hspace*{-20mm}\begin{tabular}{m{2mm}m{20cm}}
		E & \img[width=12cm,trim=0 5.5mm 0 0.0mm,clip]{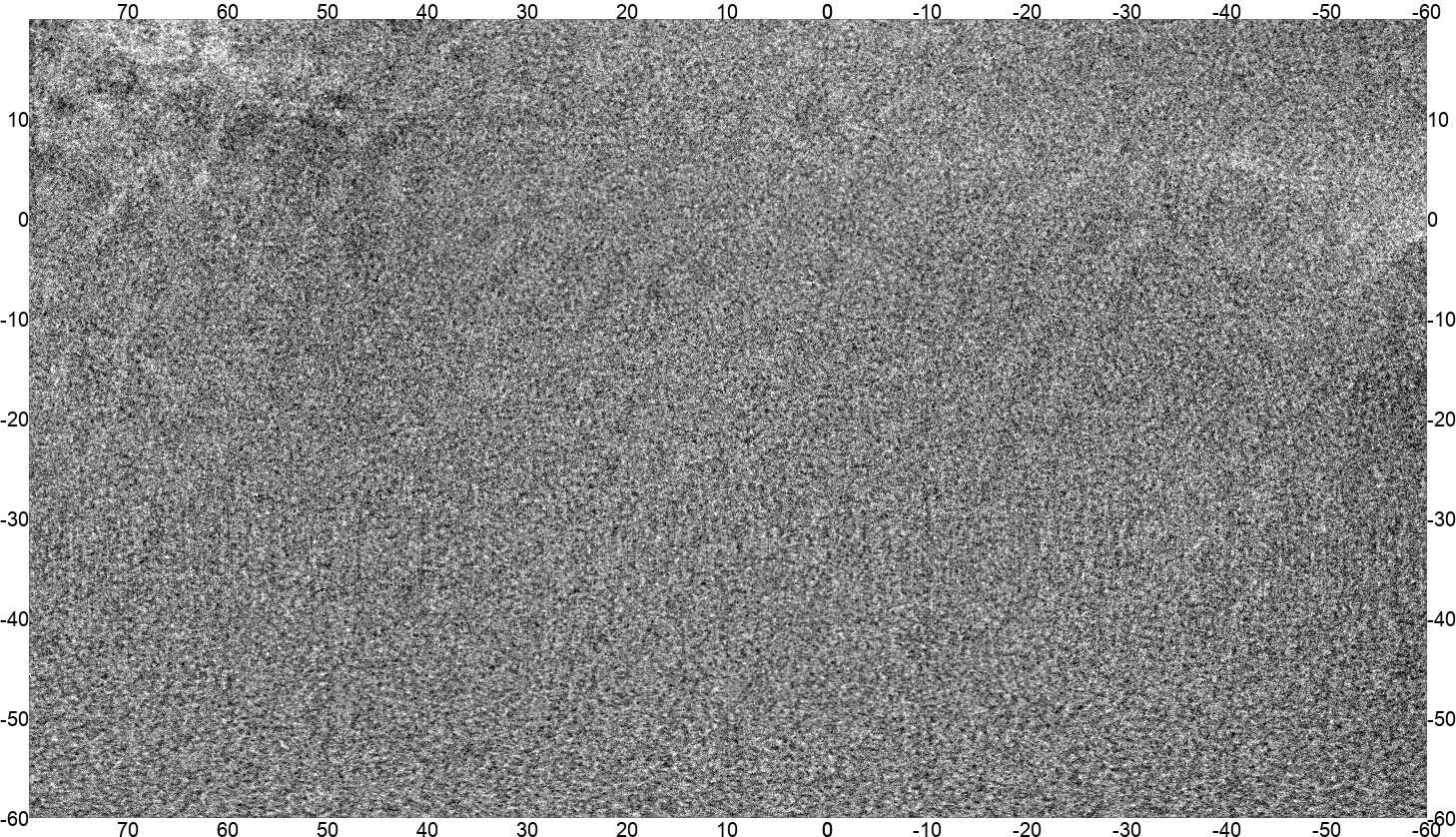} \\
		B & \img[width=12cm,trim=0 0.0mm 0 6.5mm,clip]{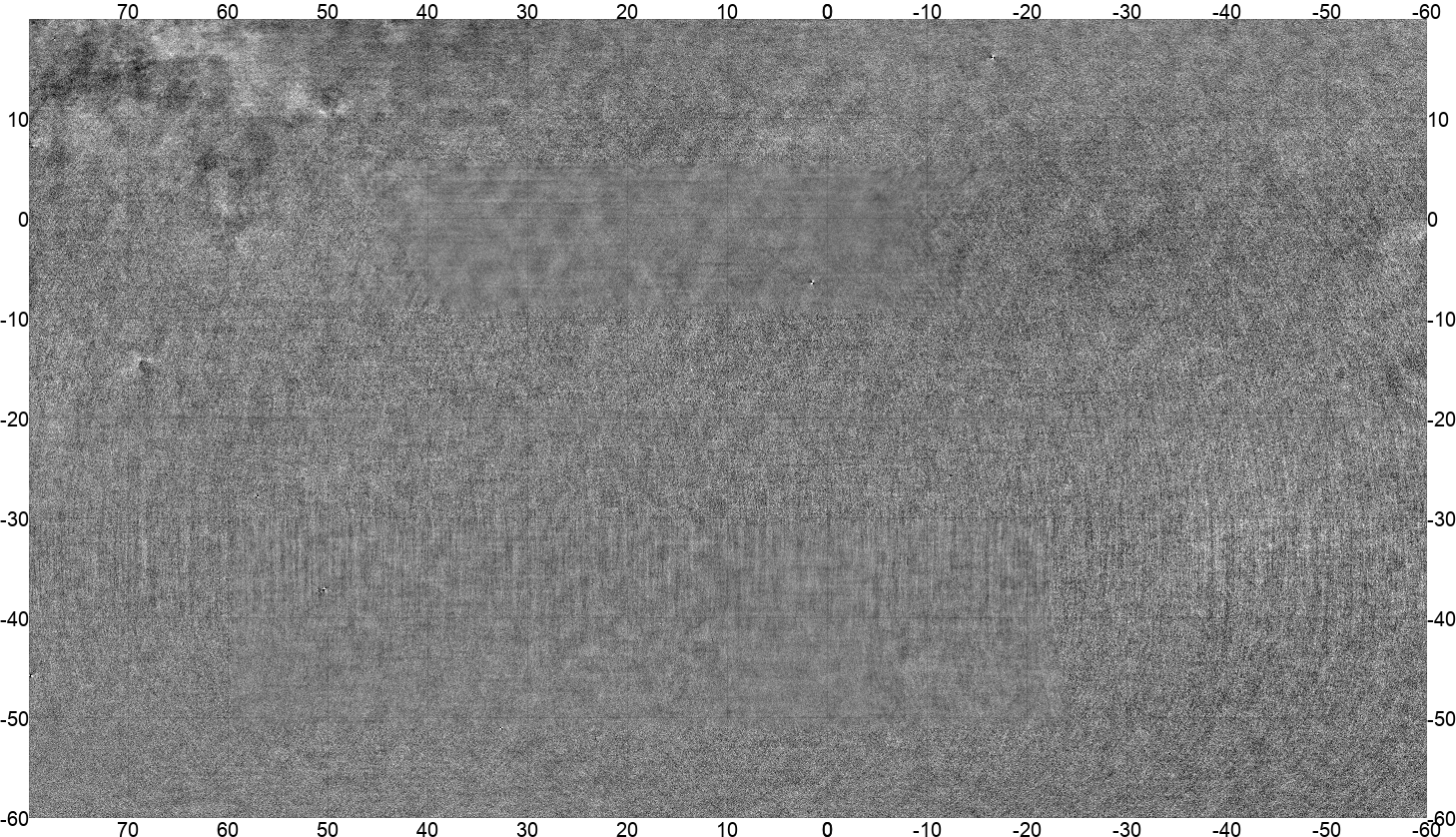}
	\end{tabular}
	\end{closetabrows}
	\caption{A 9690 square degree patch covering $-60 < \textrm{RA} < 80$ and
	$-60 < \textrm{dec} < 20$,
	23\% of the sky and a bit more than half the full ACT area. Aside from a few deeper patches, this represents
	the shallowest parts of our survey. The top/bottom panel show E/B-modes for a f090+f150 ACT+\planck\
	coadd. This includes daytime data in the areas where it exists, but 87\% of this area is night-only.
	Despite the shallowness of this area, E-modes around $\ell=500$ are still signal dominated. At this multipole
	\planck\ contributes roughly 20\% of the weight. This gradually increases at lower $\ell$, and starts to dominate
	for $\ell<150$. Some galactic dust contamination is visible near the edges of the map.}
	\label{fig:eb-wide}
\end{figure}

Figure~\ref{fig:eb-wide} shows the same f090+f150 E and B fields over a much larger area covering 9\,500
square degrees (23\% of the sky and a bit more than half the full ACT area). On this length scale the
characteristic E-mode length scale appears small enough that they could be mistaken for noise, but the
difference between the E and B maps shows that the E-modes are signal dominated even in the shallower parts
of our survey.

\begin{figure}[h]
	\centering
	\includegraphics[width=0.5\textwidth]{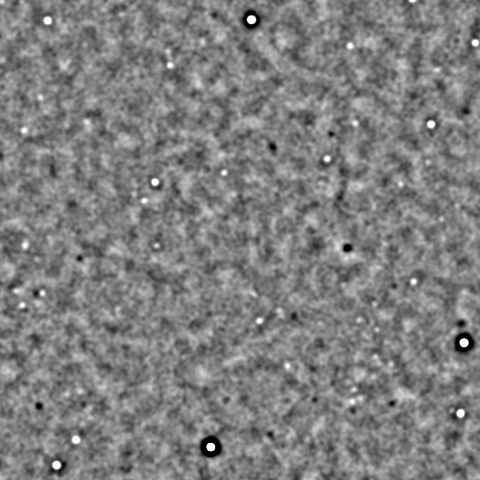}
	\caption{The ACT f090 day+night map filtered to enhance point sources, clusters and other
	small-scale features. The region shown is a $4^\circ\times4^\circ$ square centered on RA = 224.5$^\circ$,
	dec = 6$^\circ$. More than 20 clusters are visible as temperature decrements (dark) through the tSZ
	effect, and more than 45 point sources are visible as temperature increments (light). This is one
	of the deepest ACT regions, with a depth of about 9 \textmu{}K-arcmin (0.6 mJy) in this band.
	The filter used here is approximately matched to the beam profile and atmospheric correlation structure,
	and is more optimal for point sources than for clusters.}
	\label{fig:matched-filter}
\end{figure}

Figure~\ref{fig:matched-filter} applies a matched filter for point sources to the ACT f090 map in a
$4^\circ\times4^\circ$ degree sub-patch of Day-N, revealing more than 45 point sources and 20 clusters
in a 16 square degree area. The point source and cluster potential of these maps will be explored in
two upcoming papers, but a preliminary search has 4\,000 confirmed clusters \citep[in prep]{act-clusters-2020}
and 18\,500 point source
candidates at $> 5\sigma$. For comparison, the largest published point source catalog at these frequencies is
\citet{spt-clusters-2020} with 4845 point sources at $> 4.5\sigma$, and the largest published SZ-detected cluster catalog, PSZ2, has 1203 confirmed clusters \citep{psz2}.

\begin{figure}[h]
	\centering
	\begin{closetabcols}[0.2mm]
	\begin{closetabrows}[0.25]
	\begin{tabular}{>{\centering\arraybackslash}m{3mm}>{\centering\arraybackslash}m{5cm}>{\centering\arraybackslash}m{5cm}>{\centering\arraybackslash}m{5cm}}
		& \bf f090 & \bf f150 & \bf f220 \\
		\rotatebox[origin=c]{90}{\bf ACT+\planck} & \img{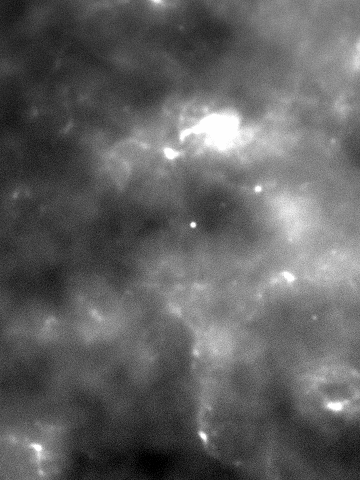} & \img{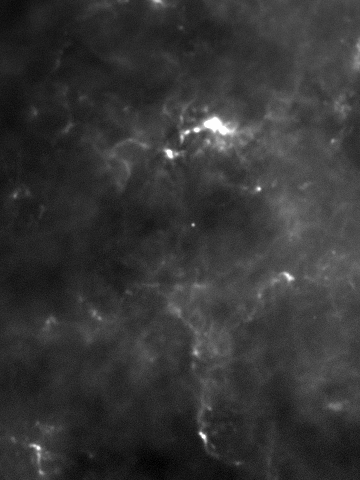}  & \img{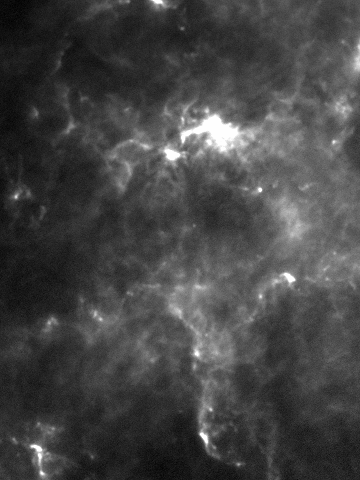}  \\
		\rotatebox[origin=c]{90}{\bf \planck} &  \img{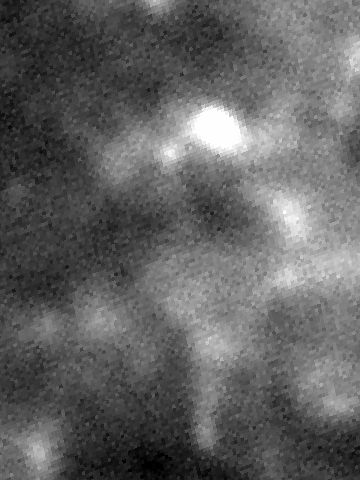} & \img{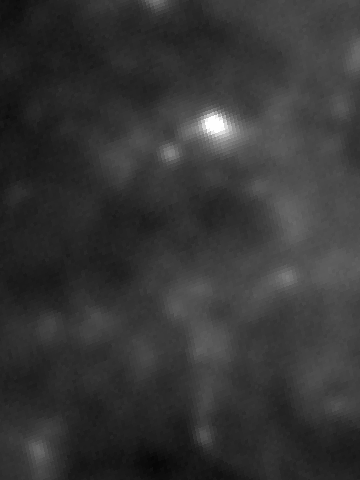}  & \img{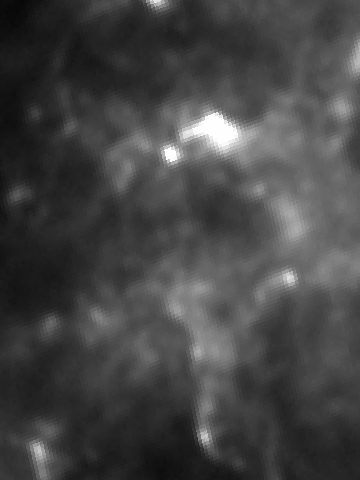}
	\end{tabular}
	\end{closetabrows}
	\end{closetabcols}
	\caption{Comparison of ACT+\planck\ and \planck\ alone in the region $95^\circ > \textrm{RA} > 93^\circ$,
	$15^\circ < \textrm{dec} < 19^\circ$ in the galactic plane. ACT's $5\times$  higher resolution than \planck\
	allows for high-resolution dust science at CMB-relevant frequencies.}
	\label{fig:galaxy-comp}
\end{figure}

As seen in figure~\ref{fig:depth2d}, ACT has partial coverage of the galactic plane. About 1/3
of the disk is covered at very shallow ($>60$ \textmu{}K-arcmin but still strongly signal-dominated) depth,
barely including the galactic center. Additionally, the area with galactic longitude $190 < \ell < 245$
(including e.g. the Orion and Rosette nebulae) is covered at depths of 16--60 \textmu{}K-arcmin typical of
shallow-to-medium CMB areas.
Compared to \planck\ alone, ACT's $5\times$ higher resolution reveals much more of the small-scale structure of the
dust (see figure~\ref{fig:galaxy-comp}) without needing to extrapolate from the much higher frequencies
of e.g. WISE \citep{unwise3}.

\begin{figure}[p]
	\centering
	\includegraphics[width=1\textwidth,trim=13mm 0 4mm 0]{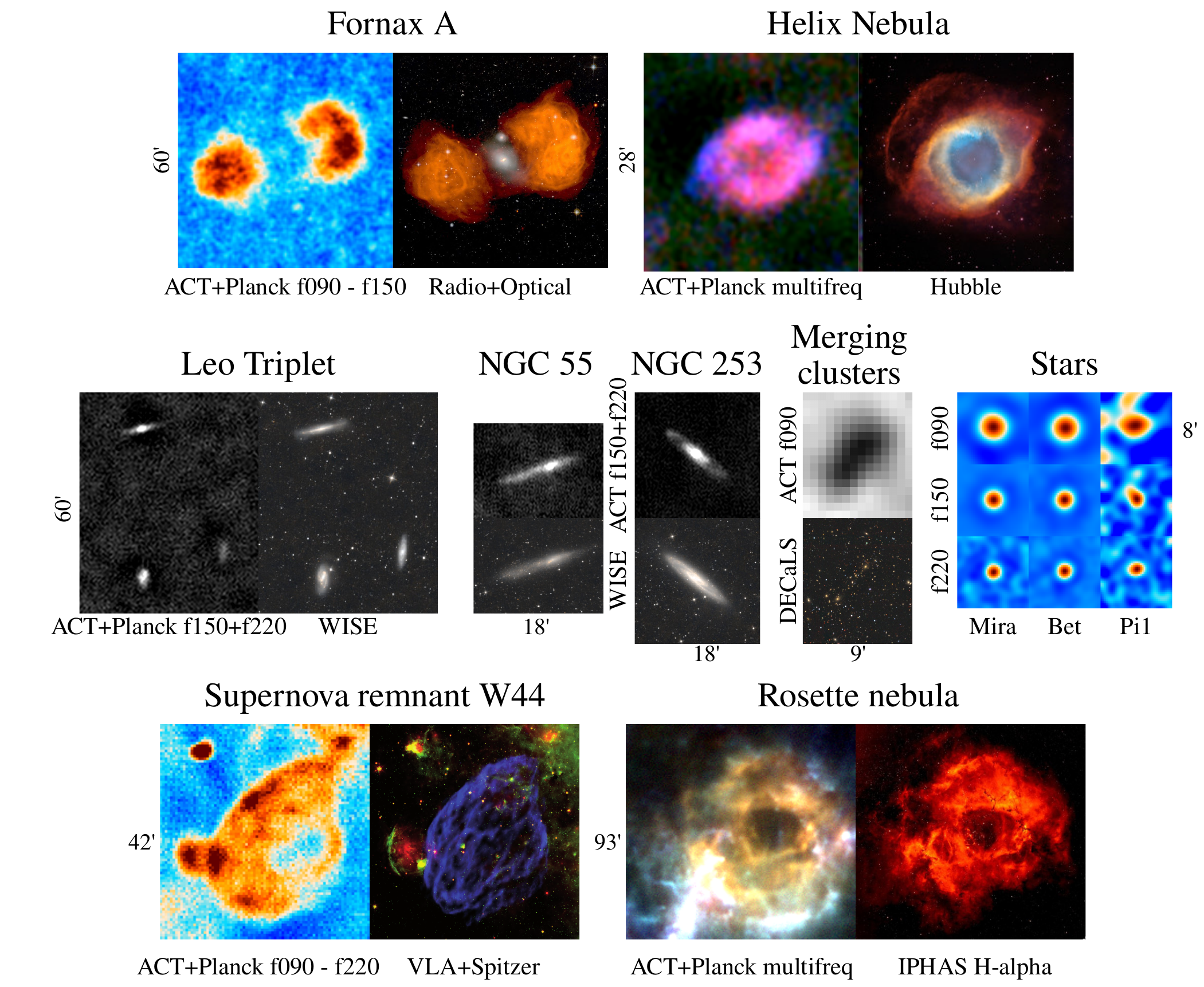}
	\caption{Some nearby objects that can be found in these maps. The small numbers next to each
	set of images indicates the length of the corresponding side in arcminutes. \dfn{Top left}: Fornax A (NGC 1316).
	The CMB was removed by subtracting f150 from f090. Compared to a radio + optical composition by
	\citet{fomalont-fornax-a,fomalont-fornax-a-color}.
	The central galaxy is visible in individual frequencies, but was cancelled in this difference map.
	\dfn{Top right}: The Helix Nebula (NGC 7293, The Eye of God/Sauron). Here f090, f150 and f220 were
	scaled until they had roughly the same noise level, and then assigned to the red, green and blue color
	channels respectively. Compared to a famous Hubble image of the same object \citep{hubble-helix}. \dfn{Middle left}:
	The Leo Triplet (M65, M66, and NGC 3628) from a pixel-space sum of ACT+\planck\ f150 and f220, compared
	to unWISE W1/W2 \citep{unwise3}. \dfn{Middle center-left}: As the Leo Triplet, but for NGC 55 (the Whale Galaxy) and NGC 253 (the
	Sculptor galaxy).
	\dfn{Middle center-right}: Example of a merging galaxy cluster pair that appears clearly elongated both to
	ACT (tSZ) and DECaLS (stars) \citep{legacy-survey}.
	\dfn{Middle right}: Several large, nearby stars are detectable as significant mm-band sources by ACT.
	Shown here are filtered images of three examples: Mira, Betelgeuse and $\pi^1$ Gruis.
	Other strong detections include V0711 Tau and RV Aqr.
	\dfn{Bottom left}: The supernova remnant W44. Galactic dust was reduced by subtracting f220 from f090.
	The comparison image shows a composite of images from radio (VLA 324
	MHz, in blue) and infrared (Spitzer 8 \textmu{}m and 24\textmu{}m, in green and red,
	respectively) from \citet{vla-w44-2007}.
	\dfn{Bottom right}: The Rosette nebula. f090, f150 and f220 are mapped to the red, green and blue channels
	respectively. Compared to an H$\alpha$ (656.28 nm) image prepared by \citet{rosette-wright} using data
	from IPHAS \citep{iphas-dr2}.
	}
	\label{fig:curious-stuff}
\end{figure}

Finally, figure~\ref{fig:curious-stuff} gives some examples of other miscellaneous objects one
can find in the maps, including radio lobes from active galaxy Fornax A, the Helix planetary nebula,
resolved nearby galaxies including the Leo Triplet, NGC 55 and NGC 253, merging clusters detected
through their asymmetric tSZ signal, and the individual stars Mira, Betelgeuse and $\pi^1$ Gruis.
These images suggest the wealth of new information that is present in these new publicly available maps.

\FloatBarrier

\section{Conclusion}
We have presented a method for combining maps with greatly varying sky coverage, depth and angular resolution
and spatially varying anisotropic noise into a near-optimal sky map.

We used this to combine
270 ACT maps representing observations from 2008 to 2018 into three frequency maps, centered at roughly 90 GHz,
150 GHz and 220 GHz. These maps cover more than 19\,000 square degrees (46\% of the sky), and include
previously unreleased preliminary data from the two first seasons of Advanced ACTPol. We also provide a second
set of maps that also include the more challenging ACT daytime data, which provide a large boost in
depth over a 3\,000 square degree subset of the survey area.

In addition to these ACT-only maps, we also produce versions that have been combined with the
nearest-frequency \planck\ HFI maps. This has the effect of filling in the large angular scales
($\ell \lesssim 1000$) that ground-based millimeter surveys like ACT have trouble measuring due to the
influence of the atmosphere, resulting in a map that covers scales from $180^\circ$ to $\sim 1$ arcmin.

We make these maps available to the public in the hope that they will be useful, but caution that
due to the preliminary nature of some of the component data sets, these maps should not be used for
precision cosmological analysis, and the version of the maps that include daytime data in particular
should only be used for cases that can tolerate a position-dependent O(10\%) beam uncertainty.
The effective band-center is also somewhat scale-dependent due to differences in the bandpasses
of the individual input maps.

In \citet{act-clusters-2020} we use these maps to find 4\,000 confirmed clusters through the thermal Sunyaev Zel'dovich
effect, and in a second upcoming publication we detect 18\,500 millimeter point sources at $>5\sigma$,
both substantial improvements on
the state of the art. Other anticipated use cases include tSZ and kSZ cluster stacking and CMB cluster lensing
measurements. The maps also include hundreds of resolved galaxies and polarized point sources;
cover about 1/3 of the galactic disk at high resolution; and also include several classes of objects one
would not normally associate with a map from a CMB survey, including radio lobes from active galactic nuclei,
planetary nebulae and even a few individual stars.

The cosmological analysis of an expanded and fully calibrated version of this
data set, including CMB and lensing power spectra and cosmological parameters,
will be the subject of a future ACT data release.

\section*{Acknowledgments}
This work was supported by the U.S. National Science Foundation through awards
AST-0408698, AST-0965625, and AST-1440226 for the ACT project, as well as
awards PHY-0355328, PHY-0855887 and PHY-1214379. Funding was also provided by
Princeton University, the University of Pennsylvania, and a Canada Foundation
for Innovation (CFI) award to UBC. ACT operates in the Parque Astron\'omico
Atacama in northern Chile under the auspices of the Comisi\'on Nacional de
Investigaci\'on (CONICYT). Flatiron Institute is supported by the Simons Foundation.
NS acknowledges support from NSF grant numbers AST-1513618 and AST-1907657.
KM acknowledges support from the National Research Foundation of South Africa.
JPH acknowledges support from NSF grant number AST-1615657
R.D. thanks CONICYT for grant BASAL CATA AFB-170002.
ZL, ES and JD are supported through NSF grant AST-1814971.
RH is a CIFAR Azrieli Global Scholar, Gravity \& the Extreme Universe
Program, 2019, and a 2020 Alfred. P. Sloan Research Fellow. RH is supported
by Natural Sciences and Engineering Research Council of Canada. The Dunlap
Institute is funded through an endowment established by the David Dunlap
family and the University of Toronto.
We made use of the healpix library as part of this analysis.
Computations were performed on the Niagara supercomputer at the SciNet HPC
Consortium. SciNet is funded by the CFI under the auspices of Compute Canada,
the Government of Ontario, the Ontario Research Fund--Research Excellence, and
the University of Toronto.
Additional computations were performed on Tiger and Feunman at Princeton. The
development of multichroic detectors and lenses was supported by NASA grants
NNX13AE56G and NNX14AB58G. Detector research at NIST was supported by the NIST
Innovations in Measurement Science program. The shops at Penn and
Princeton have time and again built beautiful instrumentation on which ACT
depends.
LP gratefully acknowledges support from the Mishrahi and Wilkinson funds. We
thank our many colleagues from ALMA, APEX, CLASS, and Polarbear/Simons Array
who have helped us at critical junctures. Colleagues at AstroNorte and RadioSky
provide logistical support and keep operations in Chile running smoothly.

\bibliographystyle{act_titles}
\bibliography{refs}

\pagebreak

\appendix

\section{The data release}
\label{sec:data-release}
\subsection{Sky maps}
The main data products in this data release are the ACT+\planck\  and ACT-only sky maps:
\begin{verbatim}
act_planck_dr5.01_s08s18_AA_f090_night_map.fits
act_planck_dr5.01_s08s18_AA_f150_night_map.fits
act_planck_dr5.01_s08s18_AA_f220_night_map.fits
act_planck_dr5.01_s08s18_AA_f090_daynight_map.fits
act_planck_dr5.01_s08s18_AA_f150_daynight_map.fits
act_planck_dr5.01_s08s18_AA_f220_daynight_map.fits
act_dr5.01_s08s18_AA_f090_night_map.fits
act_dr5.01_s08s18_AA_f150_night_map.fits
act_dr5.01_s08s18_AA_f220_night_map.fits
act_dr5.01_s08s18_AA_f090_daynight_map.fits
act_dr5.01_s08s18_AA_f150_daynight_map.fits
act_dr5.01_s08s18_AA_f220_daynight_map.fits
\end{verbatim}
These are 32-bit float FITS images with shape 43200,10320,3. The first two axes are
RA and dec in the Plate Carreé projection, covering the area $180^\circ > \textrm{RA} > -180^\circ$
and $-63^\circ < \textrm{dec} < 23^\circ$ at 0.5 arcmin resolution. The last axis represents the three
Stokes parameters I, Q and U in the Healpix/Cosmo polarization convention. The maps are in units of
\textmu{}K CMB temperature increment. Note that the axes appear in the opposite order when loaded
as a \texttt{pixell} \texttt{enmap}, since \texttt{enmap} (like \texttt{numpy}) uses row-major ordering
instead of column-major ordering like FITS does.

\subsection{Inverse variance maps}
Associated with each of these maps is a noise floor inverse variance map, which has the
same shape and contains an estimate of the non-atmospheric inverse variance in
\textmu{}K$^{-2}$ per pixel. This does not include the contribution from \planck\ due to \planck's
limited multipole range. These files are labeled \path{ivar}, e.g. \path{act_dr5.01_s08s18_AA_f090_night_ivar.fits}.

\subsection{Detailed noise model}
A more detailed noise model is provided in the files
\begin{verbatim}
act_planck_dr5.01_s08s18_AA_f090_night_fullivar.fits
act_planck_dr5.01_s08s18_AA_f150_night_fullivar.fits
act_planck_dr5.01_s08s18_AA_f220_night_fullivar.fits
act_planck_dr5.01_s08s18_AA_f090_daynight_fullivar.fits
act_planck_dr5.01_s08s18_AA_f150_daynight_fullivar.fits
act_planck_dr5.01_s08s18_AA_f220_daynight_fullivar.fits
\end{verbatim}
These are 32-bit float FITS images with shape 720,172,50,15,3, and provides the noise inverse
variance in units of \textmu{}K$^{-2}$ per square arcmin as a function of position, $\ell$ and
detector array, albeit at reduced resolution to make the file size managable.

The first two axes are RA, dec in the
same projection as the main maps, but at only 0.5$^\circ$ resolution. The third axis represents
50 exponentially spaced multipoles from 100 to 20\,000: $\ell_i = 100\cdot 200^{i/50}$. The
noise power is a smooth function of $\ell$, so these 50 sample points should suffice for most
purposes.

The fourth axis represents the 15 different detector arrays that contribute to these
coadds:
\path{pa1_f150}, \path{pa2_f150}, \path{pa3_f090}, \path{pa3_f150}, \path{pa4_f150},
\path{pa4_f220}, \path{pa5_f090}, \path{pa5_f150}, \path{pa6_f090}, \path{pa6_f150},
\path{ar1_f150}, \path{ar2_f220}, \path{planck_f090}, \path{planck_f150} and \path{planck_f220}.
This axis can be summed over if one is not interested in how much each of these contributes to
the total inverse variance. Because the sky maps are inverse variance weighted combinations
of the individual input maps, one can recover the relative weight of each array in
the combination by its inverse variance by the total, resulting in a set of per-array weights.

As with the sky maps, the axes in these files appear in the opposite order when
loaded as an \texttt{enmap}, i.e. 3,15,50,172,720.
The list of multipoles and arrays is also included in the file \path{act_planck_dr5.01_s08s18_fullivar_info.txt}.

\subsection{Bandpasses}
\label{sec:app-bandpass}
\path{act_planck_dr5.01_s08s18_bandpasses.txt} provides the bandpasses of each of the 15 detector arrays
in units of \textmu{}K/(MJy/sr)/GHz. The first column is the frequency in GHz, followed by
a column for each array, in the same order as for the detailed noise model. This can
be combined with the per-array weights to estimate the effective bandpass at any point
and multipole in the map. However, this would only be approximate because this file does
not capture the scale-dependence of the individual detector array bandpasses.

The file \path{ act_planck_dr5.01_s08s18_bandpasses_scaledep.hdf} provides the full scale-dependence of the
array bandpasses. It is a HDF file containing 4 data sets: \path{arrays}, \path{ls}, \path{freqs}
and \path{bandpass}. \path{arrays} and \path{ls} just list the array names and multipoles,
which are the sames as those given above, while \path{freqs} is the list of the 434 frequencies the
bandpasses are resampled to (66 GHz to 283 GHz with 0.5 GHz steps). \path{bandpasses} is
a data set with shape 2,15,50,434. The first axis corresponds to the bandpass value and uncertainty,
while the remaining axes are the arrays, multipoles and frequencies respectively.

The code fragment below illustrates how to compute the effective $\ell$-dependent
bandpass for total intensity component of the ACT+\planck\ f150 day+night map at
RA = 0$^\circ$, dec = 0$^\circ$, which corresponds to the noisebox pixel [126,359]:
\begin{verbatim}
noisebox    = enmap.read_map("act_planck_dr5.01_s08s18_AA_f150_night_fullivar.fits")
weights     = noisebox / np.sum(noisebox,1)[:,None]
bands_array = hget("act_planck_dr5.01_s08s18_bandpasses_scaledep.hdf", "bandpass")
band_at_pos = np.sum(weights[0,:,:,None,126,359]*bands_array[0],0)
\end{verbatim}
The result in this case is a 2D array with shape $[N_\ell,N_\textrm{freq}]$. One can similarly
get the average bandpass over the whole map. To exclude \planck, simply set the \planck\ entries in
\code{noisebox} to zero (\code{noisebox[:,-3:]=0}) before computing the weights.

\subsection{Responses and bandcenters}
The main use of the bandpasses is to find the map response to individual signal components like
the CMB, tSZ, dust and synchrotron. For convenience we provide map-averaged versions of these,
labeled \path{response_cmb}, \path{response_tsz}, \path{response_dust} and \path{response_sync}.
These are text tables with columns $\ell$, I, dI, Q, dQ, U, dU, where I, Q, U are the response
in each Stokes component\footnote{The distinction between Q and U is mostly meaningless here,
since they have the same noise properties in the maps, which translates into the same
effective bandpasses.},
and dI, dQ, dU are their spatial standard deviation. The response
is in units of \textmu{}K/\textmu{}K for the CMB\footnote{This is always 1, making the CMB response files
redundant.}, \textmu{}K/y for tSZ and \textmu{}K/arbitrary for dust and synchrotron.

We also provide band-centers in the same format, labeled \path{band_center_cmb}, \path{band_center_tsz},
\path{band_center_dust} and \path{band_center_sync}, in units of GHz, but for most purposes the response
files will be more relevant.

\subsection{Beams}
The map beams transfer functions are available in the following files.
\begin{verbatim}
act_planck_dr5.01_s08s18_f090_daynight_beam.txt  act_planck_dr5.01_s08s18_f090_night_beam.txt
act_planck_dr5.01_s08s18_f150_daynight_beam.txt  act_planck_dr5.01_s08s18_f150_night_beam.txt
act_planck_dr5.01_s08s18_f220_daynight_beam.txt  act_planck_dr5.01_s08s18_f220_night_beam.txt
\end{verbatim}
These have two columns, $\ell$ and $B_\ell$, which applies to all Stokes parameters, and to
both the ACT+\planck\ and ACT-only maps.

\subsection{Point source subtracted maps}
There is also a source-subtracted version of each map (see appendix~\ref{sec:srcsub}), labeled \path{srcfree_map} instead of \path{map},
e.g. \path{act_planck_dr5.01_s08s18_AA_f090_night_map_srcfree.fits}. These have the same format and noise properties
as the normal maps.

\section{Variability bias and point source subtraction}
\label{sec:srcsub}
\begin{figure}[h]
	\centering
	\begin{closetabcols}
	\begin{tabular}{>{\centering\arraybackslash}m{2.5cm}>{\centering\arraybackslash}m{2.5cm}>{\centering\arraybackslash}m{2.5cm}>{\centering\arraybackslash}m{2.5cm}>{\centering\arraybackslash}m{2.5cm}>{\centering\arraybackslash}m{2.5cm}}
		ACT map & \planck\ map & ACT weighted & \planck\ weighted & Coadd & Coadd (var) \\
		\img{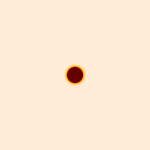} & \img{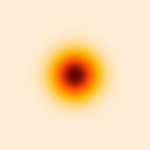} & \img{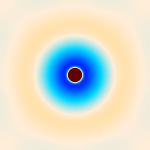} & \img{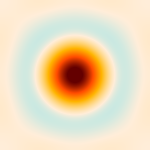} & \img{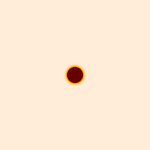} & \img{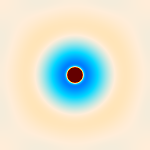}
	\end{tabular}
	\end{closetabcols}
	\caption{How variable point sources introduce bias.
	\dfn{Column~1--2}: Simulated ACT (left) and \planck\ (right) noise-free maps for a single point source.
	\dfn{Column~3--4}: Each source's contribution to the coadded map in equation~\ref{eq:maxlik},
	assuming a simple constant-covariance noise model. ACT's contribution contains a negative shadow
	that will exactly cancel \planck's broader beam, resulting in a narrow beam in the final coadd,
	shown in \dfn{column~5}. If ACT and \planck\ do not agree on the source's flux, then this cancellation
	is not exact, leaving a residual artifact approximately the size of the \planck\ beam surrounding
	the source. This is illustrated in \dfn{column~6} where the source varied in flux by a factor of
	two between when \planck\  and ACT observed it.}
	\label{fig:varsrcs}
\end{figure}

When deriving the maximum-likelihood solution for the coadded sky map, we assumed that all
individual maps saw different views of a single, consistent sky (see eq.~\ref{eq:model}).
However, this assumption breaks down if the sky changes between the time where the data for the
maps was collected, and the central engines of quasars are small enough that their flux can
change greatly over time scales of days to months. This is much less than the $\sim10$ year
time-span the data-sets we combine here cover, so we can expect the model to break down in the
vicinity of bright, variable quasars.

Figure~\ref{fig:varsrcs} illustrates how the this modelling error gives rise to artifacts roughly
the size of the largest beam involved. A prominent example of this in practice can be seen around the
bright, variable quasar PKS 0003-066 in the ACT+\planck\ map, which is surrounded by a blue shadow
with an amplitude of 0.2\% of the peak of the source extending about 0.15$^\circ$ away.
This relative faintness compared to the source itself explains why this effect is only
noticeable around the few brightest and most variable quasars in the map.

For most purposes these rare artifacts can probably be ignored or masked. However, we
still provide an alternative version of the coadded maps where a large catalog of
point sources were individually fit and subtracted from each input map before they
were combined. This was done in two steps:

\begin{enumerate}
	\item A matched filter point-source finder was run on the standard coadded ACT+\planck\
		day+night maps, resulting in a catalog of 18507/14643/4084 objects detected
		at more than $5\sigma$ at f090/f150/f220. This corresponds to a flux cut that
		depends on the local map depth, and is roughly 3.5/3.5/10 mJy in deep regions and
		15/15/50 mJy in shallow regions.
	\item The flux (but not position) for each source in these catalogs was fit
		and subtracted in each input map, resulting in source-free maps. During
		this subtraction, catalog flux measured in the previous step was used as a weak
		prior. Specifically, the catalog flux entered into the fit as an extra
		data point, which was assigned a standard deviation
		$\sigma = \sqrt{\sigma_F^2 + (\alpha F)^2}$, where $F$ is the catalog
		flux and $\sigma_F$ is its uncertainty.
		The second term is there to allow for source variability, with the degree
		of variability being controlled by the parameter $\alpha$. We set $\alpha=1$
		based on it being relatively common for a point source to double in brightness from one
		season to another.

		The reason for applying a prior was mainly to avoid oversubtracting CMB
		fluctuations in the \planck\ maps. The majority of the point sources we
		subtract are too faint to be detected by \planck. Without a prior, the fit for
		these would be dominated by the much larger CMB fluctuations, resulting in part
		of the CMB being subtracted.
\end{enumerate}

The resulting point source subtracted maps were then combined using the same process as the
standard maps, resulting in a set of point-source subtracted ACT-only and ACT-\planck\  maps.

A limitation of this subtraction process is that it relies on the sources being unresolved
points. However, a few nearby galaxies are erroneously detected as point sources. For these,
the subtraction process results in incomplete subtraction, and in some cases artifacts that are
superficially similar to the ones variable point sources cause in the standard maps. NGC 253
is an example of this.

\section{DR4 coadd}
\label{sec:coadd-dr4}
In addition to the full 2008-2018 coadded maps, we also provide a ``DR4 coadd'' that
only uses data up to and including ACT DR4 \citep{aiola/etal:2020} (i.e. 2008-2016).
This version contains
no f220 data and is 2.3/2.6 times shallower at f090/f150 (in terms of white noise RMS) over
most of the map -- see figure~\ref{fig:depth1d} -- but unlike the full version it was not
subject to the 3 month post-publication delay, and is available on LAMBDA at
\url{https://lambda.gsfc.nasa.gov/product/act/act_dr4_derived_maps_get.cfm} as part of ACT DR4.

\section{Source code}
The Python source code for this analysis is available on Github in the repositories
\url{https://github.com/amaurea/enlib} and \url{https://github.com/amaurea/tenki}.
The main coadding code is in \verb|enlib/jointmap.py|, with the top-level driving script
being \verb|tenki/auto_coadd2.py|. We warn the reader that the implementation is not very
clean, and in particular, \verb|jointmap.py| also contains the implementation of several
other unrelated projects. The most relevant parts are \verb|Mapset|, \verb|sanitize_maps|,
\verb|build_noise_model| and \verb|Coadder|.

\FloatBarrier

\section{Test on simulated maps}
\label{sec:sims}
To investigate the noise and bias bias properties of the coadding procedure, we simulated three
simple datasets covering
a the same $16^\circ \times 8^\circ$ patch at 0.5 arcmin resolution, but with different beams and noise properties.
Dataset 1 and 2 represent ACT-like behavior, with a Gaussian 1.4 arcmin FWHM beam and a low noise floor, but
strong noise correlations. Dataset 1 has vertically stripy noise which gradually increases in magnitude towards
the left of the image. Dataset 2 has horizontally stripy noise with magnitude instead increasing towards the right.
Both have $1/\ell$ noise power spectra with a spectral index of -4.5 and an $\ell_\textrm{knee}$ of 3000 in
total intensity and 300 in polarization.
Dataset 3 represents \planck-like behavior, with a Gaussian 7.0 arcmin FWHM beam and uniform, uncorrelated
noise, but on average 10 times higher white noise RMS. We used four different noise realizations for each
dataset, resulting in what are effectively 4-way split maps for each of them.

For the signal we used a Gaussian realization of a lensed $\Lambda$CDM power spectrum. This has the advantage
of being familiar and reflecting the properties of the real data, but also has the disadvantage of falling rapidly
at high $\ell$ due to Silk damping, making it hard to measure bias for $\ell \gtrsim 10\,000$. This could have
been avoided by using a simple power law spectrum like $C_\ell \propto \ell^{-2}$, but we stuck with
$\Lambda$CDM to make the results more intuitive to interpret.

To test for bias, we also made an almost noise-free variant of each dataset. This was done by simply
scaling down the noise RMS in the maps by a factor of $50\,000$. Since the noise model used in the coadding
procedure only depends on difference maps, the resulting noise models will be identical to those in
the normal, noisy case up to a common multiplicative factor, which ultimately cancels when performing
the coadd\footnote{$(P^TN^{-1}P)^{-1}P^TN^{-1}d$ is unchanged when scaling $N$.}. Hence the low-noise
coadd will have exactly the same bias as the normal coadd, but can be measured much more cleanly due to
the lack of noise.

\begin{figure}[p]
	\centering
	\begin{closetabrows}[0.6]
	\begin{closetabcols}
	\begin{tabular}{>{\centering\arraybackslash}m{2mm}>{\centering\arraybackslash}m{8.5cm}>{\centering\arraybackslash}m{8.5cm}}
		& T ($\pm 400$\textmu{}K) & Q ($\pm 40$\textmu{}K) \\
		\rotatebox[origin=c]{90}{Signal} &
		\img[trim=0mm 6mm 6mm 0mm,clip]{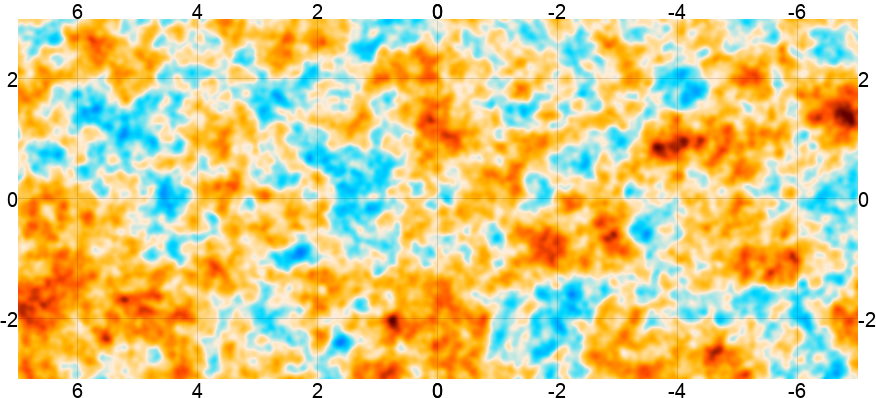} &
		\img[trim=6mm 6mm 0mm 0mm,clip]{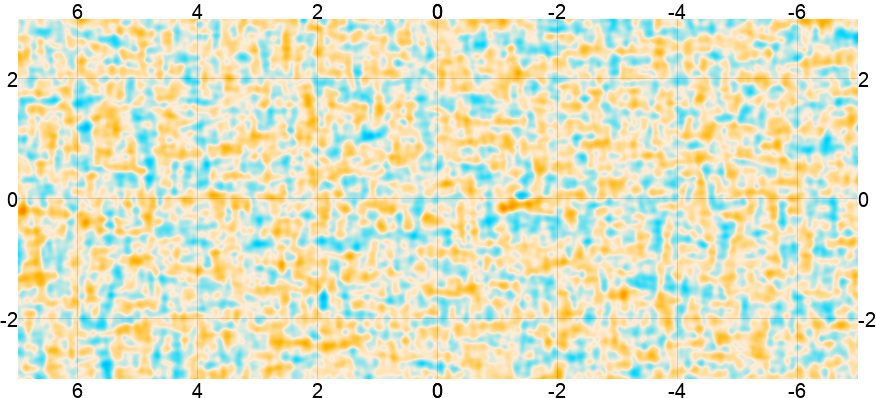} \\
		\rotatebox[origin=c]{90}{Dataset 1} &
		\img[trim=0mm 6mm 6mm 7mm,clip]{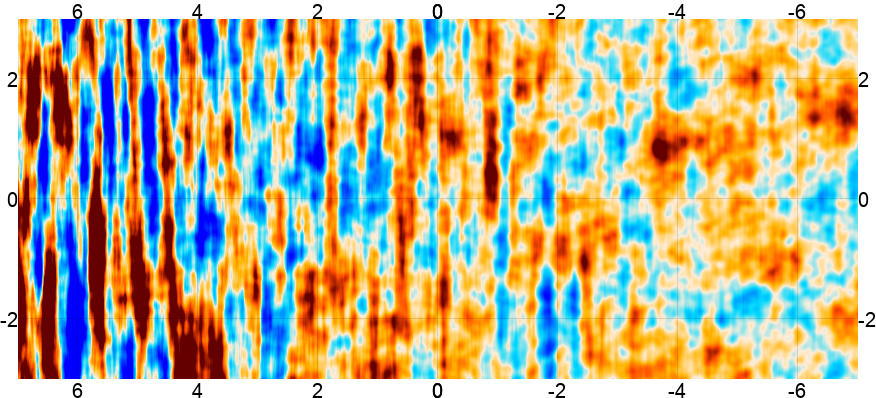} &
		\img[trim=6mm 6mm 0mm 7mm,clip]{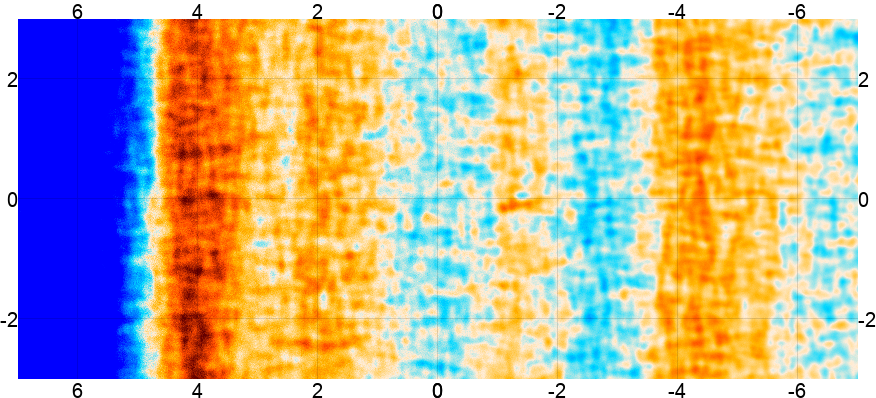} \\
		\rotatebox[origin=c]{90}{Dataset 2} &
		\img[trim=0mm 6mm 6mm 7mm,clip]{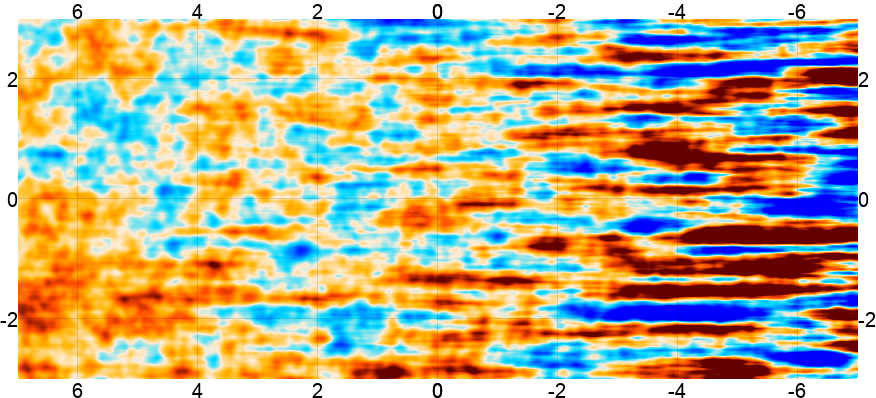} &
		\img[trim=6mm 6mm 0mm 7mm,clip]{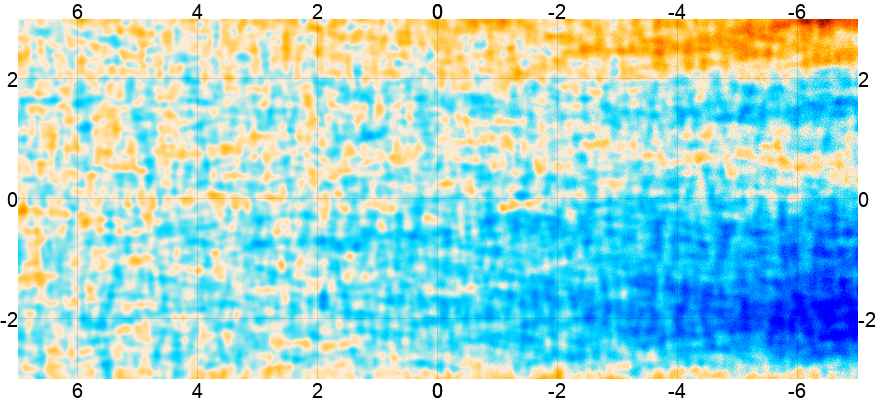} \\
		\rotatebox[origin=c]{90}{Dataset 3} &
		\img[trim=0mm 6mm 6mm 7mm,clip]{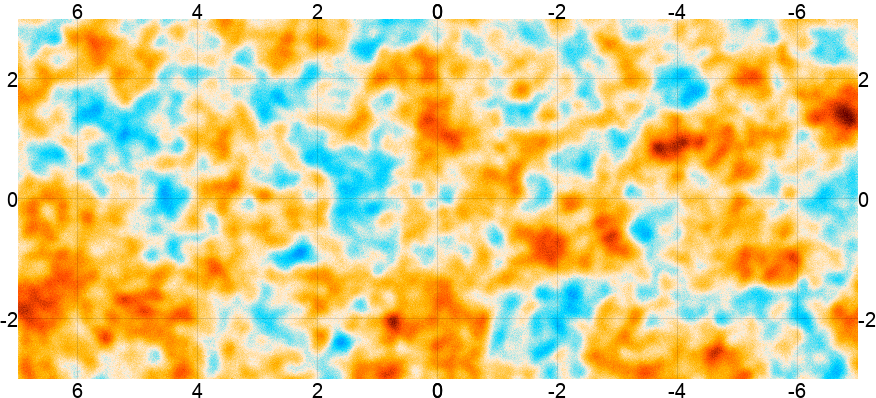} &
		\img[trim=6mm 6mm 0mm 7mm,clip]{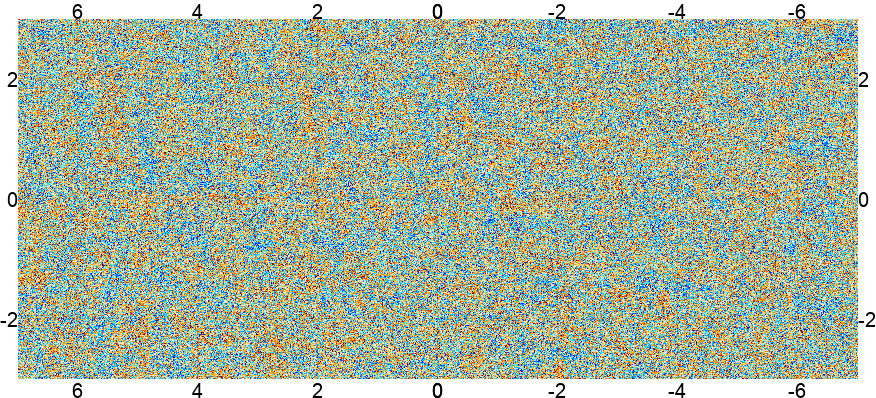} \\
		\rotatebox[origin=c]{90}{Coadd} &
		\img[trim=0mm 0mm 6mm 7mm,clip]{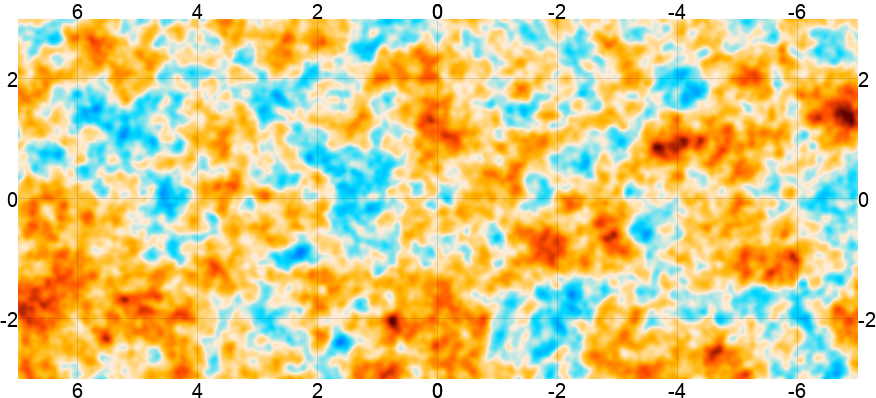} &
		\img[trim=6mm 0mm 0mm 7mm,clip]{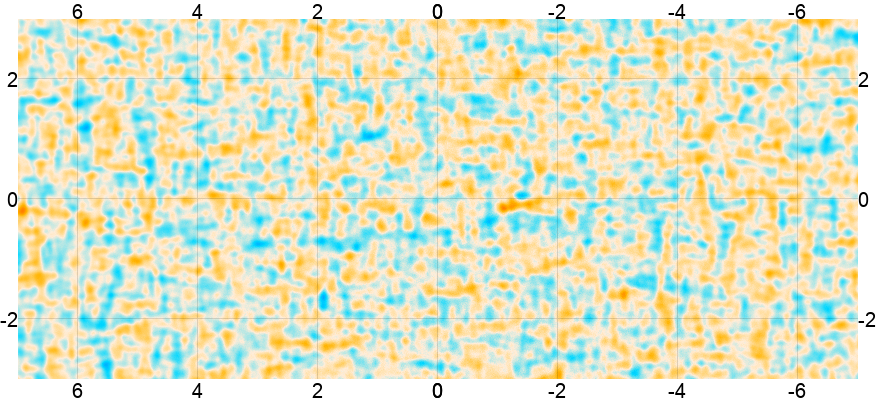} \\
	\end{tabular}
	\end{closetabcols}
	\end{closetabrows}
	\caption{
		\dfn{Top}: The simulated signal map, smoothed to 1.4 arcmin FWHM resolution.
		\dfn{Row 2-4}: The mean of the 4 data realizations for dataset 1-3. The spatially
		dependent stripy noise of the ACT-like datasets 1 and 2 is clearly visible.
		The \planck-like dataset 3 is 5 times lower resolution and 10 times noisier at
		high $\ell$, but does not suffer from excess noise at low $\ell$.
		\dfn{Bottom}: The coadded map combines the best properties of the input datasets.
		It is high resolution, deep, and has almost no stripy noise.
		The two columns are total intensity (left) and Stokes-Q (right). Stokes-U would be
		similar to Stokes-Q, but was left out to save space.}
	\label{fig:sim-maps}
\end{figure}

\begin{figure}[ht]
	\centering
	\begin{closetabrows}[0.6]
	\begin{closetabcols}
	\begin{tabular}{>{\centering\arraybackslash}m{2mm}>{\centering\arraybackslash}m{8.5cm}>{\centering\arraybackslash}m{8.5cm}}
		& T ($\pm 400$\textmu{}K) & Q ($\pm 40$\textmu{}K) \\
		\vspace*{-17mm}\rotatebox[origin=c]{90}{Coadd noise $\times 10$} &
		\img[trim=0mm 13mm 13mm 0mm,clip]{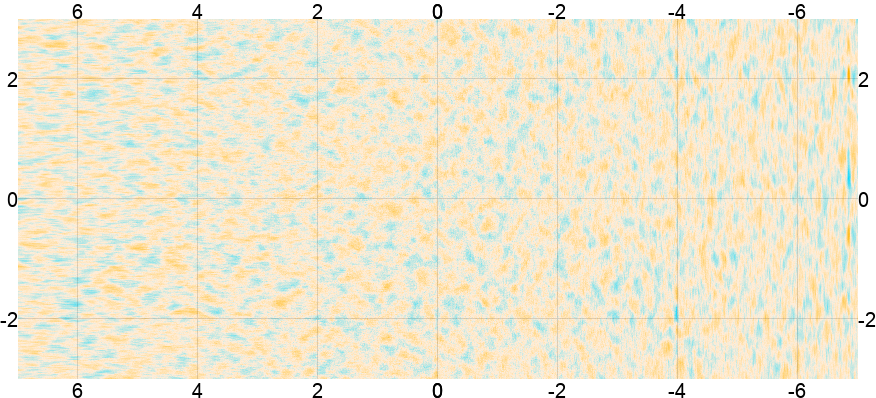} &
		\img[trim=13mm 13mm 0mm 0mm,clip]{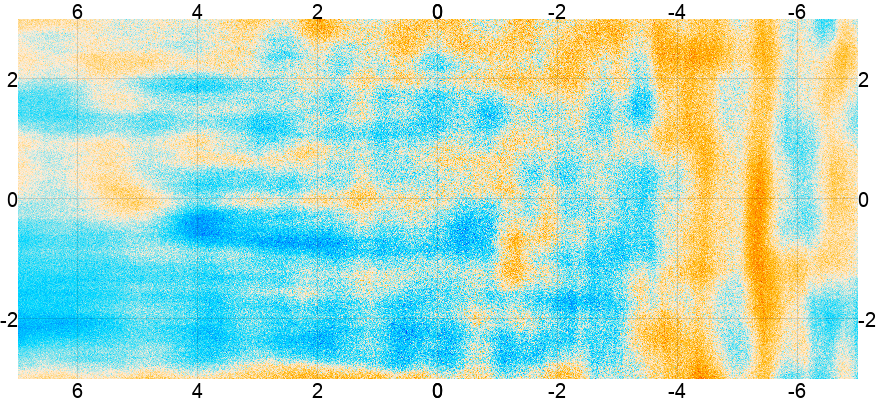} \\
		\vspace*{-20mm}\rotatebox[origin=c]{90}{Coadd bias $\times 1000$} &
		\img[trim=0mm 0mm 13mm 13mm,clip]{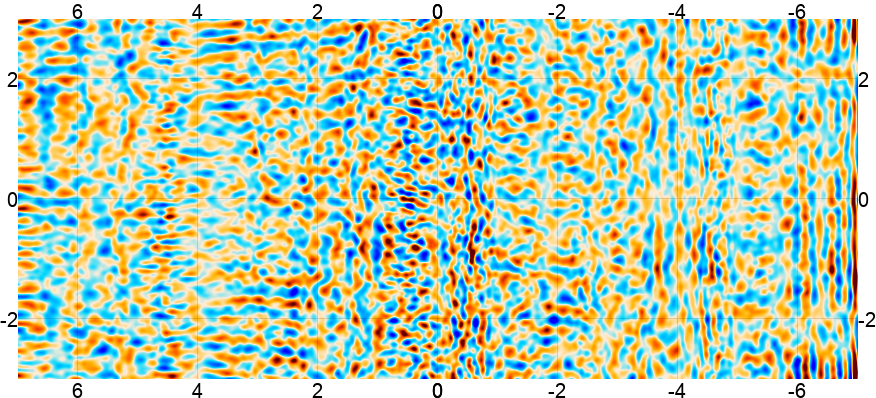} &
		\img[trim=13mm 0mm 0mm 13mm,clip]{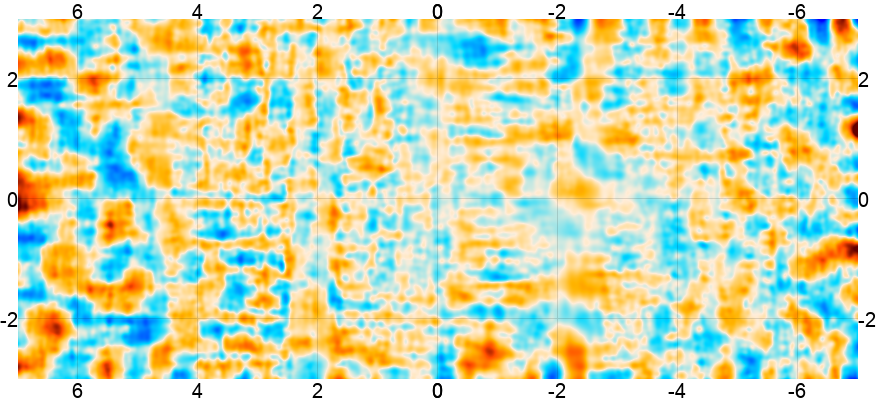} \\
	\end{tabular}
	\end{closetabcols}
	\end{closetabrows}
	\caption{Illustration of the noise and bias properties of the coadded map.
	\dfn{Top}: The difference between the coadd map and the true signal (i.e. between the
	bottom and top panel in figure~\ref{fig:sim-maps}), showing the noise
	properties of the coadd. Scaled by a factor of 10 to make it easier to compare to the
	signal maps in figure~\ref{fig:sim-maps}. The total intensity (left) noise has almost no
	stripiness or spatial dependence, unlike the input maps. The polarization noise (right)
	retains more of the large-scale correlated noise due to dataset 3 being too noisy to help
	much here.
	\dfn{Bottom}: The difference between the low-noise coadd map and the true signal, showing the
	bias properties of the coadd. Scaled by a factor 1000 for visualization purposes.
	}
	\label{fig:sim-diffs}
\end{figure}

We ran the three datasets through the full noise modelling and coadding procedure described in this paper,
with the exception that the ground filtering was turned off.
The input signal, individual dataset maps and the resulting coadded map are shown in figure~\ref{fig:sim-maps},
and examples of the noise and bias are shown in figure~\ref{fig:sim-diffs}. As expected, the coadded map has
the best properties of each of the input maps, with the high resolution and low high-$\ell$ noise of dataset 1
and 2, and the low low-$\ell$ noise of dataset 3. What correlated noise is present in the coadd is much more isotropic
than the very stripy noise in dataset 1 and 2. No trace of the tile structure is visible.

\begin{figure}[ht]
	\centering
	\begin{closetabcols}
	\hspace*{-12mm}\begin{tabular}{cc}
		TT & EE \\
		\includegraphics[width=9.5cm,trim=6mm 0 2mm 0]{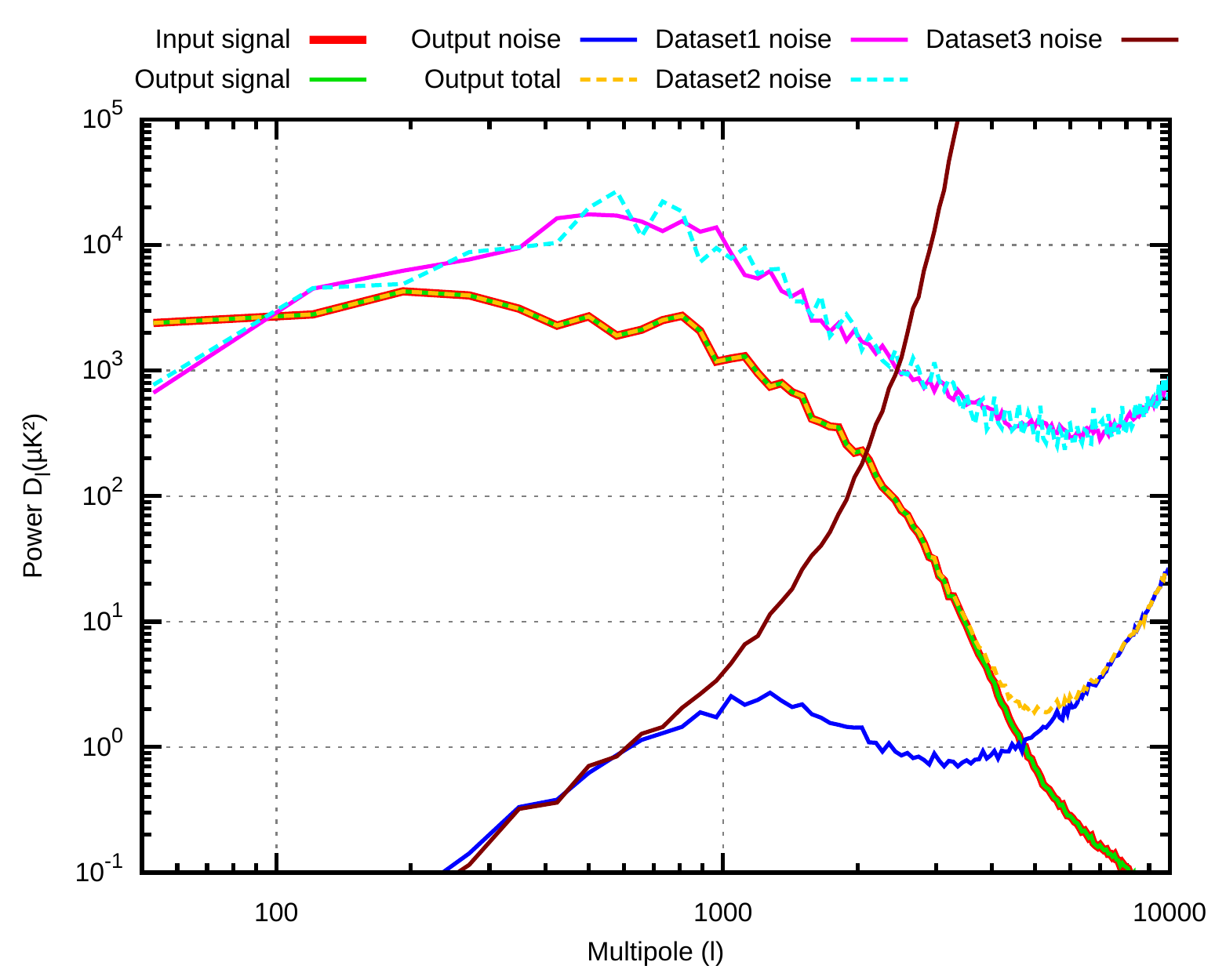} &
		\includegraphics[width=9.5cm,trim=6mm 0 2mm 0]{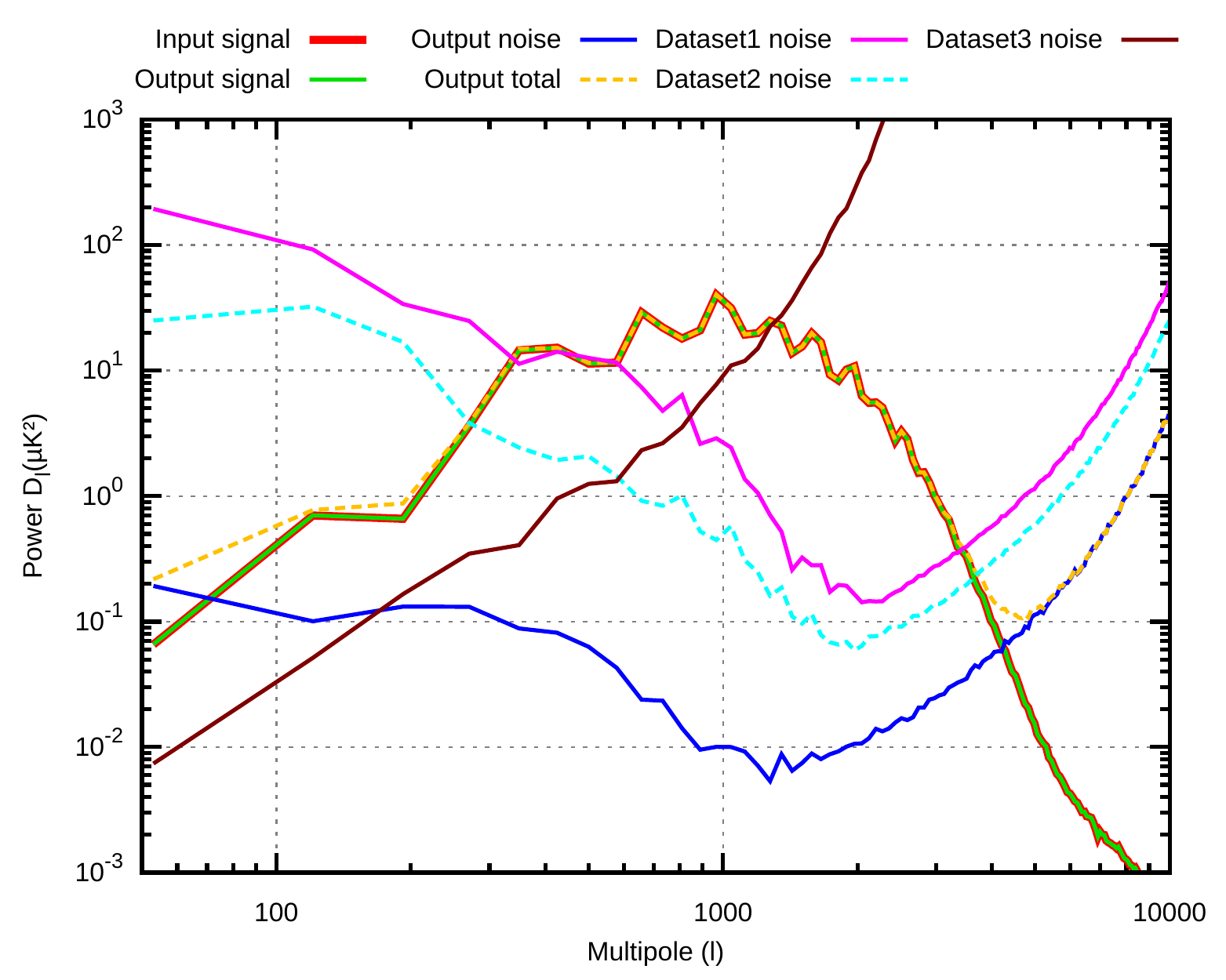}
	\end{tabular}
	\end{closetabcols}
	\caption{Power spectra of the maps. \dfn{Left}: The TT power spectrum of the true signal (red)
	is indistinguishable from that of the low-noise coadd (green). The (normal)
	coadd's noise power spectrum (blue) is always lower than those of the individual
	datasets (magenta, cyan and brown). The large improvement in the area dominated by
	dataset 1 and 2 at $\ell > 1000$ is due to the coadd greatly reducing the amount of
	stripy noise. The yellow curve (output total) is simply the sum of the coadd's signal
	and noise power spectra.
	\dfn{Right}: The EE spectra.}
	\label{fig:sim-specs}
\end{figure}

The bias is about 1000 times smaller than the signal, and is caused by the nonzero Conjugate Gradients convergence
tolerance. For this test we stopped iteration when the residual variance fell below $10^{-7}$ of the value it had
after the first iteration. Lower biases could be achived with a lower tolerance, at a significant performance cost.
For the real data coadds we used a higher tolerance of $10^{-4}$, which resulted in much faster convergence
while still keeping the bias below 1\% of the signal.
The excellent signal fidelity and low noise of the coadded map is also visible in the power spectra, as
shown in figure~\ref{fig:sim-specs}.

\begin{figure}[ht]
	\centering
	\includegraphics[width=12cm]{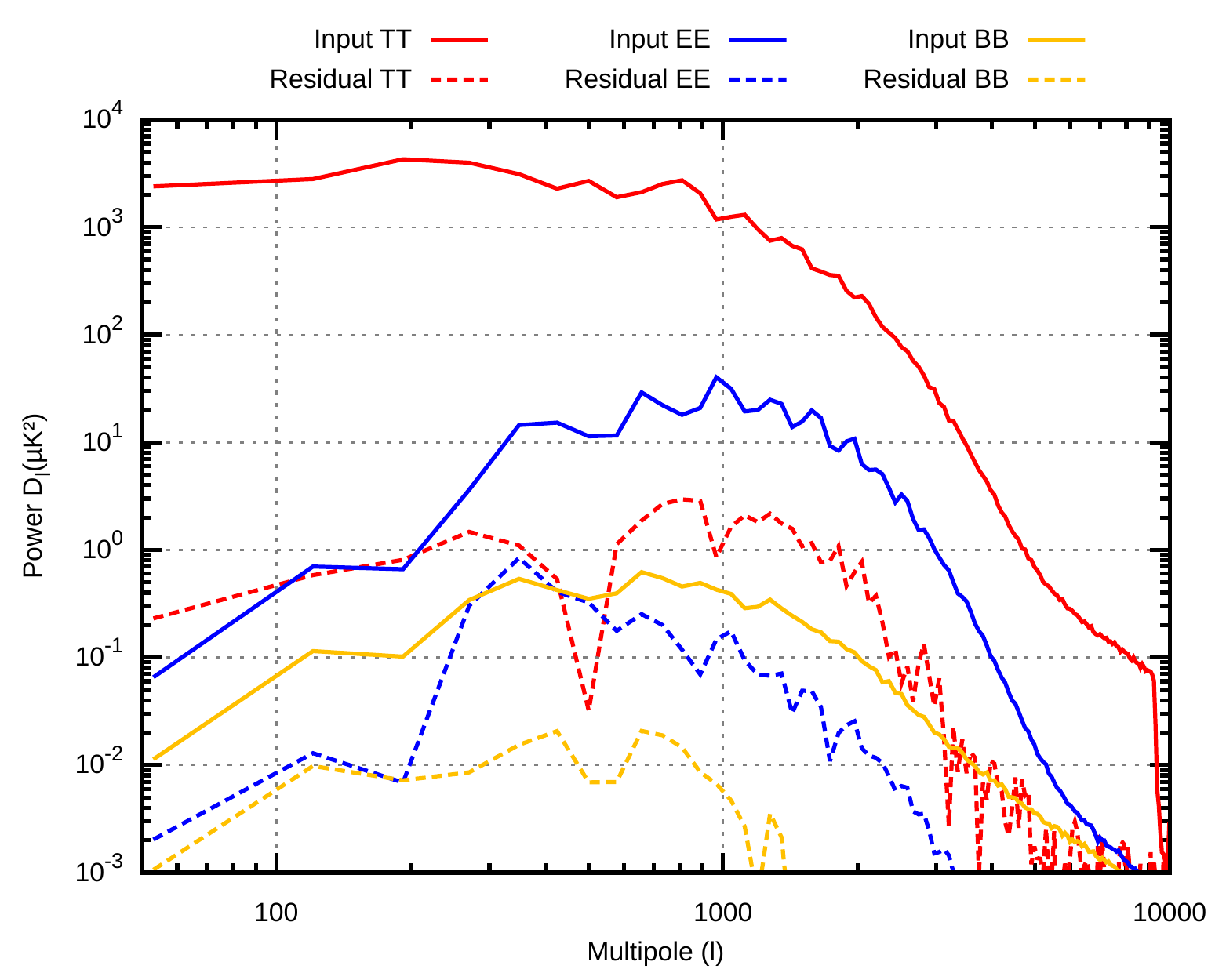}
	\caption{The bias introduced by applying ground filtering to the ACT-like datasets
	1 and 2, but not the \planck-like dataset 3. The solid lines show the spectra of the
	simulated signal. The dashed lines show the residual power, defined as the absolute
	value of the difference between the power spectra of the simulated map and the output
	map from the low-noise coadd. TT has low bias on large scales due to
	the dominance of dataset 3 there, but has a 0.4\% power deficiency for $\ell \gtrapprox 2000$.
	EE and BB behave similarly at high $\ell$, but are more strongly affected for $\ell < 1000$
	due to the relatively lower weight from the unfiltered dataset 3 there, reaching up to a
	10\% error at $\ell < 350$. These biases can be avoided by replacing simple filtering
	with maximum-likelihood weighting, which we will do in a future version of these combined maps.}
	\label{fig:sim-filter-bias}
\end{figure}

After confirming that the coadding procecure itself has negligible bias, we performed a second
run with the ground filter turned on for the ACT-like datasets 1 and 2. We expect this ground
cleaning to introduce a bias because it is implemented as a simple filter instead of being incorporated
as a weight in the maximum-likelihood framework, and this is confirmed in figure~\ref{fig:sim-filter-bias},
which shows a $\sim$ 0.4\% bias for all spectra for $\ell > 1000$. At lower $\ell$ this bias falls to
0.05\% for TT but grows to up to 10\% for EE and BB. This difference is due to filtered datasets 1 and 2 having
high noise at low $\ell$ in total intensity, leading to the unfilered dataset 3 dominating there, and
hence suppressing the bias.

A future version of these maps will replace ground filtering with maximum-likelihood downweighting,
which should eliminate this bias, but for now the user should be aware of its presence, which serves
as one of the reasons why these maps should not be used for precision cosmology.

\end{document}